\documentclass[10pt]{iopart}

\usepackage{iopams}
\usepackage{amsfonts} 
\usepackage{epsf}
\usepackage{euscript}
\usepackage{graphicx}
\usepackage{color}
\usepackage{fancyhdr} 
\usepackage[indentafter]{titlesec}
\usepackage{etoolbox}
\usepackage{hyperref} 
\usepackage{lipsum}
\usepackage{tocloft}
\usepackage{bbold}
\usepackage{bbm}
\usepackage{scalerel}
\usepackage{xcolor}
\DeclareMathAlphabet\mathbfcal{OMS}{cmsy}{b}{n}
\DeclareMathAlphabet{\mathpzc}{OT1}{pzc}{m}{it}
\usepackage{mathrsfs}
\usepackage{mathabx}
\usepackage{amsthm} 
\theoremstyle{plain} 
 
\theoremstyle{definition} 
 
\theoremstyle{remark} 
\newtheorem{remark}{Remark} 
\begin{document}
 
\title[Asymptotic Approximations for the Phase Space Schr\"{o}dinger Equation]{Asymptotic Approximations for the\\
 Phase Space Schr\"{o}dinger Equation}

\author{Panos D Karageorge$^1$\footnote{\texttt{pkarag}@\texttt{uoc.gr}} and George N Makrakis$^{1,2}$\footnote{\texttt{makrakg}@\texttt{uoc.gr} and \texttt{g.n.makrakis}@\texttt{iacm.forth.gr}}}
\address{$^1$Department of Mathematics and Applied Mathematics, University of Crete, Voutes Campus, 700 13 Heraklion, Greece}
\address{$^2$Institute of Applied and Computational Mathematics, Foundation for Research and Technology, 100 Nikolaou Plastira, 700 13 Heraklion, Greece}
\begin{abstract}
We consider semi-classical time evolution for the phase space Schr\"{o}dinger equation and present two methods of constructing short time asymptotic solutions. The first method consists of constructing a semi-classical phase space propagator in terms of semi-classical Gaussian wave packets on the basis of the Anisotropic Gaussian Approximation, related to the Nearby Orbit Approximation, by which we derive an asymptotic solution for configuration space WKB initial data. The second method consists of constructing a phase space narrow beam asymptotic solution, following the Complex WKB Theory developed by Maslov, on the basis of a canonical system in double phase space related to the Berezin-Shubin-Marinov Hamilton-Jacobi and transport equations. We illustrate the methods for sub-quadratic potentials in $\mathbb{R}$.
\end{abstract}

\normalsize

%%%%%%%%%%%%%%%%%%%%%%%%%
%\medskip
%\centerline{\today}
%%%%%%%%%%%%%%%%%%%%%%%%%

\tableofcontents

\section{Introduction}
\label{intro}

\subsection{The General Setting}

Phase space formulations of Quantum Mechanics constitute the sufficient theoretical frame for the description of microscopic physical processes strongly influenced by their external environment, by representing mixed quantum states in terms of phase space quasi-probability distributions, which are used to express expectation values or classical energy densities and fluxes as phase space averages \cite{ZFC}. Despite the increase in complexity of these formulations, compared to the Schr\"{o}dinger representation of Quantum Mechanics, in configuration space, with issues arising such as non-uniqueness of phase space quasi-densities, doubling of variables, non-locality of evolution equations, etc., they prove worth studying in themselves, even outside the context of the theory of Open Quantum Systems, for a series of reasons. 

The phase space is the conceptually natural setting of Quantum Mechanics, more so on its border with Classical Mechanics, the \textit{semi-classical regime,} the range of motions of physical systems for which the Correspondence Principle becomes manifest \cite{BeSh}, and the correspondence between the two theories becomes transparent. Besides the important conceptual reasons for a phase space formulation of Quantum Mechanics, there is a significant methodological reason which dictates such a choice, even for the study of closed physical systems. This is the inherent shortcoming of conventional semi-classical methods to provide global asymptotic solutions for the problem of semi-classical time evolution in a direct fashion, yielding local semi-classical solutions exhibiting singularities at finite times, void of physical content, due to the development of caustics \cite{MaFe}. A theory of the semi-classical Cauchy problem for the Schr\"{o}dinger equation, the theory of the \textit{canonical operator} of Maslov \cite{Mas1,MaFe,Mas2}, overcomes this problem by constructing local semi-classical solutions and `patching' them together, transforming between position and momentum spaces by means of the semi-classical Fourier transform. A semi-classical approach in the context of phase space Quantum Mechanics, placing positions and momenta on equal footing should, in principle, provide a more direct solution to this problem. Caustics become persistent obstacles toward global semi-classical asymptotic solutions of the Schr\"{o}dinger equation in position space and momentum space, alike, while in phase space formulations singularity formation due to caustics is resolved. Another technical reason is that in the context of phase space formulations, a unified approach to the two main classes of semi-classical quantum states, i.e., \textit{WKB} or \textit{Lagrangian states} and \textit{coherent states} or \textit{Gaussian wave packets,} is plausible \cite{HHL}.

The \textit{wave packet representation} is a particular phase space formulation of Quantum Mechanics, first proposed by Torres-Vega and Frederick \cite{ToFr} and thereafter by Harriman \cite{Har}, developed by Chruscinski and Mlodawski \cite{ChMl}, de Gosson \cite{Gos5} and Nazaikinskii \cite{NaSt}, to mention some contributions. In its core is the correspondence of pure classical states, i.e., single phase space points, to \textit{semi-classical isotropic Gaussian wave packets}, or \textit{coherent states}, localized at that phase space point on the Heisenberg scale, $O(\hbar^{1/2})$, 
\begin{equation*}
G_{(\boldsymbol{q},\boldsymbol{p})}(\boldsymbol{x};\hbar)=(\pi\hbar)^{-d/4}\,\exp\frac{i}{\hbar}\Big(\frac{\boldsymbol{p}\cdot\boldsymbol{q}}{2}+\boldsymbol{p}\cdot (\boldsymbol{x}-\boldsymbol{q})+\frac{i}{2}|\boldsymbol{x}-\boldsymbol{q}|^2\Big) 
\end{equation*}
where the point $(\boldsymbol{q},\boldsymbol{p})\in\mathbb{R}^{2d}$ is called the \textit{base point} of the wave packet.

In the wave packet representation, the configuration space wavefunction is expressed as a wave packet superposition for the totality of base phase space points
\begin{equation*}
\psi(\boldsymbol{x};\hbar)=\Big(\frac{1}{2\pi\hbar}\Big)^{d/2}\int_{\mathbb{R}^{2d}} \Psi(\boldsymbol{q},\boldsymbol{p};\hbar) \,G_{(\boldsymbol{q},\boldsymbol{p})}(\boldsymbol{x};\hbar)\, d\boldsymbol{q}d\boldsymbol{p}
\end{equation*}
where the coefficient of the superposition is defined as the \textit{phase space wavefunction}, $\Psi(\boldsymbol{q},\boldsymbol{p};\hbar)$, given, in turn, by
\begin{equation*}
\Psi(\boldsymbol{q},\boldsymbol{p};\hbar)=\Big(\frac{1}{2\pi\hbar}\Big)^{d/2}\int_{\mathbb{R}^{d}}{\widebar G}_{(\boldsymbol{q},\boldsymbol{p})}(\boldsymbol{x};\hbar)\psi(\boldsymbol{x};\hbar)\, d\boldsymbol{x} \ .
\end{equation*}
The above inverse of the wave packet resolution defines the \textit{wave packet transform}, also known as the \textit{Fourier-Bros-Iagolnitzer transform}, closely related to the \textit{Bargmann transform} \cite{Bar,CoFe,NaSt}. 

The dynamics of the phase space wavefunction satisfies the non-local \textit{phase space Schr\"{o}dinger equation} \cite{Gos1,Gos2,Gos5,Har,ToFr}. As a phase space formulation founded on the wave packet transform, it is simpler than the Wigner-Weyl formulation, insofar as it is a linear representation of Quantum Mechanics; it does, however, have its drawbacks, as it cannot account for microscopic systems in significant interaction with their environment, nor does it escape the essential singularity of the semi-classical limit. 

In the wave packet representation, the phase space Schr\"{o}dinger spectral problem is well understood, e.g., through the works of Luef and de Gosson \cite{Gos3}, who have derived the Schr\"{o}dinger spectral equation departing from the spectral equation of Moyal. However, understanding the corresponding \textit{Cauchy problem} for the phase space Schr\"{o}dinger equation, in particular, the problem of \textit{semi-classical} time evolution, remains terra incognita. The deeper understanding of semi-classical time evolution toward a direct theory of semi-classical dynamics in phase space is a challenging and substantial contribution to the theory of phase space formulations of Quantum Mechanics, unifying approaches taken from the theory of semi-classical Fourier integral operators \cite{NOSS} to the theory of the Maslov canonical operator \cite{Mas1,MaFe,Mas2}. Important contributions in this directions have been made, for example, by Oshmyan et al. and Nazaikinksii et al. \cite{NSS,NaSt}.

The wave packet transform has been implemented in other problems in differential equations besides the Schr\"{o}dinger equation, in a variety of settings, where technical issues of diverse nature arise. It has been applied, for example, to problems for the wave equation \cite{CoFe,GeTa}.

In this article, we focus our attention to the problem of semi-classical time evolution \textit{directly} in phase space, in particular, to semi-classical Fourier integral representations and narrow beam solutions of the Cauchy problem for the \textit{phase space Schr\"{o}dinger equation,} rather than giving phase space representations of semi-classical solutions of the Schr\"{o}dinger equation.

A fundamental semi-classical approximation of quantum dynamics is the \textit{Gaussian Approximation}, which pertains to approximating quantum evolution of position-momentum localized quantum states with the dynamics of an \textit{individual} Gaussian wave packet translated along a Hamiltonian orbit. This proposition is traced back to the foundational work of Schr\"{o}dinger \cite{Schr}. Semi-Classical Gaussian wave packets provide a natural semi-classical approximation of quantum states, for free motion, as they are localized in phase space on the Heisenberg scale $O(\hbar^{1/2})$, occupying a Planck cell centered at that point, and exhibiting oscillations at the de Broglie wavelength $O(\hbar)$. For a general account on coherent states see \cite{CoRo}. 

The theory of semi-classical wave packet dynamics has profoundly evolved thereafter. Heller \cite{Hel} and Heller et al. \cite{Hub} argued for the use of single isotropic and anisotropic Gaussian wave packets as an approximation to the propagation of initially semi-classical Gaussian wave packets, based on the semi-classical \textit{Nearby Orbit Approximation}, while Huber et al. \cite{HHL} showed that isotropic Gaussian wave packet dynamics can stand as a generalization of complex phase WKB semi-classical propagation. An analogous systematic work on isotropic Gaussian wave packet dynamics from the viewpoint of the work of Maslov, is that of Bagrov et al. \cite{BBT}, while there are other works along these lines, such as that of Robert \cite{Rob1}, Nazaikinskii et al. \cite{NSS}, de Gosson \cite{Gos6} and Faure \cite{Fau}. Hagedorn \cite{Hag} showed that an initially Gaussian state retains its Gaussian form within a certain semi-classical timescale, under quantum dynamics. A review on the subject of semi-classical wave packet dynamics, from a physical perspective, touching upon its dynamical and algebraic aspects, is given by Littlejohn \cite{Lit1}.

These ideas have been implemented in the solution of the more general problem of establishing asymptotic solutions of the Schr\"{o}dinger equation or the wave equation, along a given curve, modulated by a Gaussian profile, known as \textit{Gaussian beams} or \textit{narrow beams.} Beginning with the ground-breaking work of Babich and Danilov \cite{BaDa}, who considered asymptotic solutions of the Schr\"{o}dinger equation concentrated along a reference curve, a programme of semi-classical techniques was born. A rigorous account on Gaussian beams for hyperbolic equations is given by Ralston \cite{Ral} and Katchalov et al. \cite{KKL}. Gaussian beams find applications in a wide range of physical problems, such as in Acoustics \cite{Kat,QiYi}.

In the case of anisotropic Gaussian wave packets, we have the following ansatz for semi-classical time evolution \cite{BBT,NSS,Rob1}
\begin{eqnarray*}
\fl G^{\scriptsize{\mathbfcal{Z}}}_{(\boldsymbol{q},\boldsymbol{p})}(\boldsymbol{x},t;\hbar)=(\pi\hbar)^{-d/4}a(\boldsymbol{q},\boldsymbol{p},t)\\ \nonumber
\times\exp\frac{i}{\hbar}\Big(\frac{\boldsymbol{p}\cdot\boldsymbol{q}}{2}+ A (\boldsymbol{q},\boldsymbol{p},t)+\boldsymbol{p}_t\cdot (\boldsymbol{x}-\boldsymbol{q}_t)+\frac{1}{2}(\boldsymbol{x}-\boldsymbol{q}_t)\cdot \mathbfcal{Z}(\boldsymbol{x}-\boldsymbol{q}_t)\Big)
\end{eqnarray*} 
where $(\boldsymbol{q}_t,\boldsymbol{p}_t)$ is the image of the point $(\boldsymbol{q},\boldsymbol{p})$ under the Hamiltonian flow, demanding it to be a semi-classical asymptotic solution of the Cauchy problem \cite{Rob1}
\begin{equation*}
\Big\|\Big(i\hbar\,\frac{\partial}{\partial t}-\widehat H\Big)G^{\scriptsize{\mathbfcal{Z}}}_{(\boldsymbol{q},\boldsymbol{p})}(\bullet,t;\hbar)\Big\|_{L^2(\mathbb{R}^{d})}=O(\hbar^{3/2}) \ , \ \ \hbar\rightarrow 0^+
\end{equation*}
for a fixed time interval. The anisotropy matrix $\mathbfcal{Z}$, which satisfies the symmetry and positivity properties $\mathbfcal{Z}^{T}=\mathbfcal{Z}$ and ${\rm Im}\, \mathbfcal{Z}\succ 0$, is shown to obey a certain matrix Riccati equation, equivalent to the dynamics of initially nearby orbits. These dynamics are common to all Gaussian beam asymptotic solutions \cite{KKL,Kel,LiRa,QiYi,Zel1,Zel2}. 

In \cite{Lit1} Littlejohn generalized the works of Heller et al. on the thawed, or anisotropic, Gaussian dynamics for generic initially localized states, by constructing a phase space propagator, as an explicit composition of Weyl shifts and metaplectic operators.

The Gaussian approximation, however, breaks down as localization is lost in an irreversible spreading at a certain semi-classical timescale \cite{Rob1,SVT}, the \textit{Ehrenfest time-scale,} an effect suppressed only for quadratic scalar potentials. Added to the above, the fact that for quadratic potentials the evolution of wave packet \textit{superpositions} of phase space eigenfunctions result in expressions reminiscent of the evolution of single Gaussian wave packets, hints the method of approximating the evolution of a quantum state \textit{by a superposition of semi-classically propagated Gaussian wave packets}. 

We construct a semi-classical phase space propagator based on the \textit{Anisotropic Gaussian Approximation,} which is closely related to the \textit{Nearby Orbit Approximation.} As the Anisotropic Gaussian Approximation is applied for the totality of orbits of the underlying Hamiltonian flow, all of which are taken into account in the wave packet resolution of the phase space wavefunction, the semi-classical propagator admits generic initial data, not just localized ones.

The starting point of this approximation is the resolution of the identity in quantum state space in the over-complete set of coherent states, in particular isotropic Gaussian wave packets, 
\begin{equation*}
\Big(\frac{1}{2\pi\hbar}\Big)^{d}\int G_{(\boldsymbol{q},\boldsymbol{p})} \langle G_{(\boldsymbol{q},\boldsymbol{p})},\bullet \rangle \,d\boldsymbol{q}d\boldsymbol{p}= \mathbb{1}_{L^2} 
\end{equation*} 
by which we obtain the following representation for the Schr\"{o}dinger flow
\begin{equation*}
U_t \bullet=\Big(\frac{1}{2\pi\hbar}\Big)^{d}\int U_t G_{(\boldsymbol{q},\boldsymbol{p})}\langle G_{(\boldsymbol{q},\boldsymbol{p})},\bullet \rangle \,d\boldsymbol{q}d\boldsymbol{p} \ .
\end{equation*}
The approximation itself amounts to an explicit ansatz for the semi-classical propagation of a single wave packet under the Schr\"{o}dinger propagator \cite{NSS,Rob1}, $U_t G_{(\boldsymbol{q},\boldsymbol{p})}$.

The totality of approximations involving single Gaussian dynamics amount to the \textit{Initial Value Representations} of quantum dynamics, closely related to the theory of semi-classical Fourier Integral Operators.

The first work in this direction was that of Herman and Kluk \cite{HeKl}, who argued on the validity of approximating semi-classical evolution by analyzing wavefunctions by a multitude of non-spreading isotropic Gaussian wave packets, their form held rigid, modulated by some overall amplitude and phase factor, setting off from the van Vleck approximation of the semi-classical propagator. More recently, Rousse and Robert \cite{Rob2,RoSw} assumed a semi-classical time evolution for generic initial data of the Schr\"{o}dinger equation, in terms of a certain semi-classical Fourier integral operator, which is readily identified with the Herman-Kluk propagator, in order to justify this approximation on the basis of estimates for the asymptotic solutions. Other works in the direction of a direct theory of time evolution in the frame of linear representations of the Schr\"{o}dinger equation include the work of Almeida et al. \cite{AlBr}

Besides the traditional field of application of Initial Value Representations in semi-classical schemes, such as Quantum Chemistry, groundbreaking progress in areas such as Quantum Optics (see, e.g., \cite{ZSM}), in atomic optical trapping, bore new interest in Initial Value Representations, and phase space representations in general. In such applications, one is able to generate optical traps or scatterers by multiple LASER pulses, well approximated by linear or parabolic potentials. Dynamics are simplified by additional techniques of LASER cooling, enabling one to focus on the overall orbital motion, by suppressing internal degrees of freedom to their ground quantum states. We also note the work of \cite{CTY} in the field of Theoretical Seismology, where an asymptotic wave group was constructed for the high frequency Cauchy problem for the wave equation, as a model of high frequency acoustic wave propagation in a small depth sub-terrain inhomogeneous medium, by means of the Isotropic Gaussian Approximation.

The scope of this paper is to construct semi-classical approximations to the Cauchy problem for the Weyl-symmetrized phase space Schr\"{o}dinger equation (Section \ref{PSS})
\begin{equation*}
i\hbar\,\frac{\partial \Psi}{\partial t}=H\Big(\frac{\boldsymbol{q}}{2}+i\hbar\,\frac{\partial}{\partial {p}},\frac{\boldsymbol{p}}{2}-i\hbar\,\frac{\partial}{\partial \boldsymbol{q}}\Big)\Psi
\end{equation*}
for Weyl ordering of the non-commuting operator arguments. The initial data are prepared as the wave packet transform of a standard WKB function in configuration space.

We construct two different approximations. 

The first has the form of semi-classical Fourier integral, and it is derived by the action of an approximate semi-classical propagator on the initial data (Section \ref{solutions}). This propagator is constructed in Section \ref{AGAP} by analyzing the exact phase space propagator in a superposition of semi-classical Gaussian wave packets by the wave packet transform, and propagating each one, individually, according to the \textit{Anisotropic Gaussian Approximation}. We consider the evolution as a superposition of propagated wave packets, and consequently express the semi-classical phase space propagator in terms of propagated and non-propagated wave packets.

Further approximation of the Fourier integral by the \textit{Complex Stationary Phase methods} suggests the consideration of a WKB ansatz with complex phase in phase space, and consider such an ansatz as an independent approximation (Section \ref{nbsolution}). By the calculus of exponential asymptotic equivalence, it turns out that the amplitude and the phase of the ansatz must satisfy a canonical system that comprises of a \textit{Weyl-symmetrized Hamilton-Jacobi equation,} and a \textit{Weyl-symmetrized transport equation,} similar to the corresponding equations considered by Marinov \cite{Mar} and Berezin and Shubin \cite{BeSh}. Such a system has been studied by Maslov \cite{Mas2} and leads to the construction of a narrow beam solution in phase space.

Finally, we give detailed illustrations of the methods for the short time evolution for scalar sub-quadratic potentials on the real line, which model low energy motion of electrons influenced by simple electrostatic fields, such as free motion, scattering off a constant electrostatic field and bound motion by a parabolic optical trap.

\subsection{Assumptions and Notational Conventions}
\begin{table*}
	\centering
		\begin{tabular}{|l|l|} 
$\boldsymbol{X}=(\boldsymbol{q},\boldsymbol{p}),\boldsymbol{Y}=(\boldsymbol{\eta},\boldsymbol{\xi})$ & \scriptsize{phase space points} \\		
$(\boldsymbol{X},\boldsymbol{P})=(\boldsymbol{q},\boldsymbol{p},\boldsymbol{u},\boldsymbol{v})$ & \scriptsize{double phase space points} \\		
$H$ & \scriptsize{Hamiltonian function} \\
$g_t$ & \scriptsize{Hamiltonian flow in phase space generated by $H$} \\
$\boldsymbol{X}_t=g_t \boldsymbol{X}=(\boldsymbol{q}_t,\boldsymbol{p}_t)$ & \scriptsize{propagation of $\boldsymbol{X}=(\boldsymbol{q},\boldsymbol{p})$ under Hamiltonian flow} \\
$\mathfrak{F}$ & \scriptsize{Fock-Bargmann space} \\
$\psi$ & \scriptsize{configuration space wavefunction} \\ 
$\Psi$ & \scriptsize{phase space wavefunction} \\
$\mathcal{W}$ & \scriptsize{wave packet transform} \\
$\sigma(L)$ & \scriptsize{symbol of operator $L$ according to a certain quantization}\\
$\widehat F$ & \scriptsize{Weyl quantization of physical quantity} $F$\\
$\widehat \mathcal{F}$ & \scriptsize{wave packet representation of Weyl quantization of physical quantity} $F$ \\
$G_{\boldsymbol{X}}(\boldsymbol{x};\hbar)$ & \scriptsize{semi-classical isotropic Gaussian wave packet} \\ 
$\mathbfcal{Z}(\boldsymbol{q},\boldsymbol{p},t)$ & \scriptsize{anisotropy matrix} \\
$\textbf{A}(\boldsymbol{q},\boldsymbol{p},t),\textbf{B}(\boldsymbol{q},\boldsymbol{p},t)$ & \scriptsize{position and momentum variational matrices} \\
$\mathbfcal{Q}(\boldsymbol{q},\boldsymbol{p},t)$ & 
\scriptsize{phase space anisotropy matrix} \\
$ {A}(\boldsymbol{q},\boldsymbol{p},t)$ & \scriptsize{phase space action function}\\
$G_{\boldsymbol{X}}^{\scriptsize{\mathbfcal{Z}}}(\boldsymbol{x},t;\hbar)$ & \scriptsize{propagated semi-classical anisotropic Gaussian wave packet} \\
$U_t$ & \scriptsize{propagator}\\
$K(\boldsymbol{x},\boldsymbol{y},t;\hbar)$ & \scriptsize{kernel of} $U_t$\\
$\mathcal{U}_t$ & \scriptsize{phase space propagator} \\ 
$\mathcal{K}(\boldsymbol{q},\boldsymbol{p},\boldsymbol{\eta},\boldsymbol{\xi},t;\hbar)$ & \scriptsize{kernel of} $\mathcal{U}_t$\\
$U_{t}^{\scriptsize{\mathbfcal{Z}}}$ & \scriptsize{semi-classical approximation of $U_t$} \\ 
$K^{\scriptsize{\mathbfcal{Z}}}(\boldsymbol{x},\boldsymbol{y},t;\hbar)$ &\scriptsize{kernel of} $U_{t}^{\scriptsize{\mathbfcal{Z}}}$\\
$\mathcal{U}_{t}^{\scriptsize{\mathbfcal{Z}}}$ & \scriptsize{semi-classical phase space propagator} \\ 
$\mathcal{K}^{\scriptsize{\mathbfcal{Z}}}(\boldsymbol{q},\boldsymbol{p},\boldsymbol{\eta},\boldsymbol{\xi},t)$ & \scriptsize{kernel of} $\mathcal{U}_t^{\scriptsize{\mathbfcal{Z}}}$\\
$\Lambda_0$ & \scriptsize{Lagrangian manifold in phase space} \\
$\Lambda_t=g_t\Lambda_0$ & \scriptsize{propagated Lagrangian manifold in phase space} \\ 
$\boldsymbol{\alpha}=(\alpha_1,\ldots,\alpha_d)$ & \scriptsize{local co-ordinates on a neighborhood of a point of $\Lambda_t$} \\ 
$\mathcal{H}$ & \scriptsize{phase space Weyl symbol of $\widehat \mathcal{H}$} \\
$G_t$ & \scriptsize{Hamiltonian flow in double phase space generated by $\mathcal{H}$} \\ 
$\mathcal{S}$ & \scriptsize{invariant symplectic plane of $G_t$} \\ 
$L_0$ & \scriptsize{Lagrangian manifold in double phase space} \\ 
$L_t=G_tL_0$ & \scriptsize{propagated Lagrangian manifold in double phase space} \\ 
$\tilde \mathbfcal{Q}(\boldsymbol{q},\boldsymbol{p},t)$ & \scriptsize{phase space narrow beam anisotropy matrix} \\
$\textbf{C}(\boldsymbol{\alpha},t),\textbf{D}(\boldsymbol{\boldsymbol{\alpha}},t)$ & \scriptsize{phase space position and momentum variational matrices} \\
$\Psi^{\scriptsize{\mathbfcal{Z}}}$ & \scriptsize{asymptotic solution of the phase space Schr\"{o}dinger equation} \\ 
$\Psi^\hbar$ & \scriptsize{Fourier integral approximation of $\Psi^{\scriptsize{\mathbfcal{Z}}}$} \\
$\Psi^\hbar_B$ & \scriptsize{narrow beam asymptotic solution of the phase space Schr\"{o}dinger equation} \\

\end{tabular}	
\caption{Reference table of basic symbol notations.}
\label{tab:notation}
\end{table*}

We consider semi-classical time evolution for closed non-relativistic microscopic physical systems comprising of $n$ particles with no spin interactions, such as isolated atomic systems with electrostatic interactions, or closed systems of electron transport in mesoscopic structures under the influence of an electrostatic field.

We assume $\mathbb{R}^d$ as the configuration space with co-ordinates $\boldsymbol{x}=(x_1,\ldots,x_d)$. We denote by $\cdot $ the Euclidean inner product in $\mathbb{R}^d$ or its extension as a bi-linear map in $\mathbb{C}^d$, $|\bullet|$ the Euclidean norm in either $\mathbb{R}^d$ or $\mathbb{C}^d$ or $\mathbb{R}^{2d}$, which one being obvious in the given context, and by $\|\bullet\|$ the natural norm in $L^2(\mathbb{R}^d,\mathbb{C};d\boldsymbol{x})$. 

The phase space is taken as $\mathbb{R}^d\oplus\mathbb{R}^d\cong \mathbb{R}^{2d}$, with canonical co-ordinates $(\boldsymbol{q},\boldsymbol{p})=(q_1,\ldots,q_d,p_1,\ldots,p_d)$, or, collectively $\boldsymbol{X}=(\boldsymbol{q},\boldsymbol{p})$, while we denote for a different point the canonical co-ordinates $(\boldsymbol{\eta},\boldsymbol{\xi})=(\eta_1,\ldots,\eta_d,\xi_1,\ldots,\xi_d)$, or, collectively $\boldsymbol{Y}=(\boldsymbol{\eta},\boldsymbol{\xi})$. As a symplectic space, the phase space is equipped with the symplectic form $\omega(\boldsymbol{X},\boldsymbol{Y})=\boldsymbol{X}\cdot \textbf{J} \boldsymbol{Y}=\boldsymbol{q}\cdot \boldsymbol{\xi}-\boldsymbol{p}\cdot \boldsymbol{\eta}=\sum_{j=1}^d(q_j \xi_j-p_j\eta_j)$, where $\textbf{J}=\left(\begin{array}{ccc} \textbf{0} & \textbf{I} \\ -\textbf{I} & \textbf{0} \end{array} \right)$ is the canonical symplectic matrix with respect to the canonical basis. In the dual of the canonical basis, we denote the canonical 2-form as $\boldsymbol{\omega}^2=d\boldsymbol{p}\wedge d\boldsymbol{q}=\sum_{j=1}^ddp_j\wedge dq_j$, by $\boldsymbol{\omega}^1=\boldsymbol{p}\cdot d\boldsymbol{q}=\sum_{j=1}^dp_j \,dq_j$ the normal canonical 1-form and by $\boldsymbol{\omega}^1_W=\frac{1}{2}(\boldsymbol{p}\cdot d\boldsymbol{q}-\boldsymbol{q}\cdot d\boldsymbol{p})=\sum_{j=1}^d\frac{1}{2}(p_j\, dq_j-q_j \,dp_j)$ the Weyl-symmetrized canonical $1$-form \cite{Gos1}. 

We consider autonomous Hamiltonian systems on phase space, with smooth Hamiltonian function $H$ satisfying the growth condition 
\begin{equation*}\Big|\frac{\partial ^{\boldsymbol{\mu}} H}{\partial \boldsymbol{X}^{\boldsymbol{\mu}}}(\boldsymbol{X})\Big|\leq C_{\boldsymbol{\mu}}\Big(1+|\boldsymbol{X}|\Big)^{M_{|\boldsymbol{\mu}|}}\end{equation*}
for some constants $C_{\boldsymbol{\mu}}>0$ and $M_{|\boldsymbol{\mu}|}\in\mathbb{R}$, for any multi-index $\boldsymbol{\mu}\in\mathbb{N}^{2d}_0$.

Further, for the phase space of the phase space, we coin the term \textit{double phase space}, $\mathbb{R}^{2d}\oplus\mathbb{R}^{2d}\cong \mathbb{R}^{4d}$, with canonical co-ordinates $(\boldsymbol{X},\boldsymbol{P})=(q_1,\ldots,q_d,p_1,\ldots,p_d,u_1,\ldots,u_d,v_1,\ldots,v_d)$.

Operator symbols defined in phase space are denoted by roman lettering while symbols in double phase space by script lettering, e.g., $H$ versus $\mathcal{H}$; these are assumed to satisfy analogous smoothness and growth conditions to those of the Hamiltonian function, as above.

We denote the Hamiltonian flow generated by $H$ by $g_t$ and by $(\boldsymbol{q}_t,\boldsymbol{p}_t)=(\boldsymbol{q}_t(\boldsymbol{q},\boldsymbol{p}),\boldsymbol{p}_t(\boldsymbol{q},\boldsymbol{p})):=g_t(\boldsymbol{q},\boldsymbol{p})$ the terminal point of the orbit with duration $t\geq0$ emanating from the point $(\boldsymbol{q},\boldsymbol{p})$. As for the dynamical properties of the Hamiltonian, we make no assumptions; reversely, we consider how the validity of the semi-classical approximation is affected by such properties.

Further on notation, for complex entry matrices $\textbf{A}=(a_{jk})$ we write $\textbf{A}^ T =(a_{kj})$ for its transpose, $\bar \textbf{A}=(\bar a_{jk})$ for its complex conjugate and $\textbf{A}^*=\bar \textbf{A}^ T $ for its hermitian adjoint. For smooth phase space complex valued functions $f$ we use the notation $\frac{\partial f}{\partial \boldsymbol{q}}=\Big(\frac{\partial f}{\partial q_j}\Big)$ and $\frac{\partial f}{\partial \boldsymbol{p}}=\Big(\frac{\partial f}{\partial p_j}\Big)$, for the column vector of partial derivatives; further, we write $f _{\boldsymbol{qq}}=\Big(\frac{\partial^2f}{\partial q_j\partial q_k}\Big)$, $f _{\boldsymbol{pp}}=\Big(\frac{\partial^2f}{\partial p_j\partial p_k}\Big)$ and $f _{\boldsymbol{pq}}=\Big(\frac{\partial^2f}{\partial q_j\partial p_k}\Big)$, while $f _{\boldsymbol{qp}}=f _{\boldsymbol{pq}}^ T $; for the Hessian matrix we use the block matrix notation $\frac{\partial^2f}{\partial\boldsymbol{X}^2}=f_{\boldsymbol{X}\!\boldsymbol{X}}=\left(
\begin{array}{ccc}
f _{\boldsymbol{qq}} & f _{\boldsymbol{pq}} \\
f _{\boldsymbol{qp}} & f _{\boldsymbol{pp}} 
\end{array}
\right)$, for $\boldsymbol{X}=(\boldsymbol{q},\boldsymbol{p})$. 

Whenever the domain of an integration is not made explicit, the integral is assumed over the whole space, as no use of indefinite integrals are made. 

\section{The Phase Space Schr\"{o}dinger Equation}
\label{PSS}

\subsection{The Wave Packet Transform}
The \textit{wave packet representation} is a linear representation of Quantum Mechanics over phase space, first proposed by Torres-Vega and Frederick \cite{ToFr}, elaborated by other authors subsequently, e.g., Harriman \cite{Har} and de Gosson \cite{Gos5}. It is related to the Schr\"{o}dinger representation by means of a linear unitary operator, the \textit{wave packet transform} \cite{NSS,NaSt,ToFr}, which maps position space wavefunctions, $\psi(\boldsymbol{x})$, to phase space wavefunctions, $\Psi(\boldsymbol{q},\boldsymbol{p})$,
\begin{equation}
\mathcal{W}:L^2(\mathbb{R}^d,\mathbb{C};d\boldsymbol{x})\rightarrow L^2(\mathbb{R}^{2d},\mathbb{C}; d\boldsymbol{q}d\boldsymbol{p})\,|\,\psi\mapsto\Psi=\mathcal{W}\psi \ . 
\end{equation}
It is defined explicitly by the following integral transform
\begin{equation}
\Psi(\boldsymbol{q},\boldsymbol{p};\hbar)=(\mathcal{W}\psi)(\boldsymbol{q},\boldsymbol{p};\hbar)=\Big(\frac{1}{2\pi\hbar}\Big)^{d/2}\int{\widebar G}_{(\boldsymbol{q},\boldsymbol{p})}(\boldsymbol{x};\hbar)\psi(\boldsymbol{x};\hbar)\,d\boldsymbol{x} 
\label{eq:wpt}
\end{equation}
its kernel being the complex conjugated semi-classical isotropic Gaussian wave packet with base point $\boldsymbol{X}=(\boldsymbol{q},\boldsymbol{p})\in\mathbb{R}^{2d}$
\begin{equation}
G_{(\boldsymbol{q},\boldsymbol{p})}(\boldsymbol{x};\hbar)=(\pi\hbar)^{-d/4}\exp\frac{i}{\hbar}\Big(\frac{\boldsymbol{p}\cdot\boldsymbol{q}}{2}+\boldsymbol{p}\cdot (\boldsymbol{x}-\boldsymbol{q})+\frac{i}{2}|\boldsymbol{x}-\boldsymbol{q}|^2\Big) \ .
\label{eq:g}
\end{equation}

As a linear operator between Hilbert spaces, the wave packet transform is not a bijection; its image is a sub-space $\mathfrak{F}\subset L^2(\mathbb{R}^{2d})$,
the \textit{Fock-Bargmann space}, defined by the \textit{Fock-Bargmann constraint} \cite{Bar,Gos2,NSS,NaSt}
\begin{equation}
\Psi\in\mathfrak{F}\,\iff\,\Bigg(\Big(\frac{\boldsymbol{q}}{2}-i\hbar\,\frac{\partial}{\partial \boldsymbol{p}}\Big)-i\Big(\frac{\boldsymbol{p}}{2}+i\hbar\,\frac{\partial}{\partial \boldsymbol{q}}\Big)\Bigg)\Psi=\boldsymbol{0} \ .
\label{eq:FBc}
\end{equation}
This constraint is equivalent to the Cauchy-Riemann relations
\begin{equation}
\Big(\frac{\partial}{\partial \boldsymbol{q}}-i\frac{\partial}{\partial \boldsymbol{p}} \Big)\Big(e^{\frac{1}{2\hbar}(i\boldsymbol{p}\cdot\boldsymbol{q}+|\boldsymbol{p}|^2)}\Psi\Big)=\boldsymbol{0}
\end{equation}
and therefore only Gaussian-weighted square integrable analytic functions in the variable $\boldsymbol{q}-i\boldsymbol{p}\in\mathbb{C}^d$ are admissible phase space wavefunctions. 
See \cite{NSS} for the detailed construction and further properties of the wave packet transform.

\subsection{Weyl Operators in Phase Space}
\label{weylops}

For a symbol $F(\boldsymbol{q},\boldsymbol{p})$ defined on phase space, we consider the one-parameter family of quantizations, $F\mapsto\widehat{F}=\mathbf{{\rm Op}_\lambda}(F)$, for $\lambda\in[0,1]$, mapping the symbol to an operator defined by 
\begin{equation}
\fl \mathbf{{\rm Op} _\lambda}(F)\psi(\boldsymbol{x})=\Big(\frac{1}{2\pi\hbar}\Big)^{d}\int \!\! \int e^{\frac{i}{\hbar}\boldsymbol{p}\cdot (\boldsymbol{x}-\boldsymbol{q})}F\Big(\lambda \boldsymbol{x}+(1-\lambda)\boldsymbol{q},\boldsymbol{p}\Big)\psi(\boldsymbol{q})\,d\boldsymbol{q}d\boldsymbol{p}
\end{equation}
which acts on functions defined on configuration space.

The values $\lambda=0$ and $\lambda=1$ correspond to the \textit{normal} and \textit{anti-normal operators,}\footnote{Indices above operator arguments, termed \textit{Feynman indices,} indicate relative order of action, while the index $\boldsymbol{\omega}$ denotes uniformly ordered action. They are necessary in order to make functions of non-commuting arguments well defined as operators \cite{Fey,Mas1}.} $F\Big(\stackrel{2}{\boldsymbol{x}},-i\hbar\,\stackrel{1}{\frac{\partial}{\partial \boldsymbol{x}}}\Big):=\mathbf{{\rm Op}_n}(F)=\mathbf{{\rm Op}} _0(F)$ 
and $F\Big(\stackrel{1}{\boldsymbol{x}},-i\hbar\,\stackrel{2}{\frac{\partial}{\partial \boldsymbol{x}}}\Big):=\mathbf{{\rm Op}_{an}}(F)= \mathbf{{\rm Op}}_1(F)$, respectively, while the intermediate value $\lambda=\frac{1}{2}$ corresponds to the \textit{Weyl operator,} $F\Big(\stackrel{\boldsymbol{\omega}}{\boldsymbol{x}},-i\hbar\,\stackrel{\boldsymbol{\omega}}{\frac{\partial}{\partial \boldsymbol{x}}}\Big):=\mathbf{{\rm Op}_w}(F)=\mathbf{{\rm Op}} _{1/2}(F)$ (see, e.g., \cite{BeSh,Gos5,Mart}). 
Also, by using the wave packet transform, the \textit{wave packet quantization}, $F \mapsto \mathbf{{\rm Op}_{wp}}(F)$, defines the operator acting on configuration space 
\begin{equation}
\mathbf{{\rm Op}_{wp}}(F):=\mathcal{W}^* F \mathcal{W}
\end{equation}
where care must be taken in the definition of the adjoint, $\mathcal{W}^*$ \cite{NSS},\cite{NaSt}.

Moreover, operators $\widehat F$, resulting as the Weyl quantization of a symbol $F$, acting on some subspace of $L^2(\mathbb{R}^d)$, can be represented in phase space by means of the \textit{wave packet representation,} 
\begin{equation}
\widehat F\mapsto \widehat{\mathcal{F}}=\mathcal{W}\widehat F\mathcal{W}^{-1}
\end{equation}
which will be used in the sequel for posing the phase space Schr\"{o}dinger equation.

For the canonical pair $\boldsymbol{x}$ and $-i\hbar\,\partial/\partial\boldsymbol{x}$, we have
\begin{eqnarray}
\mathcal{W} \boldsymbol{x} \mathcal{W}^{-1}=\frac{\boldsymbol{q}}{2}+i\hbar\,\frac{\partial}{\partial \boldsymbol{p}} \\ \nonumber 
\mathcal{W}\Big(-i\hbar\,\frac{\partial}{\partial \boldsymbol{x}}\Big) \mathcal{W}^{-1}=\frac{\boldsymbol{p}}{2}-i\hbar\,\frac{\partial}{\partial \boldsymbol{q}}
\end{eqnarray}
the resulting pair closely related to the \textit{Bopp shifts} \cite{Bop}. 

Operators defined as functions of Bopp shifts have been thoroughly studied by de Gosson \cite{Gos1,Gos2,Gos7} in the framework of phase space Quantum Mechanics. In particular, the wave packet representation $\widehat{\mathcal{H}}=\mathcal{W}\widehat H \mathcal{W}^{-1}$
of $\widehat H=H\Big(\stackrel{\boldsymbol{\omega}}{\boldsymbol{x}},-i\hbar\,\stackrel{\boldsymbol{\omega}}{\frac{\partial}{\partial \boldsymbol{x}}}\Big)$ has been identified by de Gosson and Luef \cite{Gos3} as the Weyl quantization of the deformed symbol defined in \textit{double phase space}
\begin{equation}
\mathcal{H} (\boldsymbol{X},\boldsymbol{P}):=H\Big(\frac{\boldsymbol{q}}{2}-\boldsymbol{v},\frac{\boldsymbol{p}}{2}+\boldsymbol{u}\Big)= H\Big(\frac{\boldsymbol{X}}{2}-\textbf{J}\boldsymbol{P}\Big) \ .
\label{eq:phsham}
\end{equation}
where $\boldsymbol{X}=(\boldsymbol{q},\boldsymbol{p})$, $\boldsymbol{P}=(\boldsymbol{u},\boldsymbol{v})$, that is
\begin{equation}
\widehat{\mathcal{H}}=\mathcal{W}\widehat H \mathcal{W}^{-1}=\mathbf{{\rm Op}_w}(\mathcal{H}) \ .
\label{eq:hweyl}
\end{equation}

The action of $\widehat{\mathcal{H}}$ on the phase space wavefunction $\Psi$ is given in integral form by
\begin{equation}
\fl \widehat{\mathcal{H}}\Psi(\boldsymbol{X})=\Big(\frac{1}{2\pi\hbar}\Big)^{2d}\int e^{\frac{i}{\hbar}\boldsymbol{P}\cdot (\boldsymbol{X}-\boldsymbol{Y})}\mathbf{\sigma_w}(\widehat{\mathcal{H}})\Big(\frac{\boldsymbol{X}+\boldsymbol{Y}}{2},\boldsymbol{P}\Big)\Psi(\boldsymbol{Y})\, d\boldsymbol{Y}d\boldsymbol{P}
\end{equation}
where the Weyl symbol of the operator $\widehat{\mathcal{H}}$ is 
\begin{equation}
\mathbf{\sigma_w}(\widehat{\mathcal{H}})(\boldsymbol{X},\boldsymbol{P})=\mathcal{H} (\boldsymbol{X},\boldsymbol{P}) \ .
\end{equation}

Thus we can write symbolically
\begin{equation}
\fl \widehat{\mathcal{H}}=\mathcal{H}\Big(\stackrel{\boldsymbol{\omega}}{\boldsymbol{X}},-i\hbar \stackrel{\boldsymbol{\omega}}{\frac{\partial}{\partial \boldsymbol{X}}}\Big)=H\Big(\frac{\boldsymbol{q}}{2}+i\hbar \,\frac{\partial}{\partial \boldsymbol{p}},\frac{\boldsymbol{p}}{2}-i\hbar \,\frac{\partial}{\partial \boldsymbol{q}}\Big)= H\Big(\frac{\boldsymbol{X}}{2}+i\hbar\, \textbf{J}\frac{\partial}{\partial \boldsymbol{X}}\Big) \ .
\end{equation}

For the Weyl and normal operators corresponding to the symbols $\mathbf{\sigma_w}(\widehat{\mathcal{H}})$ and $\mathbf{\sigma_n}(\widehat{\mathcal{H}})$, respectively, to coincide, that is,
\begin{equation}
\widehat{\mathcal{H}}=\mathbf{{\rm Op}_w}(\mathbf{\sigma_w}(\widehat{\mathcal{H}}))=\mathbf{{\rm Op}_n}(\mathbf{\sigma_n}(\widehat{\mathcal{H}}))
\end{equation}
the symbols must satisfy the relation \cite{BeSh,NSS} 
\begin{equation}\label{wnsymb}
\mathbf{\sigma_n} (\widehat{\mathcal{H}})(\boldsymbol{X},\boldsymbol{P})= \exp\Big(-\frac{i\hbar}{2}\frac{\partial^2}{\partial \boldsymbol{X}\partial\boldsymbol{P}}\Big)\mathbf{\sigma_w}(\widehat{\mathcal{H}})(\boldsymbol{X},\boldsymbol{P}) \ .
\end{equation}

By direct computation, we incur the identity 
\begin{equation}
\mathbf{\sigma_n}(\widehat{\mathcal{H}})(\boldsymbol{X},\boldsymbol{P})=\mathbf{\sigma_w}(\widehat{\mathcal{H}})(\boldsymbol{X},\boldsymbol{P})=\mathcal{H}(\boldsymbol{X},\boldsymbol{P})
\end{equation}
by virtue of the special dependence of $\mathcal{H}$ on $(\boldsymbol{X},\boldsymbol{P})$. Thus, we have
\begin{equation}
\widehat{\mathcal{H}}=H\Big(\stackrel{2}{\frac{\boldsymbol{X}}{2}}+i\hbar \,\textbf{J}\stackrel{1}{\frac{\partial}{\partial \boldsymbol{X}}}\Big) \ . 
\end{equation}

\subsection{Derivation of the Phase Space Schr\"{o}dinger Equation}

We begin by considering the semi-classical Cauchy problem for the Schr\"{o}dinger equation in the Weyl quantization, which reads
\begin{equation}
\Big(i\hbar\,\frac{\partial}{\partial t}- \widehat H\Big)\psi(t)=0 \ , \ \ t\in [0,T] \ , \ \ \psi(0)=\psi_0\in L^2(\mathbb{R}^d)
\label{eq:ivp}
\end{equation}
where the Hamiltonian operator $\widehat H =\mathbf{{\rm Op}_{w}}(H)$ is the Weyl quantization of the Hamiltonian function $H$ (see \ref{QuRep}).
 
The solution $\psi$ of the problem is given by the action of the \textit{Schr\"{o}dinger propagator} $U_t$ on the initial data $\psi_0$, through the integral representation \cite{BeSh}
\begin{equation}
\psi (\boldsymbol{x},t;\hbar)= (U_t \psi _0 )(\boldsymbol{x},t;\hbar)=\int K(\boldsymbol{x},\boldsymbol{y},t;\hbar)\psi _0(\boldsymbol{y};\hbar)\,d\boldsymbol{y} \ .
\label{eq:Kprop}
\end{equation}
The kernel $K$ of $U_t $ is a fundamental solution of the Schr\"{o}dinger equation \cite{BeSh} in the sense that 
\begin{equation}
\Big(i\hbar\,\frac{\partial}{\partial t}-\widehat H \Big)K(\boldsymbol{x},\boldsymbol{y},t;\hbar)=i\hbar\,\delta(t)\delta(\boldsymbol{x}-\boldsymbol{y}) \ , \ \ t\in[0,T]
\label{eq:fund}
\end{equation}
where the operator acts on the first argument, $\boldsymbol{x}$, satisfying the initial condition
\begin{equation}
K(\boldsymbol{x},\boldsymbol{y},0;\hbar)=\delta(\boldsymbol{x}-\boldsymbol{y}) \ . 
\end{equation}

The Schr\"{o}dinger propagator itself evolves according to the dynamics \cite{BeSh} 
\begin{equation}
\Big(i\hbar\,\frac{d}{dt}-\widehat H \Big)U_t=0 \ , \ \ t\in [0,T] \ , \ \ U_0=\mathbb{1}_{L^2}
\label{eq:eqprop}
\end{equation}
comprising a unitary group\cite{BeSh} 
\begin{equation}
\{U_t \}_{t\in\mathbb{R}}=\{e^{-\frac{i}{\hbar}t\widehat H}\}_{t\in\mathbb{R}} 
%\label{eq:Ut}
\end{equation}
in the sense that it satisfies the group composition property $U_t U_{s}=U_{t+s}$ for $t,s\in\mathbb{R}$, and the unitarity property, $U_t^*=U_t^{-1}= U_{-t},$ for $t\in\mathbb{R}$. This is true as the Hamiltonian flow is autonomous.

In order to pose the Schr\"{o}dinger equation and the problem of time evolution in phase space, we introduce the \textit{wave packet representation} of the operator $\widehat H$, that is the phase space operator
\begin{equation}
\widehat{\mathcal{H}}=\mathcal{W}\widehat H \mathcal{W}^{-1} \ .
\label{eq:psH}
\end{equation}
This is a Weyl-symmetrized pseudo-differential operator acting on phase space wavefunctions. Some basic results concerning the definition and the symbols of such operators are given in the next section.

By conjugating problem ($\ref{eq:ivp}$) with $\mathcal{W}$, we obtain the Cauchy problem for the \textit{phase space Schr\"{o}dinger equation}, governing the evolution of the phase space wavefunction $\Psi=\mathcal{W}\psi$
with initial data $\Psi _0=\mathcal{W}\psi_0$. This problem reads 
\begin{equation}
\Big(i\hbar\,\frac{\partial}{\partial t}- \widehat{\mathcal{H}} \Big)\Psi(t)=0 \ , \ \ t\in [0,T] \label{eq:pivp}\ , \ \ \Psi(0)=\Psi _0 \ .
\label{eq:pscauchy}
\end{equation}

The wave packet transform retains the essential singularity of the equation itself, as well as of the initial data, as will be shown in Section \ref{solutions} for WKB states as prototype semi-classical states. This renders the phase space image of the initial problem semi-classically singular as well.

A notable difference from the Schr\"{o}dinger equation, in the case of the standard form Hamiltonian, $H(\boldsymbol{q},\boldsymbol{p})=|\boldsymbol{p}|^2+V(\boldsymbol{q})$, arises from the potential term; for non-polynomial potentials, this term introduces non-locality as well as transport effects in phase space dynamics, both common features in phase space evolution equations \cite{ChMl}, such as the von Neumann equation \cite{ZFC}. 

It should be also noted that while semi-classical asymptotic solutions of the Schr\"{o}dinger equation have a common semi-classical limit, sub-leading order terms may substantially differ with respect to choice of quantization. Such differences become manifest, e.g., between the treatment of semi-classical wave packet dynamics, in the work of Robert \cite{Rob1}, assuming the Weyl quantization, and in the work of Nazaikinskii et al. \cite{NSS}, assuming normal quantization.

\subsection{The Phase Space Schr\"{o}dinger Propagator}

We now proceed to the construction of the phase space Schr\"{o}dinger propagator $\mathcal{U}_t $, such that $\Psi(t)= \mathcal{U}_t \Psi _0$. Formally, by $\psi(t)=U_t\psi_0$ we have $\mathcal{W}^{-1}\Psi(t)=U_t \mathcal{W}^{-1}\Psi_0$. Hence, we have $\Psi(t)=\mathcal{W}U_t\mathcal{W}^{-1}\Psi_0$, which implies 
\begin{equation}
\mathcal{U}_t = \mathcal{W} U_t \mathcal{W}^{-1} \ .
\label{eq:psUt}
\end{equation}
More precisely, by the completeness relation of the semi-classical isotropic Gaussian wave packets, a resolution of the identity in $L^2(\mathbb{R}^d)$ \cite{Rob1}, 
\begin{equation}
\Big(\frac{1}{2\pi\hbar}\Big)^{d}\int {\widebar G}_{(\boldsymbol{q},\boldsymbol{p})}(\boldsymbol{x};\hbar) G_{(\boldsymbol{q},\boldsymbol{p})}(\boldsymbol{y};\hbar)\,d\boldsymbol{q}d\boldsymbol{p}=\delta(\boldsymbol{x}-\boldsymbol{y}) 
\label{eq:completeness}
\end{equation}
we acquire the following phase space resolution of the kernel $K$ of the propagator $U_t $,
\begin{equation}
K(\boldsymbol{x},\boldsymbol{y},t;\hbar)=\Big(\frac{1}{2\pi\hbar}\Big)^{d}\int U_t G_{(\boldsymbol{q},\boldsymbol{p})}(\boldsymbol{x};\hbar){\widebar G}_{(\boldsymbol{q},\boldsymbol{p})}(\boldsymbol{y};\hbar)\,d\boldsymbol{q}d\boldsymbol{p} \ .
\label{eq:wpek}
\end{equation}

Then, by applying the wave packet transform ($\ref{eq:wpt}$) on the representation formula ($\ref{eq:Kprop}$), and using ($\ref{eq:wpek}$), we derive the integral representation of the phase space wavefunction
\begin{equation}
\fl \Psi (\boldsymbol{q},\boldsymbol{p},t;\hbar)=\Big(\frac{1}{2\pi\hbar}\Big)^{d}\int\!\!\int {\widebar G}_{(\boldsymbol{q},\boldsymbol{p})}(\boldsymbol{x};\hbar) U_t G_{(\boldsymbol{\eta},\boldsymbol{\xi})}(\boldsymbol{x},t;\hbar)\Psi_0(\boldsymbol{\eta},\boldsymbol{\xi};\hbar)\,d\boldsymbol{x}d\boldsymbol{\eta}d\boldsymbol{\xi} \ .
\label{eq:pswf1}
\end{equation}
This integral representation ($\ref{eq:pswf1}$) can be compactly written in the form
\begin{equation}
\fl \Psi (\boldsymbol{q},\boldsymbol{p},t;\hbar)=(\mathcal{U}_t \Psi _0)(\boldsymbol{q},\boldsymbol{p},t;\hbar)=\int \mathcal{K}(\boldsymbol{q},\boldsymbol{p},\boldsymbol{\eta},\boldsymbol{\xi},t;\hbar)\Psi _0(\boldsymbol{\eta},\boldsymbol{\xi;\hbar})\,d\boldsymbol{\eta}d\boldsymbol{\xi}
\label{eq:pswf}
\end{equation}
which defines the action of the \textit{phase space Schr\"{o}dinger propagator} $\mathcal{U}_t $, and it is obviously in formal agreement with ($\ref{eq:psUt}$). The kernel $\mathcal{K}$ is expressed in terms of $K$ by the formula
\begin{equation}
\fl \mathcal{K}(\boldsymbol{q},\boldsymbol{p},\boldsymbol{\eta},\boldsymbol{\xi},t;\hbar)=\Big(\frac{1}{2\pi\hbar}\Big)^{d} \int\!\!\int {\widebar G}_{(\boldsymbol{q},\boldsymbol{p})} (\boldsymbol{x};\hbar)G_{(\boldsymbol{\eta},\boldsymbol{\xi})} (\boldsymbol{y};\hbar)K(\boldsymbol{x},
\boldsymbol{y},t;\hbar)\,d\boldsymbol{x} d\boldsymbol{y}
\label{eq:psprop}
\end{equation}
and it can be shown by direct computation that the following inversion formula holds
\begin{equation}
\fl K(\boldsymbol{x},\boldsymbol{y},t;\hbar)=\Big(\frac{1}{2\pi\hbar}\Big)^{d} \int\!\!\int {\widebar G}_{(\boldsymbol{q},\boldsymbol{p})} (\boldsymbol{x};\hbar)G_{(\boldsymbol{\eta},\boldsymbol{\xi})} (\boldsymbol{y};\hbar)\mathcal{K}(\boldsymbol{q},\boldsymbol{p},\boldsymbol{\eta},\boldsymbol{\xi},t;\hbar)\,d\boldsymbol{q}d\boldsymbol{p}d\boldsymbol{\eta}d\boldsymbol{\xi} \ .
\end{equation}
It can be easily checked that ($\ref{eq:pswf}$) solves the problem ($\ref{eq:pivp}$).

Formally, by $(\ref{eq:eqprop})$, $(\ref{eq:psH})$ and $(\ref{eq:psUt})$, it follows that the phase space Schr\"{o}dinger propagator $\mathcal{U}^t $ evolves according to the equation
\begin{equation}
\Big(i\hbar\,\frac{d}{dt}-\widehat{\mathcal{H}} \Big)\mathcal{U}_t =0 \ \ t\in [0,T] \ , \ \ \mathcal{U}_0=\mathbb{1}_{\mathfrak{F}}
\end{equation}
comprising a unitary group 
\begin{equation}
\{\mathcal{U}_t \}_{t\in\mathbb{R}}=\{e^{-\frac{i}{\hbar}t \widehat{\mathcal{H}} }\}_{t\in\mathbb{R}}
\label{eq:Ut}
\end{equation}
in the sense that it satisfies the group composition property, $\mathcal{U}_t\mathcal{U}_{s}=\mathcal{U}_{t+s}$, for $t,s\in\mathbb{R}$,
and the unitarity property $\mathcal{U}_t^*=\mathcal{U}_t^{-1}= \mathcal{U}_{-t}$, for $t\in\mathbb{R}$.

Conjugating (\ref{eq:fund}) by the wave packet transform, we get the analogous equation for the kernel $\mathcal{K}$ of the phase space propagator $\mathcal{U}^t $,
\begin{equation}
\fl \Big(i\hbar\,\frac{\partial}{\partial t}- \widehat{\mathcal{H}} \Big)\mathcal{K}(\boldsymbol{q},\boldsymbol{p},\boldsymbol{\eta},\boldsymbol{\xi},t;\hbar)=i\hbar\,\delta(t)\,\langle G_{(\boldsymbol{q},\boldsymbol{p})},G_{(\boldsymbol{\eta},\boldsymbol{\xi})}\rangle \ , \ \ t\in[0,T]
\end{equation}
where the operator $\widehat{\mathcal{H}}$ acts on the arguments $(\boldsymbol{q},\boldsymbol{p})$, satisfying the initial condition
\begin{equation} 
\fl b(\boldsymbol{q},\boldsymbol{p},\boldsymbol{\eta},\boldsymbol{\xi};\hbar):=\mathcal{K}(\boldsymbol{q},\boldsymbol{p},\boldsymbol{\eta},\boldsymbol{\xi},0;\hbar)=\Big(\frac{1}{2\pi\hbar}\Big)^{d} \int {\widebar G}_{(\boldsymbol{q},\boldsymbol{p})}(\boldsymbol{x};\hbar)G_{(\boldsymbol{\eta},\boldsymbol{\xi})}(\boldsymbol{x};\hbar)\,d\boldsymbol{x} 
\end{equation}
the so-called \textit{Bergmann kernel} \cite{Bar,NSS}. 

Due to the Gaussian integration in ($\ref{eq:psprop}$), the kernel $\mathcal{K}$ has stronger smoothness properties than $K(\boldsymbol{q},\boldsymbol{p},\boldsymbol{\eta},\boldsymbol{\xi},t;\hbar)$ across the hyper-plane $(\boldsymbol{q},\boldsymbol{p})=(\boldsymbol{\eta},\boldsymbol{\xi})$, while for $t\rightarrow 0^+$ it does not converge weakly to a Dirac distribution, but rather, it is a Dirac mollifier on the Heisenberg scale; in addition, the Bergmann kernel possesses the reproducing property, 
\begin{equation} 
\int b(\boldsymbol{q},\boldsymbol{p},\boldsymbol{\eta},\boldsymbol{\xi};\hbar)\Psi(\boldsymbol{\eta},\boldsymbol{\xi})\,d\boldsymbol{\eta}d\boldsymbol{\xi}= \Psi(\boldsymbol{q},\boldsymbol{p}) \ , \ \ \textrm{for \ any} \ \Psi\in\mathfrak{F} 
\end{equation}
due to the analytic structure of $\mathfrak{F}$.

\section{The Anisotropic Gaussian Approximation for the Propagator}
\label{AGAP}

\subsection{The Anisotropic Gaussian Approximation for the Schr\"{o}dinger Propagator}
\label{GAP}

We aim to construct a semi-classical asymptotic approximation of the phase space propagator $\mathcal{U}_t $, on the basis of a semi-classical approximation of the evolving wave packet $U_t G_{(\boldsymbol{q},\boldsymbol{p})}$ in the integral representation ($\ref{eq:pswf1}$). To this end, we consider the \textit{Anisotropic Gaussian Approximation} of the evolved wave packet in configuration space, following, basically, the work of Robert \cite{Rob1}, Nazaikinskii et al. \cite{NSS}, Belov et al. \cite{BBT}, based on the independent contributions of others (see section \ref{intro}). The semi-classical evolution of Gaussian wave packets in configuration space relies on the \textit{variational system} of the Hamiltonian flow, which is the basis for the \textit{Nearby Orbit Approximation} \cite{BaDa}.

In the \textit{Anisotropic Gaussian Approximation}, $U_t G_{(\boldsymbol{q},\boldsymbol{p})}$ is approximated by the anisotropic wave packet $G^{\scriptsize{\mathbfcal{Z}}} _{(\boldsymbol{q},\boldsymbol{p})}(\boldsymbol{x},t;\hbar)$
which is a specific semi-classical asymptotic solution of the Cauchy problem ($\ref{eq:ivp})$ with initial data $ \psi_0=G_{(\boldsymbol{q},\boldsymbol{p})}$ (see \ref{WPTr}). This solution is given by
\begin{eqnarray}
\fl G^{\scriptsize{\mathbfcal{Z}}}_{(\boldsymbol{q},\boldsymbol{p})}(\boldsymbol{x},t;\hbar)=(\pi\hbar)^{-d/4}a(\boldsymbol{q},\boldsymbol{p},t)\\ \nonumber
\fl \times \exp\frac{i}{\hbar}\Big(\frac{\boldsymbol{p}\cdot\boldsymbol{q}}{2}+{A} (\boldsymbol{q},\boldsymbol{p},t)+\boldsymbol{p}_t\cdot (\boldsymbol{x}-\boldsymbol{q}_t)+\frac{1}{2}(\boldsymbol{x}-\boldsymbol{q}_t)\cdot \mathbfcal{Z}
(\boldsymbol{q},\boldsymbol{p},t)(\boldsymbol{x}-\boldsymbol{q}_t)\Big) \ .
\label{eq:MSwp}
\end{eqnarray}
The anisotropy matrix $\mathbfcal{Z}(\boldsymbol{q},\boldsymbol{p},t)$ is determined by the matrix Riccati equation
\begin{equation}
\frac{d\mathbfcal{Z}}{dt}+\mathbfcal{Z}\,H_{\boldsymbol{pp}}\,\mathbfcal{Z}+H_{\boldsymbol{qp}}\,\mathbfcal{Z}+\mathbfcal{Z}\,H_{\boldsymbol{pq}}+H_{\boldsymbol{qq}}=\textbf{0} \ , \ \ \mathbfcal{Z}(0)=i\textbf{I} 
\end{equation}
and the amplitude is given by
\begin{equation}
\fl a(\boldsymbol{q},\boldsymbol{p},t)=\exp\Bigg(-\frac{1}{2}\int\displaylimits_0^t{\rm tr}\Big(H_{\boldsymbol{pp}}(\boldsymbol{q},\boldsymbol{p})\,\mathbfcal{Z}(\boldsymbol{q},\boldsymbol{p},\tau)+H_{\boldsymbol{pq}}(\boldsymbol{q},\boldsymbol{p})\Big)\,d\tau\Bigg)
\end{equation}
while the phase $ A $ is the \textit{phase space action,} related to Hamilton's principal function
\begin{equation}
{A} (\boldsymbol{q},\boldsymbol{p},t)=\int\displaylimits_0^t\boldsymbol{p}_\tau\cdot \frac{d \boldsymbol{q}_\tau}{d\tau} \,d\tau-H(\boldsymbol{q},\boldsymbol{p})\,t \ .
\label{eq:action}
\end{equation}
where $(\boldsymbol{q}_\tau,\boldsymbol{p}_\tau)=g_\tau (\boldsymbol{q},\boldsymbol{p})$, for $0\leq \tau\leq t$.

The wave packet $G^{\scriptsize{\mathbfcal{Z}}}_{(\boldsymbol{q},\boldsymbol{p})}$ moves along the orbit emanating from $(\boldsymbol{q},\boldsymbol{p})$ at $t=0$ and is semi-classically concentrated on that point, on the Heisenberg scale \cite{BaDa,NSS}.

Based on the above asymptotic solution, we define the approximate propagator $U_{t}^{\mathbfcal{Z}}$, such that
\begin{equation}
U_{t}^{\mathbfcal{Z}} G_{(\boldsymbol{q},\boldsymbol{p})}(\boldsymbol{x};\hbar):=G^{\scriptsize{\mathbfcal{Z}}} _{(\boldsymbol{q},\boldsymbol{p})}(\boldsymbol{x},t;\hbar) \ .
\label{eq:Uth}
\end{equation}

By using this approximation for the propagator $U_{t} $ into the representation $(\ref{eq:wpek})$, we obtain the following wave packet approximation of the kernel of the Schr\"{o}dinger propagator 
\begin{equation}
\fl K(\boldsymbol{x},\boldsymbol{y},t;\hbar)\sim K^{\scriptsize{\mathbfcal{Z}}}(\boldsymbol{x},\boldsymbol{y},t;\hbar):=\Big(\frac{1}{2\pi\hbar}\Big)^{d}\int G^{\scriptsize{\mathbfcal{Z}}} _{(\boldsymbol{q},\boldsymbol{p})}(\boldsymbol{x},t;\hbar)\widebar G_{(\boldsymbol{q},\boldsymbol{p})}(\boldsymbol{y};\hbar) \,d\boldsymbol{q}d\boldsymbol{p}
\label{eq:kh}
\end{equation}
 
\subsection{The Anisotropic Gaussian Approximation for the Phase Space Schr\"{o}dinger Propagator}
\label{GAPSP}

We can construct an asymptotic approximation 
\begin{equation}
\mathcal{U}_{t}^{\mathbfcal{Z}}=\mathcal{W} U_{t}^{\mathbfcal{Z}} \mathcal{W}^{-1} \ .
\end{equation}
of the phase space propagator $\mathcal{U}_t $ by substituting the approximate propagator $(\ref{eq:Uth})$ into the definition of the phase space propagator $(\ref{eq:psUt})$.
Therefore, the approximate kernel $\mathcal{K}^{\scriptsize{\mathbfcal{Z}}}$ of $\mathcal{U}_{t}^{\mathbfcal{Z}}$
is derived by substituting the approximate kernel $K^{\scriptsize{\mathbfcal{Z}}}$ into $(\ref{eq:psprop})$, and using $(\ref{eq:completeness})$,
and it reads as follows 
\begin{equation}
\mathcal{K}^{\scriptsize{\mathbfcal{Z}}}(\boldsymbol{q},\boldsymbol{p},\boldsymbol{\eta},\boldsymbol{\xi},t;\hbar):=\Big(\frac{1}{2\pi\hbar}\Big)^{d}\int\widebar G_{(\boldsymbol{q},\boldsymbol{p})}(\boldsymbol{x};\hbar) G^{\scriptsize{\mathbfcal{Z}}} _{(\boldsymbol{\eta},\boldsymbol{\xi})}(\boldsymbol{x},t;\hbar)\,d\boldsymbol{x} \ .
\label{eq:scprop}
\end{equation}
Then, by the representation formula ($\ref{eq:pswf}$), we obtain the following approximate wave packet representation of the phase space wavefunction
\begin{equation}
\fl \Psi^{\scriptsize{\mathbfcal{Z}}}(\boldsymbol{q},\boldsymbol{p},t;\hbar):=(\mathcal{U}_{t}^{\mathbfcal{Z}}\Psi_0 )(\boldsymbol{q},\boldsymbol{p},t)=
\int \mathcal{K}^{\scriptsize{\mathbfcal{Z}}}(\boldsymbol{q},\boldsymbol{p},\boldsymbol{\eta},\boldsymbol{\xi},t;\hbar)\Psi_0(\boldsymbol{\eta},\boldsymbol{\xi};\hbar)\,d\boldsymbol{\eta} d\boldsymbol{\xi}
\label{eq:sclsol}
\end{equation}

The kernel $\mathcal{K}^{\scriptsize{\mathbfcal{Z}}}(\boldsymbol{q},\boldsymbol{p},\boldsymbol{\eta},\boldsymbol{\xi},t;\hbar)$ can be calculated explicitly, while it follows that it has the form of an anisotropic Gaussian wave packet.
By direct integration, we have 
\begin{eqnarray}
\fl \mathcal{K}^{\scriptsize{\mathbfcal{Z}}}(\boldsymbol{q},\boldsymbol{p},\boldsymbol{\eta},\boldsymbol{\xi},t;\hbar)=\Big(\frac{1}{2\pi\hbar}\Big)^{d}\frac{2^{d/2}}{\sqrt{\det(\textbf{A}(\boldsymbol{\eta},\boldsymbol{\xi},t)-i\textbf{B}(\boldsymbol{\eta},\boldsymbol{\xi},t))}} \nonumber \\
\times \exp\frac{i}{\hbar}\Biggl(A (\boldsymbol{\eta},\boldsymbol{\xi},t)+\frac{\boldsymbol{\xi}\cdot\boldsymbol{\eta}-\boldsymbol{\xi}_t\cdot \boldsymbol{\eta}_t}{2}+\frac{1}{2}(\boldsymbol{q},\boldsymbol{p})\cdot \textbf{J}(\boldsymbol{\eta}_t,\boldsymbol{\xi}_t)\nonumber \\
+\frac{1}{2}\left(\begin{array}{ccc}
\boldsymbol{q}-\boldsymbol{\eta}_t\\
\boldsymbol{p}-\boldsymbol{\xi}_t 
\end{array}\right)^{T}
\mathbfcal{Q}
 \left(
\begin{array}{ccc}
\boldsymbol{q}-\boldsymbol{\eta}_t\\
\boldsymbol{p}-\boldsymbol{\xi}_t 
\end{array}\right)\Biggr) \ .
\end{eqnarray}
Equivalently,
\begin{eqnarray}
\fl \mathcal{K}^{\scriptsize{\mathbfcal{Z}}}(\boldsymbol{q},\boldsymbol{p},\boldsymbol{\eta},\boldsymbol{\xi},t)=\Big(\frac{1}{2\pi\hbar}\Big)^{d}\Big(\det\,\frac{\partial(\boldsymbol{\eta}_t-i\boldsymbol{\xi}_t)}{\partial(\boldsymbol{\eta}-i\boldsymbol{\xi})}\Big)^{-1/2} \nonumber \\
\fl \times \exp\frac{i}{\hbar}\Bigg\{A (\boldsymbol{\eta},\boldsymbol{\xi},t)+\frac{\boldsymbol{\xi}\cdot\boldsymbol{\eta}-\boldsymbol{\xi}_t\cdot \boldsymbol{\eta}_t}{2}+\frac{1}{2}(\boldsymbol{q},\boldsymbol{p})\cdot \textbf{J}(\boldsymbol{\eta}_t,\boldsymbol{\xi}_t)\nonumber \\
+\frac{1}{2}\left(\begin{array}{ccc}
\boldsymbol{q}-\boldsymbol{\eta}_t\\
\boldsymbol{p}-\boldsymbol{\xi}_t 
\end{array}\right)^{T}
\mathbfcal{Q}
 \left(
\begin{array}{ccc}
\boldsymbol{q}-\boldsymbol{\eta}_t\\
\boldsymbol{p}-\boldsymbol{\xi}_t 
\end{array}\right)\Bigg\}
\label{eq:Lwp}
\end{eqnarray}
where 
\begin{equation}
\mathbfcal{Q}(\boldsymbol{\eta},\boldsymbol{\xi},t)=\left(
\begin{array}{ccc}
i\textbf{I}-i(\textbf{I}-i\mathbfcal{Z})^{-1} & \frac{1}{2}\textbf{I}-(\textbf{I}-i\mathbfcal{Z})^{-1}\\
\frac{1}{2}\textbf{I}-(\textbf{I}-i\mathbfcal{Z})^{-1} & i(\textbf{I}-i\mathbfcal{Z})^{-1} 
\end{array}\right) \ .
\end{equation}
is the \textit{double phase space anisotropy matrix}.
This matrix is an element of Siegel upper half-space, $\mathbfcal{Q}\in \Sigma_{2d}$ \cite{Fol} and it is proportional to a complex symplectic matrix, in the sense that $\frac{2}{i}\mathbfcal{Q}\in{\rm Sp}(2d,\mathbb{C})$.

At $t=0$, the semi-classical wave packet phase space propagator shares the reproducing property of the exact phase space propagator,
\begin{eqnarray} 
\fl \mathcal{K}^{\scriptsize{\mathbfcal{Z}}} (\boldsymbol{q},\boldsymbol{p},\boldsymbol{\eta},\boldsymbol{\xi},0;\hbar)=\mathcal{K} (\boldsymbol{q},\boldsymbol{p},\boldsymbol{\eta},\boldsymbol{\xi},0;\hbar)=\Big(\frac{1}{2\pi\hbar}\Big)^{d}\int\widebar G_{(\boldsymbol{q},\boldsymbol{p})}(\boldsymbol{x};\hbar) G^{\scriptsize{\mathbfcal{Z}}} _{(\boldsymbol{\eta},\boldsymbol{\xi})}(\boldsymbol{x},0;\hbar)\,d\boldsymbol{x} \\ \nonumber
=\Big(\frac{1}{2\pi\hbar}\Big)^{d}\int\widebar G_{(\boldsymbol{q},\boldsymbol{p})}(\boldsymbol{x};\hbar) G_{(\boldsymbol{\eta},\boldsymbol{\xi})}(\boldsymbol{x};\hbar)\,d\boldsymbol{x}=b(\boldsymbol{q},\boldsymbol{p},\boldsymbol{\eta},\boldsymbol{\xi};\hbar)
\end{eqnarray}
the Bergmann reproducing kernel.

\begin{remark}
As is the case, in general, with semi-classical approximations for quantum evolution, $U_{t}^{\scriptsize{\mathbfcal{Z}}}$ and $\mathcal{U}_{t}^{\scriptsize{\mathbfcal{Z}}}$ have an approximate group character (semi-group, in the case of open systems), meaning that, as $\hbar\rightarrow 0^+$, in the appropriate weak sense, for $t,s\in\mathbb{R}$, 
\begin{equation}
U_{t}^{\scriptsize{\mathbfcal{Z}}} U_{s}^{\scriptsize{\mathbfcal{Z}}}\sim U_{t+s}^{\scriptsize{\mathbfcal{Z}}} \ , \ \ \hbar\rightarrow 0^+
\end{equation}
and
\begin{equation}
(U_{t}^{\scriptsize{\mathbfcal{Z}}})^*\sim (U_{t}^{\scriptsize{\mathbfcal{Z}}})^{-1}\sim U_{-t}^{\scriptsize{\mathbfcal{Z}}} \ , \ \ \hbar\rightarrow 0^+ \ .
\end{equation}

Similarly, the semi-classical phase space propagator, $\mathcal{U}_{t}^{\mathbfcal{Z}}$, defines semi-classically unitary Schr\"{o}dinger flow in $\mathfrak{F}$, as it preserves the Fock-Bargmann analyticity constraints (\ref{eq:FBc}), while for $t,s\in\mathbb{R}$, 
\begin{equation}
\mathcal{U}_{t}^{\mathbfcal{Z}}\mathcal{U}_{s}^{\mathbfcal{Z}}\sim \mathcal{U}_{t+s}^{\mathbfcal{Z}} \ , \ \ \hbar\rightarrow 0^+ \ .
\end{equation}
and
\begin{equation}
(\mathcal{U}_{t}^{\scriptsize{\mathbfcal{Z}}})^*\sim (\mathcal{U}_{t}^{\scriptsize{\mathbfcal{Z}}})^{-1}\sim \mathcal{U}_{-t}^{\scriptsize{\mathbfcal{Z}}} \ , \ \ \hbar\rightarrow 0^+ \ .
\end{equation}
\end{remark}

\begin{remark}
Concerning the validity of the approximation, the minimal time-scale which marks its limits is the semi-classical time-scale up to which $G^{\scriptsize{\mathbfcal{Z}}}_{(\boldsymbol{q},\boldsymbol{p})}$ remains a semi-classical solution of the given order, beyond which its Gaussian wave packet form, its micro-localization on the reference orbit, is irreversibly lost. This is the \textit{Ehrenfest time-scale} \cite{Fau,SVT}, $T_E(\hbar)$, defined for the given base point, $(\boldsymbol{q},\boldsymbol{p})$, as
\begin{equation}
|\tilde g_{T_E(\hbar)}(\boldsymbol{q},\boldsymbol{p})|\asymp \frac{1}{\hbar^{1/2}} \ , \ \ \hbar\rightarrow 0^+
\end{equation}
where $\tilde{g}_t$ is the linearized Hamiltonian flow \cite{SVT}. The Ehrenfest time-scale is sensitive to the dynamical properties of the flow, be it global or local \cite{Fau,SVT}; in the case the flow is chaotic, in particular globally hyperbolic, it reads
\begin{equation}
T_E(\hbar)\asymp \log\,\hbar \ , \ \ \hbar\rightarrow 0^+ 
\end{equation}
while in the case the flow is completely integrable, it reads
\begin{equation}
T_E(\hbar)\asymp \frac{1}{\hbar^{1/2}} \ , \ \ \hbar\rightarrow 0^+ \ . 
\end{equation}

By the analysis in \cite{Rob1}, one may expect the approximation 
\begin{equation}
U_tG_{(\boldsymbol{q},\boldsymbol{p})}(t)\sim G^{\scriptsize{\mathbfcal{Z}}}_{(\boldsymbol{q},\boldsymbol{p})}(t) \ , \ \ \hbar\rightarrow 0^+
\end{equation}
for 
\begin{equation}
t=o(\log\,\hbar) \ , \ \ \hbar\rightarrow 0^+ \ ,
\end{equation}
for a a broader class of Hamiltonian functions.
\end{remark}

\begin{remark} 
By a formal application of the Stationary Complex Phase Theorem, the approximation of $K^{\scriptsize{\mathbfcal{Z}}}$ defined in $(\ref{eq:kh})$, yields, to leading order, the Van Vleck approximation of $K$ (see also \cite{BiRo,Lit1})
\begin{equation}
\fl K{(\boldsymbol{x},\boldsymbol{y},t;\hbar})\sim\Big(\frac{1}{2\pi i\hbar}\Big)^{d/2}\sum_{r=1}^{N}\sqrt{\Big|\det\,\frac{\partial \boldsymbol{p}}{\partial \boldsymbol{q}_t}(\boldsymbol{y},\boldsymbol{p}_r(\boldsymbol{x},\boldsymbol{y},t),t)\Big|}\,e^{\frac{i}{\hbar} A (\boldsymbol{y},\boldsymbol{p}_r(\boldsymbol{x},\boldsymbol{y},t),t)-\frac{\pi i}{2}\nu_r}
\end{equation}
where $r=1,\ldots,N$ indexes all trajectories emanating from point $\boldsymbol{x}$ at time $0$ with momentum $\boldsymbol{p}_r$ and terminating at point $\boldsymbol{y}$ at time $t$, while $\nu_r$ is the index of the monodromy matrix of the trajectory.
\end{remark}

\subsection{Relation to the Littlejohn Approximation and Initial Value Representations}

Semi-classical approximations are, of course, no novelty even in phase space representations. To the author's knowledge, a first construction of a semi-classical approximation to the propagator for the Schr\"{o}dinger equation, in an explicit phase space integral representation, was given by Littlejohn \cite{Lit1} and Hermann and Kluk \cite{HeKl}, independently, in different frame-works and heading from different starting points.

In \cite{Lit1}, Littlejohn constructed a semi-classical phase space propagator based on the Nearby Orbit Approximation, as an explicit action of Weyl shifts and metaplectic operators, generalizing the approximation for the dynamics of Liouville densities in the quantum mechanical framework. 

The semi-classical Littlejohn flow reads (\cite{Lit2}, eqs. (7.14), (7.27))
\begin{equation}
\mathcal{U}_{t}^{L}\Psi(\boldsymbol{q},\boldsymbol{p};\hbar):=\int \mathcal{K}^L(\boldsymbol{q},\boldsymbol{p},\boldsymbol{\eta},\boldsymbol{\xi},t;\hbar)\Psi(\boldsymbol{\eta},\boldsymbol{\xi};\hbar)\,d\boldsymbol{\eta} d\boldsymbol{\xi}
\end{equation}
where the kernel is 
\begin{eqnarray}
\fl \mathcal{K}^L(\boldsymbol{q},\boldsymbol{p},\boldsymbol{\eta},\boldsymbol{\xi},t;\hbar)=\Big(\frac{1}{2\pi\hbar}\Big)^d\\ \nonumber 
\int \widebar G _{(\boldsymbol{q},\boldsymbol{p})}(\boldsymbol{x};\hbar)\,e^{\frac{i}{\hbar}({A} (\boldsymbol{\eta},\boldsymbol{\xi},t)-\frac{\boldsymbol{\xi}_t\cdot \boldsymbol{\eta}_t}{2})}(\mathcal{T}_{g_{_{t}}(\boldsymbol{\eta},\boldsymbol{\xi})} M(\boldsymbol{\eta},\boldsymbol{\xi},t;\hbar)G_{\boldsymbol{0}})(\boldsymbol{x};\hbar)\,d\boldsymbol{x}
\end{eqnarray}
where $\mathcal{T}_{(\boldsymbol{q},\boldsymbol{p})}$ is the Weyl shift \cite{CoRo} and $M$ is a metaplectic operator whose dynamics is governed by the equation
\begin{equation}
\fl \frac{dM}{dt}=\frac{i\hbar}{2}\Big(\boldsymbol{\eta}\cdot H_{\boldsymbol{\eta\eta}}\boldsymbol{\eta}+\frac{\partial}{\partial\boldsymbol{\eta}}\cdot H_{\boldsymbol{\xi\eta}}\boldsymbol{\eta}+\eta\cdot H_{\boldsymbol{\eta\xi}}\frac{\partial}{\partial\boldsymbol{\eta}}+\frac{\partial}{\partial\boldsymbol{\eta}}\cdot H_{\boldsymbol{\xi\xi}}\frac{\partial}{\partial\boldsymbol{\eta}}\Big)M \ , \ \ M(0)=\mathbb{1}_{L^2}
\end{equation}
the Hessian elements evaluated along the flow, at $(\boldsymbol{\eta}_t,\boldsymbol{\xi}_t)$. The propagator fixes the initial state from $(\boldsymbol{q},\boldsymbol{p})$ to the origin, symplectically `rotates' it in phase space by the action of the metaplectic operator $M(t)$, and shifts it along the reference orbit modulating by adding the action phase.
 
A different approach, on the lines of which we constructed the semi-classical phase space propagator, is that of Initial Value Representations for solutions of the Schr\"{o}dinger equation\cite{HeKl}. These representations involve superpositions of initial semi-classical states in Gaussian wave packets, $G_{(\boldsymbol{q},\boldsymbol{p})}(\boldsymbol{x};\hbar)$, propagating them, in a certain approximation along the Hamiltonian orbit emanating from point $(\boldsymbol{q},\boldsymbol{p})$ and subsequently superposing the evolved wave packets with respect to the totality of base points, i.e., initial phase space points $(\boldsymbol{q},\boldsymbol{p})$. 

In this context someone begins with the kernel $(\ref{eq:wpek})$ of the propagator $K$, 
\begin{equation}
K(\boldsymbol{x},\boldsymbol{y},t;\hbar)=\Big(\frac{1}{2\pi\hbar}\Big)^{d} \int U_tG_{(\boldsymbol{q},\boldsymbol{p})}(\boldsymbol{x};\hbar)\widebar G_{(\boldsymbol{q},\boldsymbol{p})}(\boldsymbol{y};\hbar)\,d\boldsymbol{q}d\boldsymbol{p} \ .
\end{equation}
and constructs, for small $\hbar$, three different approximations, by adopting different approximations of $U_tG_{(\boldsymbol{q},\boldsymbol{p})}$:
 
1. the \textit{Thawed} (or \textit{Anisotropic Gaussian Approximation})
\begin{equation}
\fl K(\boldsymbol{x},\boldsymbol{y},t;\hbar)\sim K^{\scriptsize{\mathbfcal{Z}}}(\boldsymbol{x},\boldsymbol{y},t):=\Big(\frac{1}{2\pi\hbar}\Big)^{d} \int G^{\scriptsize{\mathbfcal{Z}}}_{(\boldsymbol{q},\boldsymbol{p})}(\boldsymbol{x},t;\hbar)\widebar G_{(\boldsymbol{q},\boldsymbol{p})}(\boldsymbol{y};\hbar)\,d\boldsymbol{q}d\boldsymbol{p}
\end{equation}

2. the \textit{Frozen Gaussian Approximation}
\begin{equation}
K(\boldsymbol{x},\boldsymbol{y},t;\hbar)\sim \Big(\frac{1}{2\pi\hbar}\Big)^{d} \int e^{\frac{i}{\hbar}{A}(\boldsymbol{q},\boldsymbol{p},t)}G_{(\boldsymbol{q},\boldsymbol{p})}(\boldsymbol{x};\hbar)\widebar G_{(\boldsymbol{q},\boldsymbol{p})}(\boldsymbol{y};\hbar)\,d\boldsymbol{q}d\boldsymbol{p}
\end{equation}

3. the \textit{Herman-Kluk Approximation} \cite{HeKl,Rob2,RoSw}
\begin{equation}
\fl K(\boldsymbol{x},\boldsymbol{y},t;\hbar)\sim \Big(\frac{1}{2\pi\hbar}\Big)^{d} \int c(\boldsymbol{q},\boldsymbol{p},t;\hbar)\,e^{\frac{i}{\hbar}{A}(\boldsymbol{q},\boldsymbol{p},t)}G_{(\boldsymbol{q},\boldsymbol{p})}(\boldsymbol{x};\hbar)\widebar G_{(\boldsymbol{q},\boldsymbol{p})}(\boldsymbol{y};\hbar)\, d\boldsymbol{q}d\boldsymbol{p} \ . 
\end{equation}

It must be emphasized that the construction of the Littlejohn propagator does not fall into the scheme of Initial Value Representations, but instead it utilizes a single reference orbit, emanating from a given phase space point on which the initial data $\psi_0 $ is assumed to be centered at and localized, without necessarily being Gaussian.

\section{Fourier Integral Representation of the Phase Space Wavefunction}
\label{solutions}

We now turn to the issue of central interest, the semi-classical Cauchy problem $(\ref{eq:pscauchy})$ for the phase space Schr\"{o}dinger equation, 
\begin{equation}
\Big(i\hbar\, \frac{\partial}{\partial t}- \widehat{\mathcal{H}} \Big)\Psi(t)=0 \ , \ \ t\in[0,T] \ , \ \ \ \Psi(0)=\Psi_0 \in\mathfrak{F}
\end{equation}
where $T$ remains fixed. We construct an asymptotic approximation $\Psi^{\hbar}$ of the phase space wavefunction $\Psi$, on the basis of the integral representation $(\ref{eq:sclsol})$, starting with an appropriate approximation $\Psi_{0}^{\hbar}$ of the initial data $\Psi_0$ corresponding to WKB initial data $\psi_{0}^{\hbar}=R_0\,e^{\frac{i}{\hbar}S_0}$ for the Schr\"{o}dinger equation in configuration space. The approximation arises from an intermediate Fourier integral with quadratic complex phase.

The main tool for the construction of the approximation is the Complex Stationary Phase Theorem (see \cite{MeSj}, \cite{MSS}, \cite{NSS}, \cite{Tre} for a detailed exposition; also \ref{CSPT} for the basic formula), and this poses a technical difficulty, as, in general, one has to work with almost analytic extensions of the phases and the amplitudes of the integrals. 

\subsection{Approximation of the Initial Wavefunction}

We consider the phase space image 
\begin{eqnarray}
\fl \Psi_{0}(\boldsymbol{q},\boldsymbol{p};\hbar)=(\mathcal{W}\psi_{0})(\boldsymbol{q},\boldsymbol{p};\hbar)=\Big(\frac{1}{2\pi\hbar}\Big)^{d/2}\int{\widebar G}_{(\boldsymbol{q},\boldsymbol{p})}(\boldsymbol{x};\hbar)\psi_{0}(\boldsymbol{x};\hbar)\,d\boldsymbol{x} \nonumber \\
\fl =(\pi\hbar)^{-d/4}\Big(\frac{1}{2\pi\hbar}\Big)^{d/2}\int R_0(\boldsymbol{x})\,
\exp\frac{i}{\hbar}\Big(S_0(\boldsymbol{x}) -\boldsymbol{p}\cdot (\boldsymbol{x}-\boldsymbol{q})-\frac{i}{2}|\boldsymbol{x}-\boldsymbol{q}|^2 - \frac{\boldsymbol{p}\cdot\boldsymbol{q}}{2}\Big)d\boldsymbol{x}
\label{eq:pswkb0}
\end{eqnarray}
of the WKB initial state
\begin{equation}
\label{eq:WKB}
\psi_{0}^{\hbar}(\boldsymbol{x})=R_0(\boldsymbol{x})\,e^{\frac{i}{\hbar}S_0(\boldsymbol{x})} \ .
\end{equation}
We assume that 
\begin{eqnarray}
S_0\in C^\infty (\mathbb{R}^d,\mathbb{R}) \ , \ \ \det \,\frac{\partial^2 S_0}{\partial\boldsymbol{x}^2}\neq 0 \ \ {\rm everywhere}, \\
R_0\in C_0^\infty (\mathbb{R}^d,\mathbb{R}) \ , \ \ \int_{\mathbb{R}^d} R^2_0\, d\boldsymbol{x}=1
\end{eqnarray}
We also assume that the Lagrangian manifold corresponding to the initial state $\psi_{0}^{\hbar}$, i.e., generated by the initial phase, 
\begin{equation}
\Lambda_0:=\{(\boldsymbol{q},\boldsymbol{p})\in\mathbb{R}^{2d}\,|\,\boldsymbol{p}=\frac{\partial S_0}{\partial \boldsymbol{x}}(\boldsymbol{q})\}
\label{eq:LM}
\end{equation}
is smooth and connected, and is of maximal dimensionality, ${\rm dim}\,\Lambda_0=d$.

Following \cite{NSS}, we apply the \textit{Stationary Complex Phase Theorem} (see \ref{CSPT}) to the integral $(\ref{eq:pswkb0})$,
and we obtain the approximation
\begin{equation}
\Psi_{0}(\boldsymbol{q},\boldsymbol{p};\hbar)\sim\Psi_{0}^{\hbar}(\boldsymbol{q},\boldsymbol{p}):=
\chi_{0}(\boldsymbol{q},\boldsymbol{p};\hbar)\exp\Biggl(\frac{i}{\hbar}\theta_0(\boldsymbol{q},\boldsymbol{p})\Biggl)
\label{eq:wkb0}
\end{equation}
where $\chi_{0}$ is the complex amplitude
\begin{equation}
\chi_{0}(\boldsymbol{q},\boldsymbol{p};\hbar)=(\pi\hbar)^{-d/4}\frac{R_0(\boldsymbol{z}(\boldsymbol{q},\boldsymbol{p}))}{\sqrt{\det\Big(\textbf{I}-i\, \frac{\partial^2S_0}{\partial\boldsymbol{z}^2}(\boldsymbol{z}(\boldsymbol{q},\boldsymbol{p}))\Big)}}
\label{eq:chi0}
\end{equation}
and $\theta_0$ is the complex phase 
\begin{equation}
\fl \theta_0(\boldsymbol{q},\boldsymbol{p})=S_0(\boldsymbol{z}(\boldsymbol{q},\boldsymbol{p}))-\boldsymbol{p}\cdot \Big(\boldsymbol{z}(\boldsymbol{q},\boldsymbol{p})-\boldsymbol{q}\Big)+\frac{i}{2}\Big(\boldsymbol{z}(\boldsymbol{q},\boldsymbol{p})-\boldsymbol{q}\Big)^2-\frac{\boldsymbol{p}\cdot\boldsymbol{q}}{2}
\label{eq:theta0}
\end{equation}
and
$S_0(\boldsymbol{z})$ and $R_0(\boldsymbol{z})$ stand for the almost analytic extension (see \ref{CSPT}) to the complex variable $\boldsymbol{z}=\boldsymbol{x}+i\boldsymbol{y}\in\mathbb{C}^d$ of $S_0(\boldsymbol{x}),R_0(\boldsymbol{x})$ respectively, and $\boldsymbol{z}=\boldsymbol{z}(\boldsymbol{q},\boldsymbol{p})$ stands for the complex solution of the stationary equation 
\begin{equation}
\frac{\partial S_0}{\partial \boldsymbol{z}}(\boldsymbol{z})-\boldsymbol{p}+i(\boldsymbol{z}-\boldsymbol{q})=\boldsymbol{0}
\label{eq:stat0}
\end{equation}
(see \ref{CSPT}) with ${\rm Re}\,\boldsymbol{z}(\boldsymbol{q},\boldsymbol{p}) \in{\rm supp}\,R_0$. The solution is given approximately by 
\begin{equation}
\fl \boldsymbol{z}(\boldsymbol{q},\boldsymbol{p})=\boldsymbol{q}-i\Big(\textbf{I}-i\frac{\partial^2S_0}{\partial\boldsymbol{x}^2}(\boldsymbol{q})\Big)^{-1}\Big(\boldsymbol{p}-\frac{\partial S_0}{\partial\boldsymbol{x}}(\boldsymbol{q})\Big)+O\Big(\Big|\boldsymbol{p}-\frac{\partial S_0}{\partial \boldsymbol{x}}(\boldsymbol{q})\Big|^2\Big)
\end{equation}
and, on the Lagrangian manifold $\Lambda_0$ it reduces to the simple real expression
\begin{equation}
\boldsymbol{z}|_{\Lambda_0}(\boldsymbol{q},\boldsymbol{p})=\boldsymbol{q} \ .
\end{equation}
We stress that the solution is real only on the Lagrangian manifold $\Lambda_0$.

The imaginary part of the phase of the initial data vanishes along the Lagrangian manifold, $(\boldsymbol{q},\boldsymbol{p})\in\Lambda_0$, so that 
\begin{equation}
\Psi_{0}^{\hbar}|_{\Lambda_0}(\boldsymbol{q},\boldsymbol{p})=(\pi\hbar)^{-d/4}\frac{R_0(\boldsymbol{q})}{\sqrt{\det\,\Big(\textbf{I}-i\,\frac{\partial^2S_0}{\partial\boldsymbol{x}^2}(\boldsymbol{q})\Big)}}\,e^{\frac{i}{\hbar}(-\frac{\boldsymbol{p}\cdot\boldsymbol{q}}{2}+S_0(\boldsymbol{q}))} \ .
\end{equation}

\subsection{The Fourier Integral Representation}

By substituting the approximate initial data $(\ref{eq:wkb0})$ into $(\ref{eq:sclsol})$, we obtain an approximation 
$\Psi^{\hbar}$ of $\Psi^{\scriptsize{\mathbfcal{Z}}}$, in the form of a semi-classical Fourier integral with quadratic complex phase
\begin{equation}
\Psi^{\hbar}(\boldsymbol{q},\boldsymbol{p},t)=\Big(\frac{1}{2\pi\hbar}\Big)^d\int\varphi(\boldsymbol{\eta},\boldsymbol{\xi},t;\hbar)\,e^{\frac{i}{\hbar}F(\boldsymbol{q},\boldsymbol{p},\boldsymbol{\eta},\boldsymbol{\xi},t)}\,d\boldsymbol{\eta} d\boldsymbol{\xi}
\label{psfourier}
\end{equation}
where
\begin{eqnarray}
\label{psfourierphase}
\fl F(\boldsymbol{q},\boldsymbol{p},\boldsymbol{\eta},\boldsymbol{\xi},t)= \theta_0 (\boldsymbol{\eta},\boldsymbol{\xi}) +{A}(\boldsymbol{\eta},\boldsymbol{\xi},t)
+\frac{\boldsymbol{\xi}\cdot \boldsymbol{\eta}-\boldsymbol{\xi}_t\cdot \boldsymbol{\eta}_t}{2} \\ \nonumber 
+\frac{1}{2}(\boldsymbol{q},\boldsymbol{p})\cdot \textbf{J}(\boldsymbol{\eta}_t,\boldsymbol{\xi}_t)+\frac{1}{2}\left(\begin{array}{ccc}
\boldsymbol{q}-\boldsymbol{\eta}_t\\
\boldsymbol{p}-\boldsymbol{\xi}_t 
\end{array}\right)^{T}
\mathbfcal{Q}(\boldsymbol{\eta},\boldsymbol{\xi},t)
 \left(
\begin{array}{ccc}
\boldsymbol{q}-\boldsymbol{\eta}_t\\
\boldsymbol{p}-\boldsymbol{\xi}_t 
\end{array}\right)
\end{eqnarray}
and 
\begin{eqnarray}
\varphi(\boldsymbol{\eta},\boldsymbol{\xi},t;\hbar)=(\pi\hbar)^{-d/4}\Big(\det\,\frac{\partial(\boldsymbol{\eta}_t-i\boldsymbol{\xi}_t)}{\partial(\boldsymbol{\eta}-i\boldsymbol{\xi})}\Big)^{-1/2}\chi_{0}(\boldsymbol{z}(\boldsymbol{\eta},\boldsymbol{\xi}));\hbar)
\label{psfourierampl}
\end{eqnarray}

The approximation $(\ref{psfourier})$ is quite similar to the Fourier integral representation that has been constructed for the Schr\"{o}dinger wavefunction in configuration space in \cite{LaSi}, as the complex phase $F$ is quadratic and the ampitude $\varphi$ does not depend on $\boldsymbol{X}=(\boldsymbol{q},\boldsymbol{p})$ but only on the integrated variable $\boldsymbol{Y} =(\boldsymbol{\eta},\boldsymbol{\xi})$. Such a similarity is anticipated since our phase space construction departs from a wave packet with complex quadratic phase.

The Fourier integral $(\ref{psfourier})$ 
\begin{equation}
\Psi^{\hbar}(\boldsymbol{X},t)=\Big(\frac{1}{2\pi\hbar}\Big)^d\int\varphi(\boldsymbol{Y},t;\hbar)\,e^{\frac{i}{\hbar}F(\boldsymbol{X},\boldsymbol{Y},t)}\,d\boldsymbol{Y}
\end{equation}
can be further approximated by applying the \textit{Stationary Complex Phase Theorem} to derive the leading term of the phase space wavefunction.

A point $\boldsymbol{Y}$ which satisfies the conditions 
 \begin{equation}
{\rm Im}\,F(\boldsymbol{X},\boldsymbol{Y},t)=0 \ , \ \ \frac{\partial F}{\partial\boldsymbol{\eta}}(\boldsymbol{X},\boldsymbol{Y},t)=\boldsymbol{0} \ , \ \ \frac{\partial F}{\partial\boldsymbol{\xi}}(\boldsymbol{X},\boldsymbol{Y},t)=\boldsymbol{0} 
\end{equation}
is called a \textit{real stationary point} of the phase function $F$.

Let now $\varphi(\boldsymbol{Z},t;\hbar)$ and $F(\boldsymbol{X},\boldsymbol{Z},t)$ be the almost analytic extensions of $\varphi(\boldsymbol{Y},t;\hbar)$ and $F(\boldsymbol{X},\boldsymbol{Y},t)$ with respect to $\boldsymbol{Y}\in\mathbb{R}^{2d}$, to the complex variable $\boldsymbol{Z}=\boldsymbol{Y}+i\boldsymbol{\Xi}\in\mathbb{C}^{2d}$.
\footnote{In particular, $\mathbfcal{Q}(\boldsymbol{Z},t)$ and $\boldsymbol{\eta}_t, \boldsymbol{\xi}_t$ entering the expression of $F$ are defined as almost analytic extensions in an analogous way.} Then, in some neighborhood of $\Lambda_t$, the equation
\begin{equation}
\frac{\partial}{\partial \boldsymbol{Z}}{F}(\boldsymbol{X},\boldsymbol{Z},t)=\boldsymbol{0} 
\end{equation}
has a unique complex solution $\boldsymbol{Z}=\boldsymbol{Z}(\boldsymbol{X},t)$, and by the \textit{Stationary Complex Phase Theorem} \cite{MeSj,MSS, NSS}, the leading approximation of the Fourier integral is
\begin{equation}
\fl \Psi^{\hbar}(\boldsymbol{X},t)=\frac{\varphi(\boldsymbol{Z}(\boldsymbol{X},t),t;\hbar)}{\sqrt{\det\,\frac{\partial^2 F}{\partial\boldsymbol{Z}^2}(\boldsymbol{X},\boldsymbol{Z}(\boldsymbol{X},t),t)}}\,e^{\frac{i}{\hbar}F(\boldsymbol{X},\boldsymbol{Z}(\boldsymbol{X},t),t))} \Big(1+o(\hbar)\Big) \ .
\label{eq:AsSolution}
\end{equation}
Outside this neighborhood, $\Psi^{\hbar}(\boldsymbol{X},t)=O(\hbar^\infty)$ for any fixed $(\boldsymbol{X},t)$.

For $\boldsymbol{X}=(\boldsymbol{q},\boldsymbol{p})\in\Lambda_t$, $\Lambda_t=g_t\Lambda_0$ being the propagated Lagrangian manifold, $\boldsymbol{Z}(\boldsymbol{X},t)=\boldsymbol{Y}=(\boldsymbol{\eta},\boldsymbol{\xi})=g_{-t}\boldsymbol{X}\in\Lambda_0$, becomes a real stationary point, and it is unique. Then the asymptotic solution $\Psi^{\hbar}$ is concentrated on the Lagrangian manifold; its restriction on $\Lambda_t$ is given by
\begin{eqnarray}
\fl \Psi^{\hbar}|_{\Lambda_t\times \mathbb{R}} (\boldsymbol{X},t)=\varphi(g_{-t}\boldsymbol{X},t;\hbar)\Big(\det\,\frac{\partial^2 F}{\partial \boldsymbol{Y}^2}(\boldsymbol{X},g_{-t}\boldsymbol{X},t)\Big)^{-1/2}\,e^{\frac{i}{\hbar}F(\boldsymbol{X}, g_{-t}\boldsymbol{X},t)}\Big(1+o(\hbar)\Big) \nonumber \\
=(\pi\hbar)^{-d/4}\Big(\det\,\frac{\partial(\boldsymbol{\eta}_t-i\boldsymbol{\xi}_t)}{\partial(\boldsymbol{\eta}-i\boldsymbol{\xi})}\Big)^{-1/2}_{(\boldsymbol{\eta},\boldsymbol{\xi})=g_{-t}(\boldsymbol{q},\boldsymbol{p})}\,R_0(\boldsymbol{q}_{-t}) \nonumber \\
\times \frac{\exp\frac{i}{\hbar}\Big(-\frac{\boldsymbol{p}\cdot\boldsymbol{q}}{2}+S_0(\boldsymbol{q}_{-t})+ A (g_{-t}(\boldsymbol{q},\boldsymbol{p}),t)\Big)}{\sqrt{\det \,\Big(\textbf{I}-i\,\frac{\partial^2S_0}{\partial \boldsymbol{x}^2}(\boldsymbol{q}_{-t})\Big)\,{\rm det}\,\frac{\partial^2F}{\partial \boldsymbol{Y}^2}(\boldsymbol{q},\boldsymbol{p},g_{-t}(\boldsymbol{q},\boldsymbol{p}),t)}}\Big(1+o(\hbar)\Big) \ .
\label{eq:asympsol}
\end{eqnarray}

We note that the transported phase, 
\begin{equation}
S(\boldsymbol{q},t)=S_0(\boldsymbol{q}_{-t})+ A (\boldsymbol{q}_{-t},\boldsymbol{p}_{-t},t) 
\end{equation}
where $\boldsymbol{q}_{-t}=\boldsymbol{q}_{-t}(\boldsymbol{q},\frac{\partial S}{\partial\boldsymbol{x}}(\boldsymbol{q}))$, is the solution of the corresponding Cauchy problem for the \textit{Hamilton-Jacobi equation}
\begin{equation}
\frac{\partial S}{\partial t}+H\Big(\boldsymbol{q},\frac{\partial S}{\partial \boldsymbol{x}}\Big)=0 
\end{equation}
with initial data 
\begin{equation}
S(\boldsymbol{q},0)=S_0(\boldsymbol{q})
\end{equation}
inducing the Lagrangian manifold $\Lambda_t$.

Thus, by means of the Stationary Complex Phase Theorem, we have constructed the explicit asymptotic approximation $(\ref{eq:asympsol})$ along the transported manifold $\Lambda_t$, and the rather involved approximation $(\ref{eq:AsSolution})$ in a neighborhood of $\Lambda_t$. The later is of little practical importance at fixed $t$ and $(\boldsymbol{q},\boldsymbol{p})\notin\Lambda_t$, as there we have that $\Psi^\hbar(\boldsymbol{q},\boldsymbol{p},t)=O(\hbar^\infty)$. It is, however, important to construct an asymptotic solution in a semi-classically `narrow' neighborhood of the manifold $\Lambda_t$, for which we utilize the Complex WKB Theory developed by Maslov \cite{Mas2}. 

\section{Narrow Beam Solutions in Phase Space}
\label{nbsolution}

In this section we proceed to construct an asymptotic solution in a semi-classically narrow tubular neighborhood of the transported Lagrangian manifold $\Lambda_t$, in a more direct and practical form by exploiting the Complex WKB Method as underlined by Maslov \cite{Mas2}, in phase space.

The approximation (\ref{eq:AsSolution}) indicates a narrow beam ansatz
\begin{equation}
\Psi_{\scriptscriptstyle B}^{\hbar}(\boldsymbol{X},t)=\chi(\boldsymbol{X},t;\hbar)\,e^{\frac{i}{\hbar}\,\Phi(\boldsymbol{X},t)} 
\label{psansatz}
\end{equation}
with \textit{complex phase}, as an asymptotic solution of the semi-classical Cauchy problem $(\ref{eq:pscauchy})$, which must satisfy the equation
\begin{equation}
i\hbar\,\frac{\partial\Psi_{\scriptscriptstyle B}^{\hbar}}{\partial t}-\mathcal{H}\Big(\stackrel{\boldsymbol{\omega}}{\boldsymbol{X}},-i\hbar \stackrel{\boldsymbol{\omega}}{\frac{\partial}{\partial \boldsymbol{X}}}\Big)\Psi_{\scriptscriptstyle B}^{\hbar}=O(\hbar^2)
\label{eq:psweyl}
\end{equation}
and the initial conditions 
\begin{equation}
\Phi(\boldsymbol{X}, 0)= \theta_0(\boldsymbol{X}) \ , \ \ \chi(\boldsymbol{X},0;\hbar)= \chi_{0}(\boldsymbol{X};\hbar)
\label{eq:phi0} 
\end{equation}
as given by $(\ref{eq:chi0}), (\ref{eq:theta0})$. 

The choice of such initial data is justified by $(\ref{eq:AsSolution})$,
$(\ref{psfourierphase}), (\ref{psfourierampl})$ for $t=0$, since by the estimate 
\begin{equation}
\boldsymbol{Z}(\boldsymbol{X},0)=\boldsymbol{X}+O\Big(\Big|\frac{1}{2}\textbf{J}\boldsymbol{X}-\frac{\partial\theta_0}{\partial \boldsymbol{X}}(\boldsymbol{X})\Big|\Big)
\end{equation}
we have
 
(i) for the phase
\begin{eqnarray}
F (\boldsymbol{X},\boldsymbol{Z}(\boldsymbol{X},0),0)-\theta_0(\boldsymbol{X})\\ \nonumber
= \theta_0(\boldsymbol{Z}(\boldsymbol{X},0))+\frac{1}{2}\boldsymbol{X}\cdot \textbf{J}\boldsymbol{Z}(\boldsymbol{X},0)+\frac{i}{4}(\boldsymbol{Z}(\boldsymbol{X},0)-\boldsymbol{X})^2-\theta_0(\boldsymbol{X})\\ \nonumber 
=O\Big(\Big|\frac{1}{2}\textbf{J}\boldsymbol{X}-\frac{\partial\theta_0}{\partial \boldsymbol{X}}(\boldsymbol{X})\Big|^2\Big)
\end{eqnarray}

(ii) for the amplitude
\begin{equation}
\frac{\varphi(\boldsymbol{Z}(\boldsymbol{X},0),0;\hbar)}{\sqrt{\det\,\frac{\partial^2F}{\partial\boldsymbol{Z}^2}(\boldsymbol{X},\boldsymbol{Z}(\boldsymbol{X},0),0)}}-\chi_{0}(\boldsymbol{X})
=O\Big(\Big|\frac{1}{2}\textbf{J}\boldsymbol{X}-\frac{\partial\theta_0}{\partial \boldsymbol{X}}(\boldsymbol{X})\Big|\Big) \ .
\end{equation}

\subsection{Derivation of the Canonical System in Double Phase Space}

By substituting $(\ref{psansatz})$ into $(\ref{eq:psweyl})$ we have that 
\begin{equation}
\fl \Bigg(i\hbar\,\chi^{-1}\frac{\partial\chi}{\partial t}-\frac{\partial \Phi}{\partial t}-\chi^{-1}\,e^{-\frac{i}{\hbar}\Phi}
H\Big(\frac{\boldsymbol{X}}{2}+i\hbar\, \textbf{J}\frac{\partial}{\partial \boldsymbol{X}}\Big)
\Big(\chi\,e^{\frac{i}{\hbar}\Phi}\Big)\Bigg)\chi\,e^{\frac{i}{\hbar}\Phi}=O(\hbar^2) 
\end{equation}
and by using the Commutation Formula (\ref{QuRep}) we get
\begin{eqnarray}
\label{eq:sclexp}
\fl \Bigg(i\hbar\,\chi^{-1}\frac{\partial\chi}{\partial t}-\frac{\partial \Phi}{\partial t}-H\Big(\frac{\boldsymbol{X}}{2}-\textbf{J}\frac{\partial \Phi}{\partial \boldsymbol{X}}\Big)-i\hbar\,\chi^{-1}\textbf{J}\frac{\partial H}{\partial \boldsymbol{X}}\Big(\frac{\boldsymbol{X}}{2}-\textbf{J}\frac{\partial \Phi}{\partial \boldsymbol{X}}\Big)\cdot \frac{\partial\chi}{\partial \boldsymbol{X}}\\ \nonumber 
-\frac{i\hbar}{2}\,{\rm tr}\,\textbf{J}\frac{\partial^2 \Phi}{\partial \boldsymbol{X}^2}\textbf{J}\,\!\Big(\frac{\partial^2 H}{\partial \boldsymbol{X}^2}\Big)\Big(\frac{\boldsymbol{X}}{2}-\textbf{J}\frac{\partial \Phi}{\partial \boldsymbol{X}}\Big)\Bigg)\chi\,e^{\frac{i}{\hbar}\Phi}=O(\hbar^2) \ .
\end{eqnarray}

For the separation of the orders of magnitude and induction of the hierarchy of equations comprising the \textit{canonical system,} one must use the calculus of exponential asymptotic equivalence. This is due to the fact that the above asymptotic relation is not uniform in a `narrow' neighborhood of $\Lambda_t$, since, for $\boldsymbol{X}\in\mathbb{R}^{2d}\backslash \Lambda_t$, we have that ${\rm Im} \,\Phi(\boldsymbol{X},t) >0$, and at any fixed point $\boldsymbol{X}$, for fixed $t$, outside $\Lambda_t$ we have $e^{\frac{i}{\hbar}\,\Phi(\boldsymbol{X},t)}=O(\hbar^{\infty})$.

The \textit{Hamilton-Jacobi equation} is
\begin{equation}
\frac{\partial\Phi}{\partial t}+\mathcal{H}\Big(\boldsymbol{X},\frac{\partial\Phi}{\partial \boldsymbol{X}}\Big)=0
\label{eq:phshj}
\end{equation}
which, due to the special form of the symbol $\mathcal{H}(\boldsymbol{X},\boldsymbol{P})=H\Big(\frac{\boldsymbol{X}}{2}-\textbf{J}\boldsymbol{P}\Big)$ 
(see $(\ref{eq:phsham})$), takes the form of the \textit{Weyl-symmetrized Hamilton-Jacobi equation} 
\begin{equation}
\frac{\partial\Phi}{\partial t}+H\Big(\frac{\boldsymbol{X}}{2}-\textbf{J}\frac{\partial\Phi}{\partial \boldsymbol{X}}\Big)=0 \ , \ \ t\geq 0
\label{eq:phshj1}
\end{equation}
with initial condition $\Phi(\boldsymbol{X}, 0)= \theta_0(\boldsymbol{X})$.

By differentiating $(\ref{eq:theta0})$, for $ \boldsymbol{X}\in \Lambda_0 $ we get
\begin{equation}
\frac{\partial^2\theta_0}{\partial \boldsymbol{X}^2}=\left(\begin{array}{ccc} i\textbf{I}+(S_{0 \, \boldsymbol{x}\!\boldsymbol{x}}+i\textbf{I})^{-1} & S_{0 \, \boldsymbol{x}\!\boldsymbol{x}}(S_{0 \, \boldsymbol{x}\!\boldsymbol{x}}+i\textbf{I})^{-1} \\ S_{0 \, \boldsymbol{x}\!\boldsymbol{x}}(S_{0 \, \boldsymbol{x}\!\boldsymbol{x}}+i\textbf{I})^{-1} & \textbf{0} \end{array} \right) 
\end{equation}
Thus, by the assumption ${\rm det}\,\frac{\partial^2 S_0}{\partial\boldsymbol{x}^2}\neq 0$, it follows
that the initial phase $\theta_0$ satisfies the condition
\begin{equation}
\fl {\rm rank}\Big({\rm Im}\, \frac{\partial^{2} \theta_0}{\partial \boldsymbol{X}^2}(\boldsymbol{X})\Big)= d \ , \ \ \rm{for} \ \boldsymbol{X} \in \{\boldsymbol{X}\in\mathbb{R}^{2d}\,|\, \rm{Im}\, \theta_0 (\boldsymbol{X})=0\}= \Lambda_0
\end{equation}
which is a necessary condition for the validity of Maslov's construction of a narrow beam asymptotic solution.

The corresponding initial value problem for the \textit{phase space transport equation} is
\begin{equation}
\fl \frac{\partial\chi}{\partial t}-\textbf{J}\frac{\partial H}{\partial \boldsymbol{X}}\Big(\frac{\boldsymbol{X}}{2}-\textbf{J}\frac{\partial \Phi}{\partial \boldsymbol{X}}\Big)\cdot \frac{\partial\chi}{\partial \boldsymbol{X}}-\frac{1}{2}\,{\rm tr}\,\textbf{J}\frac{\partial^2 \Phi}{\partial \boldsymbol{X}^2}\textbf{J}\,\!\Big(\frac{\partial^2 H}{\partial \boldsymbol{X}^2}\Big)\Big(\frac{\boldsymbol{X}}{2}-\textbf{J}\frac{\partial \Phi}{\partial \boldsymbol{X}}\Big)\,\chi=0
\end{equation}
with initial condition $(\ref{eq:chi0})$.

\begin{remark}
Berezin and Shubin \cite{BeSh} have constructed a similar canonical system for the WKB asymptotic expansion of the symbol of the Schr\"{o}dinger propagator (\ref{eq:eqprop}) in the Weyl quantization. This expansion stands as a semi-classical correction to the Hamiltonian flow as canonical transformation and the Liouville transport equation. Subsequently, Marinov \cite{Mar} developed a purely classical geometric argument for the deduction of the same variant of the Hamilton-Jacobi equation as a generator of canonical transformations.
\end{remark}

\subsection{The Hamiltonian System in Double Phase Space}

The Hamiltonian system generated by the Hamiltonian $\mathcal{H}(\boldsymbol{X},\boldsymbol{P})=H\Big(\frac{\boldsymbol{X}}{2}-\textbf{J}\boldsymbol{P}\Big)$ in double phase space, is given by the equations 
\begin{equation}
\frac{d\boldsymbol{X}}{d t}=\frac{\partial\mathcal{H}}{\partial \boldsymbol{P}} \ , \ \ \frac{d\boldsymbol{P}}{dt}=-\frac{\partial\mathcal{H}}{\partial \boldsymbol{X}} \ , \ \ t\geq 0
\label{eq:phhameq}
\end{equation}
with initial conditions
\begin{equation}
\fl \boldsymbol{X}|_{t=0}=\boldsymbol{X}_0(\boldsymbol{\alpha})\in \Lambda_0 \ , \ \ \boldsymbol{P}|_{t=0}=\boldsymbol{P}_0(\boldsymbol{\alpha}):=\frac{\partial\theta_0}{\partial \boldsymbol{X}}(\boldsymbol{X}_0(\boldsymbol{\alpha}))=\frac{1}{2}\textbf{J}\boldsymbol{X}_0(\boldsymbol{\alpha}) \ , \ \boldsymbol{\alpha}\in \mathbb{R}^d
\label{eq: phtrindata}
\end{equation}
where $\boldsymbol{X}=\boldsymbol{X}_0(\boldsymbol{\alpha})$ is a parametrization of $\Lambda_0$.

The initial data $(\ref{eq: phtrindata})$ define the double phase Lagrangian manifold
\begin{equation}
L_0=\{(\boldsymbol{X},\boldsymbol{P})\in\mathbb{R}^{4d}\,|\,\boldsymbol{X}=\boldsymbol{X}_0(\boldsymbol{\alpha}) \ , \ \ \boldsymbol{P}=\boldsymbol{P}_0(\boldsymbol{\alpha}) \ , \ \ \boldsymbol{\alpha}\in\mathbb{R}^d\}
\end{equation}
which has dimension ${\rm dim}\,L_0=d$.
The characteristic systems $(\ref{eq:phhameq})$ induces a Hamiltonian flow in double phase space with an invariant manifold the symplectic plane\footnote{The symplectic plane $\mathcal{S}$ is symplectomorphic to the phase space $\mathbb{R}^{2d}$. We note that the restriction of the canonical 2-form $\boldsymbol{\Omega}^2=d\boldsymbol{P}\wedge d\boldsymbol{X}$ and the parametrized canonical 1-form $\boldsymbol{\Omega}^1_\lambda=\lambda\,\boldsymbol{P}\cdot d\boldsymbol{X}+(\lambda-1)\,\boldsymbol{X}\cdot d\boldsymbol{P}$ of double phase space on the invariant symplectic plane $\mathcal{S}$ are
\begin{equation*}
\boldsymbol{\Omega}^2|_{\mathcal{S}}=d\boldsymbol{p}\wedge d\boldsymbol{q}=\boldsymbol{\omega}^2 \ , \ \ \boldsymbol{\Omega}^1_\lambda|_{\mathcal{S}}=\frac{1}{2}(\boldsymbol{p}\cdot d\boldsymbol{q}-\boldsymbol{q}\cdot d\boldsymbol{p})=\boldsymbol{\omega}^1_W
\end{equation*}
the last for all $\lambda\in[0,1]$.}
\begin{equation}
\mathcal{S}=\{(\boldsymbol{X}, \boldsymbol{P})\in\mathbb{R}^{4d} \,|\,\boldsymbol{P}=\frac 12\textbf{J}\boldsymbol{X}\} \ .
\end{equation}
This Hamiltonian flow moves $L_0$ to
\begin{equation}
L_t=\{(\boldsymbol{X},\boldsymbol{P})\in\mathbb{R}^{4d} \,|\,\boldsymbol{X}=\boldsymbol{X}_t(\boldsymbol{\alpha}),
\, \boldsymbol{P}= \boldsymbol{P}_t(\boldsymbol{\alpha})=\frac{1}{2}\textbf{J}\boldsymbol{X}_t(\boldsymbol{\alpha})\} \ ,
\end{equation}
and for all $t\geq 0$, $\Lambda_t$ is the canonical projections of $L_t$ onto phase space.

The equations $(\ref{eq:phhameq})$ can be simplified in the form
\begin{equation}
 \frac{d\boldsymbol{X}}{dt}=\textbf{J}\frac{\partial H}{\partial \boldsymbol{X}}\Big(\frac{\boldsymbol{X}}{2}-\textbf{J}\boldsymbol{P}\Big) \ , \ \ \frac{d\boldsymbol{P}}{dt}=-\frac{1}{2}\frac{\partial H}{\partial \boldsymbol{X}}\Big(\frac{\boldsymbol{X}}{2}-\textbf{J}\boldsymbol{P}\Big)
\end{equation}
from which we deduce that the $2d$ integrals of the motion
\begin{equation}
\boldsymbol{c}(\boldsymbol{X},\boldsymbol{P})=\frac{\boldsymbol{X}}{2}+ \textbf{J}\boldsymbol{P}
\end{equation}
take the value zero
\begin{equation}
\boldsymbol{c}(\boldsymbol{X}_0(\boldsymbol{\alpha}),\boldsymbol{P}_0(\boldsymbol{\alpha}))=\frac{1}{2}\boldsymbol{X}_0(\boldsymbol{\alpha})+\textbf{J}\boldsymbol{P}_0(\boldsymbol{\alpha})=\boldsymbol{0} \ .
\end{equation}
Since $\boldsymbol{c}=\boldsymbol{0}$, we have $\frac{\boldsymbol{X}}{2}-\textbf{J}\boldsymbol{P}=\boldsymbol{X}$, and so the equations $(\ref{eq:phhameq})$ are further simplified to
\begin{equation}
\frac{d\boldsymbol{X}}{dt}=\textbf{J}\frac{\partial H}{\partial \boldsymbol{X}}(\boldsymbol{X}) \ , \ \ \frac{d\boldsymbol{P}}{dt}=-\frac{1}{2}\frac{\partial H}{\partial \boldsymbol{X}}(\boldsymbol{X}) \ .
\end{equation}

Therefore, the canonical projection of the Hamiltonian flow with respect to $\mathcal{H}$ from
$\mathbb{R}^{4d}$ onto $\mathbb{R}^{2d}$ is in fact the Hamiltonian flow with respect to $H$ 
\begin{equation}
\frac{d\boldsymbol{X}}{dt}=\textbf{J}\frac{\partial H}{\partial \boldsymbol{X}}(\boldsymbol{X})
\end{equation}
with initial condition 
\begin{equation}
\boldsymbol{X}|_{t=0}=\boldsymbol{X}_0(\boldsymbol{\alpha}) \ .
\end{equation}

The corresponding \textit{variational system} for the system $(\ref{eq:phhameq})$ is given by the equations
\begin{equation}
\frac{d}{dt}\left(
\begin{array}{ccc}
\textbf{C} \\
\textbf{D}
\end{array}\right)=\left(\begin{array}{ccc}
\mathcal{H}_{\boldsymbol{P}\!\boldsymbol{X}} & \mathcal{H}_{\boldsymbol{P}\!\boldsymbol{P}} \\
-\mathcal{H}_{\boldsymbol{X}\! \boldsymbol{X}} & -\mathcal{H}_{\boldsymbol{X}\!\boldsymbol{P}} 
\end{array}\right)\left(
\begin{array}{ccc}
\textbf{C} \\
\textbf{D} 
\end{array}\right) \ .
\label{eq:phvarsys}
\end{equation}

Since ${\rm Im}\,\frac{\partial\theta_0}{\partial \boldsymbol{X}}(\boldsymbol{X}_0(\boldsymbol{\alpha}))=\textbf{0}$, the solutions of the systems $(\ref{eq:phhameq})$, $(\ref{eq:phvarsys})$ exist on any interval $0\leq t \leq T$.

Moreover, by $\mathcal{H}(\boldsymbol{X},\boldsymbol{P})=H\Big(\frac{\boldsymbol{X}}{2}-\textbf{J}\boldsymbol{P}\Big)$, the system $(\ref{eq:phvarsys})$ is reduced to 
\begin{equation}
\frac{d}{dt}\left(
\begin{array}{ccc}
\textbf{C} \\
\textbf{D}
\end{array}\right)=\left(\begin{array}{ccc}
\frac{1}{2}\textbf{J}H_{\boldsymbol{X}\!\boldsymbol{X}} & -\textbf{J}H_{\boldsymbol{X}\!\boldsymbol{X}}\textbf{J} \\
-\frac{1}{4}H_{\boldsymbol{X}\!\boldsymbol{X}} & \frac{1}{2}H_{\boldsymbol{X}\!\boldsymbol{X}}\textbf{J}
\end{array}\right)\left(
\begin{array}{ccc}
\textbf{C} \\
\textbf{D} 
\end{array}\right)
\end{equation}
with initial conditions
\begin{equation} 
\textbf{C}(\boldsymbol{X}_0(\boldsymbol{\alpha}),\boldsymbol{P}_0(\boldsymbol{\alpha}),0)=\textbf{I} \ , \ \ \ \textbf{D}(\boldsymbol{X}_0(\boldsymbol{\alpha}),\boldsymbol{P}_0(\boldsymbol{\alpha}),0)=\frac{\partial^{2} \theta_0}{\partial \boldsymbol{X}^2}(\boldsymbol{X}_0(\boldsymbol{\alpha})) \ . 
\end{equation}
The argument of the Hessian elements is $(\boldsymbol{X}_t(\boldsymbol{\alpha}),\boldsymbol{P}_t(\boldsymbol{\alpha}))$ and therefore the variational matrices $\textbf{C}, \textbf{D}$ depend on $(\boldsymbol{\alpha},t)$. We shall use the notation $\textbf{C}(\boldsymbol{\alpha},t):=\textbf{C}(\boldsymbol{X}_t(\boldsymbol{\alpha}),\boldsymbol{P}_t(\boldsymbol{\alpha}),t)$ and $\textbf{D}(\boldsymbol{\alpha},t):=\textbf{D}(\boldsymbol{X}_t(\boldsymbol{\alpha}),\boldsymbol{P}_t(\boldsymbol{\alpha}),t)$.

It follows that the matrix of the variational system is singular and the corresponding dynamics is constrained by the linear relation
\begin{equation}
\frac{d}{dt}\Big(\textbf{D}-\frac{1}{2}\textbf{J}\textbf{C}\Big)=\textbf{0} \ .
\end{equation}
Thus the system de-couples the system into independent equations 
\begin{equation}
\fl \frac{d\textbf{C}}{dt}= \textbf{J}H_{\boldsymbol{X}\!\boldsymbol{X}}\, \textbf{C}-\frac{1}{2}\textbf{J}H_{\boldsymbol{X}\!\boldsymbol{X}}(i\textbf{J}+\textbf{I}) \ , \ \ \frac{d\textbf{D}}{dt}= H_{\boldsymbol{X}\!\boldsymbol{X}}\textbf{J}\, \textbf{D}-\frac{1}{4}H_{\boldsymbol{X}\!\boldsymbol{X}}(i\textbf{J}+\textbf{I}) \ .
\end{equation}

The variational system is equivalently put in the form of a Riccati equation for the phase space narrow beam anisotropy matrix $\tilde\mathbfcal{Q}(\boldsymbol{\alpha},t)=\textbf{D}(\boldsymbol{\alpha},t)\textbf{C}(\boldsymbol{\alpha},t)^{-1}$, in particular 
\begin{equation}
\frac{d\tilde \mathbfcal{Q}}{dt}+\tilde \mathbfcal{Q}\,\mathcal{H}_{\boldsymbol{P}\!\boldsymbol{P}}\,\tilde\mathbfcal{Q}+\mathcal{H}_{\boldsymbol{X}\!\boldsymbol{P}}\,\tilde\mathbfcal{Q}+\tilde\mathbfcal{Q}\,\mathcal{H}_{\boldsymbol{P}\!\boldsymbol{X}}+\mathcal{H}_{\boldsymbol{X}\!\boldsymbol{X}}=\textbf{0}
\label{eq:phricatti0}
\end{equation}
or 
\begin{equation}
\frac{d\tilde\mathbfcal{Q}}{dt}-\tilde\mathbfcal{Q}\,\textbf{J}H_{\boldsymbol{X}\!\boldsymbol{X}}\textbf{J}\,\tilde\mathbfcal{Q}-\frac{1}{2}H_{\boldsymbol{X}\!\boldsymbol{X}}\textbf{J}\,\tilde\mathbfcal{Q}+\frac{1}{2}\tilde\mathbfcal{Q}\,\textbf{J}H_{\boldsymbol{X}\!\boldsymbol{X}}+\frac{1}{4}H_{\boldsymbol{X}\!\boldsymbol{X}}=\textbf{0}
\label{eq:phricatti}
\end{equation}
with initial condition
\begin{equation}
\tilde\mathbfcal{Q}(0)=\frac{\partial^{2} \theta_0}{\partial \boldsymbol{X}^2} (\boldsymbol{X}_0(\boldsymbol{\alpha})) \ .
\end{equation}

\subsection{Asymptotic Solution in a neighborhood of $\Lambda_t$ (Complex Phase)}

In the sequel we assume that for all $t \in [0,T]$, the manifold $\Lambda_t$ is simply connected and smooth, and that
\begin{equation}
{\rm rank}\Big(\frac{\partial\boldsymbol{X}_t}{\partial\boldsymbol{\alpha}}(\boldsymbol{\alpha})\Big)=d \ .
\end{equation}
In case that the last condition is not satisfied, we need to consider appropriate atlases on $\Lambda_t$ as in the general theory of Maslov's canonical operator in the complex situation. 

Then, we have the following facts (\cite{Mas2}, Theorem 1.1):
\begin{enumerate}
\item
for all $t \in [0,T]$, $\det \,\textbf{C}\ne 0$, and the system 
\begin{equation}\label{eq:orthog}
(\boldsymbol{X}-\boldsymbol{X}_t(\boldsymbol{\alpha}))\cdot \frac{\partial \boldsymbol{X}_t}{\partial \alpha_j}(\boldsymbol{\alpha})=0 \ , \ \ j=1,\ldots,d
\end{equation}
has a unique solution with respect to $\boldsymbol{\alpha}$, smooth with respect to $\boldsymbol{X}$, namely $\boldsymbol{\alpha}=\boldsymbol{\alpha}(\boldsymbol{X},t)$ in a closed neighborhood $\Delta_t$ of $\Lambda_t$. This condition implies that to every point $\boldsymbol{X}$ external, but in proximity to $\Lambda_t$, there exists a unique closest point 
$\boldsymbol{X}_t(\boldsymbol{\alpha}(\boldsymbol{X},t))$ on $\Lambda_t$.

\item
in any interval $t \in [0 \, T]$, an asymptotic solution of the problem for the Weyl-symmetrized Hamilton-Jacobi equation $(\ref{eq:phshj})$ exists and is defined in $\Delta_t$ by
 
\begin{eqnarray}
\fl \Phi(\boldsymbol{X},t)=\Big\{\theta_0(\boldsymbol{X}_0(\boldsymbol{\alpha}))+\boldsymbol{P}_t(\boldsymbol{\alpha})\cdot (\boldsymbol{X}-\boldsymbol{X}_t(\boldsymbol{\alpha}))+\int\displaylimits_0^t\boldsymbol{P}_\tau(\boldsymbol{\alpha})\cdot \frac{d\boldsymbol{X}_\tau(\boldsymbol{\alpha})}{d\tau}\,d\tau\\ \nonumber
\fl -\mathcal{H}(\boldsymbol{X}_0(\boldsymbol{\alpha}),\boldsymbol{P}_0(\boldsymbol{\alpha}))\,t+\frac{1}{2}(\boldsymbol{X}-\boldsymbol{X}_t(\boldsymbol{\alpha}))\cdot \tilde \mathbfcal{Q}(\boldsymbol{\alpha},t)(\boldsymbol{X}-\boldsymbol{X}_t(\boldsymbol{\alpha}))\Big\}_{\boldsymbol{\alpha}=\boldsymbol{\alpha}(\boldsymbol{X},t)}
\end{eqnarray}
where $\tilde \mathbfcal{Q}(\boldsymbol{\alpha},t)$ is the solution of the initial value problem for the phase space Riccati equation $(\ref{eq:phricatti})$.

Since $\boldsymbol{P}_t(\boldsymbol{\alpha})=\frac{1}{2}\textbf{J}\boldsymbol{X}_t(\boldsymbol{\alpha})$, the above formula is further simplified to
\begin{eqnarray}
\fl \Phi(\boldsymbol{X},t)=\Big\{\theta_0(\boldsymbol{X}_0(\boldsymbol{\alpha}))+\frac{1}{2}\textbf{J}\boldsymbol{X}_t(\boldsymbol{\alpha})\cdot (\boldsymbol{X}-\boldsymbol{X}_t(\boldsymbol{\alpha}))
+A_w(\boldsymbol{X}_0(\boldsymbol{\alpha}),t)\\ \nonumber 
+\frac{1}{2}(\boldsymbol{X}-\boldsymbol{X}_t(\boldsymbol{\alpha}))\cdot \tilde \mathbfcal{Q}{(\boldsymbol{\alpha},t)}(\boldsymbol{X}-\boldsymbol{X}_t(\boldsymbol{\alpha}))\Big\}_{\boldsymbol{\alpha}=\boldsymbol{\alpha}(\boldsymbol{X},t)}
\end{eqnarray}
where $ A_w$ is the symmetrized phase space action(see \cite{Lit1}, eq (7.25); also \cite{Gos5}), Section 10.1.1)
\begin{equation}
A_w(\boldsymbol{X}_0(\boldsymbol{\alpha}),t)=
\int\displaylimits_0^t \Big(\frac{1}{2}\textbf{J}\boldsymbol{X}_\tau(\boldsymbol{\alpha})\cdot \frac{d\boldsymbol{X}_\tau(\boldsymbol{\alpha})}{d\tau} -H(\boldsymbol{X}_0(\boldsymbol{\alpha}))\Big)d\tau
\end{equation}

Note that $A_w$ is expressed in terms of the classical action $(\ref{eq:action})$ as
\begin{equation}
\fl A_w(\boldsymbol{X}_0(\boldsymbol{\alpha}),t)=A(\boldsymbol{X}_0(\boldsymbol{\alpha}),t)- \frac{\boldsymbol{p}_t(\boldsymbol{\alpha}) \cdot \boldsymbol{q}_t(\boldsymbol{\alpha})-\boldsymbol{p}_0(\boldsymbol{\alpha}) \cdot \boldsymbol{q}_0(\boldsymbol{\alpha})}{2} \ .
\end{equation}

\item 
on any compact set $\mathcal{K}$ outside the neighborhood $\Delta_t$, ${\rm Im}\,\Phi(\boldsymbol{X},t)> C(\mathcal{K})$, where $C(\mathcal{K})$ positive constant.

\item
in the neighborhood $\Delta_t$ there is defined an asymptotic solution of the initial value problem for the phase space transport equation, which is given by
\begin{eqnarray}
\fl \chi(\boldsymbol{X},t;\hbar)=\Bigg\{\frac{\chi_0(\boldsymbol{X}_0(\boldsymbol{\alpha});\hbar)}{\sqrt{{\rm det}\,\textbf{C}_{(\boldsymbol{\alpha},t)}}}\,\exp\frac{1}{2}\int\displaylimits_0^t{\rm tr}\,\frac{\partial^2\mathcal{H}}{\partial \boldsymbol{P}\partial \boldsymbol{X}}(\boldsymbol{X}_\tau(\boldsymbol{\alpha}),\boldsymbol{P}_\tau(\boldsymbol{\alpha}))\,d\tau\Bigg\}_{\boldsymbol{\alpha}=\boldsymbol{\alpha}(\boldsymbol{X},t)}\\ \nonumber 
=\Bigg\{\frac{\chi_0(\boldsymbol{X}_0(\boldsymbol{\alpha});\hbar)}{\sqrt{{\rm det}\,\textbf{C}_{(\boldsymbol{\alpha},t)}}}\Bigg\}_{\boldsymbol{\alpha}=\boldsymbol{\alpha}(\boldsymbol{X},t)}
\end{eqnarray}
by virtue of the variational system $(\ref{eq:phvarsys})$ and the fact that ${\rm tr}\,\mathcal{H}_{\boldsymbol{X}\!\boldsymbol{P}}=0$. 

\end{enumerate}

\subsection{Asymptotic Solution on $\Lambda_t$ (Real Phase)}

We particularize the previous analysis to the construction of the asymptotic solution of the canonical system on the Lagrangian manifold $\Lambda_t$. By differentiating $ \theta_0$, we obtain, for $\boldsymbol{X}\in\Lambda_0$,
\begin{equation}
\frac{\partial \theta_0}{\partial \boldsymbol{X}}(\boldsymbol{X})=\frac{1}{2}\textbf{J}\boldsymbol{X} \ .
\end{equation}

Then, applying Prop. 3.8 in\cite{MaFe}, for the Hamilton-Jacobi equation $(\ref{eq:phshj})$, it follows that if
\begin{equation}
\boldsymbol{P}_0(\boldsymbol{\alpha})=\frac{\partial\Phi}{\partial \boldsymbol{X}}(\boldsymbol{X}_0(\boldsymbol{\alpha}),0)
\end{equation}
then for $t>0$ holds
\begin{equation}
\boldsymbol{P}_t(\boldsymbol{\alpha})=\frac{\partial\Phi}{\partial \boldsymbol{X}}(\boldsymbol{X}_t(\boldsymbol{\alpha}),t) \ .
\end{equation}
Therefore
\begin{equation}
\frac{\partial\Phi}{\partial\boldsymbol{X}}(\boldsymbol{X}_t(\boldsymbol{\alpha}),t)=\frac{1}{2}\textbf{J}\boldsymbol{X}_t(\boldsymbol{\alpha})
\label{eq:derphi}
\end{equation}
along the orbits of the Hamiltonian system $(\ref{eq:phhameq})$.

This result can be also formally derived as follows. By differentiating the Hamilton-Jacobi equation $(\ref{eq:phshj})$ with respect to $\boldsymbol{X}$ we have
\begin{equation}
\frac{\partial}{\partial t}\frac{\partial\Phi}{\partial\boldsymbol{X}}+\frac{\partial\mathcal{H}}{\partial\boldsymbol{X}}+\frac{\partial^2\Phi}{\partial\boldsymbol{X}^2}\frac{\partial\mathcal{H}}{\partial \boldsymbol{P}}=\boldsymbol{0} \ .
\end{equation}
Along the trajectory we have
\begin{equation}
\frac{\partial}{\partial t}\frac{\partial\Phi}{\partial\boldsymbol{X}}=\frac{d}{d t}\frac{\partial\Phi}{\partial\boldsymbol{X}}-\frac{\partial^2\Phi}{\partial\boldsymbol{X}^2}\frac{d\boldsymbol{X}}{dt}
=\frac{d}{d t}\frac{\partial\Phi}{\partial\boldsymbol{X}}-\frac{\partial^2\Phi}{\partial\boldsymbol{X}^2}\frac{\partial \mathcal{H}}{\partial \boldsymbol{P}}
\end{equation}
and thus
\begin{equation}
\frac{d}{d t}\frac{\partial\Phi}{\partial\boldsymbol{X}}=-\frac{\partial \mathcal{H}}{\partial\boldsymbol{X}}\Big(\boldsymbol{X},\frac{\partial\Phi}{\partial\boldsymbol{X}}\Big) \ .
\end{equation}
Then, by the Hamiltonian system $(\ref{eq:phhameq})$, for any $\boldsymbol{X}\in\mathcal{S}$, we identify $\frac{\partial\Phi}{\partial\boldsymbol{X}}$ with the momentum and we derive $(\ref{eq:derphi})$.

We can also formally show that $\frac{\partial^2\Phi}{\partial\boldsymbol{X}^2}$ satisfies a Riccati equation in double phase space. Indeed, by differentiating two times the Hamilton-Jacobi equation $(\ref{eq:phshj})$ with respect to $\boldsymbol{X}$ we obtain
\begin{equation}
\fl \frac{\partial}{\partial t}\frac{\partial^2\Phi}{\partial\boldsymbol{X}^2}+\frac{\partial^2\mathcal{H}}{\partial\boldsymbol{X}^2}+\frac{\partial^2\Phi}{\partial\boldsymbol{X}^2}\frac{\partial^2\mathcal{H}}{\partial \boldsymbol{P}\partial\boldsymbol{X}}+\frac{\partial^2\mathcal{H}}{\partial\boldsymbol{X}\partial \boldsymbol{P}}\frac{\partial^2\Phi}{\partial\boldsymbol{X}^2}+\frac{\partial^2\Phi}{\partial\boldsymbol{X}^2}\frac{\partial^2\mathcal{H}}{\partial \boldsymbol{P}^2}\frac{\partial^2\Phi}{\partial\boldsymbol{X}^2}+\frac{\partial \mathcal{H}}{\partial \boldsymbol{P}}\cdot \frac{\partial^3\mathcal{H}}{\partial\boldsymbol{X}^3}=\textbf{0}
\end{equation}
and since along the trajectory 
\begin{equation}
\frac{d}{dt}\frac{\partial^2\Phi}{\partial\boldsymbol{X}^2}=\frac{\partial}{\partial t}\frac{\partial^2\Phi}{\partial\boldsymbol{X}^2}+\frac{d\boldsymbol{X}}{dt}\frac{\partial^3\Phi}{\partial\boldsymbol{X}^3} \ , \ \ \frac{d\boldsymbol{X}}{dt}=\frac{\partial\mathcal{H}}{\partial \boldsymbol{P}}
\end{equation}
we derive the Riccati equation $(\ref{eq:phricatti0})$
\begin{equation}
\fl \frac{d}{dt}\frac{\partial^2\Phi}{\partial\boldsymbol{X}^2}+\frac{\partial^2\Phi}{\partial\boldsymbol{X}^2}\,\mathcal{H}_{\boldsymbol{P}\!\boldsymbol{P}}\,\frac{\partial^2\Phi}{\partial\boldsymbol{X}^2}+\frac{\partial^2\Phi}{\partial\boldsymbol{X}^2}\,\mathcal{H}_{\boldsymbol{P}\!\boldsymbol{X}}+\mathcal{H}_{\boldsymbol{X}\!\boldsymbol{P}}\,\frac{\partial^2\Phi}{\partial\boldsymbol{X}^2}+\mathcal{H}_{\boldsymbol{X}\!\boldsymbol{X}}=\textbf{0}
\end{equation}
if we identify that along the trajectories 
\begin{equation}
\frac{\partial^2\Phi}{\partial \boldsymbol{X}^2}(\boldsymbol{X}_t(\boldsymbol{\alpha}),t)=\tilde \mathbfcal{Q}(\boldsymbol{\alpha},t)
\label{eq:der2phi}
\end{equation}
since they both satisfy the same initial condition for $\boldsymbol{X}\in\Lambda_0$.

Let
\begin{equation}
S(\boldsymbol{X},t):=\Phi|_{\Lambda_t\times \mathbb{R}}(\boldsymbol{X},t) \ \ {\rm and} \ \ R(\boldsymbol{X},t;\hbar):=\chi|_{\Lambda_t\times \mathbb{R}}(\boldsymbol{X},t;\hbar) \ .
\end{equation}
be the restriction of the phase and amplitudes on the manifold $\Lambda_t$.

For every given $\boldsymbol{X}\in\Delta_t\backslash\Lambda_t$, due to the condition $(\ref{eq:orthog})$, there exists unique closest point of $\boldsymbol{X}$, say $\boldsymbol{X}'\in\Lambda_t$. Considering $\boldsymbol{X}'$ as a function of $t$ and $\boldsymbol{X}$, we have that 
\begin{equation}
{\rm det}\,\frac{\partial \boldsymbol{X}'}{\partial \boldsymbol{X}}(\boldsymbol{X},t)\neq 0
\end{equation}
while 
\begin{equation}
\frac{\partial \boldsymbol{X}'}{\partial \boldsymbol{X}}\Big|_{\Lambda_t\times \mathbb{R}}(\boldsymbol{X},t)=\textbf{I} \ .
\end{equation}
Thus, we have 
\begin{eqnarray}
\fl \chi(\boldsymbol{X},t;\hbar)=R(\boldsymbol{X}',t;\hbar)+(\boldsymbol{X}-\boldsymbol{X}')\cdot \frac{\partial \chi}{\partial \boldsymbol{X}}(\boldsymbol{X}',t;\hbar)+O(|\boldsymbol{X}-\boldsymbol{X}'|^2)\\ \nonumber 
\fl \Phi(\boldsymbol{X},t)=S(\boldsymbol{X}',t)+(\boldsymbol{X}-\boldsymbol{X}')\cdot \frac{\partial \Phi}{\partial \boldsymbol{X}}(\boldsymbol{X}',t)+O(|\boldsymbol{X}-\boldsymbol{X}'|^2) \ .
\end{eqnarray}
The above yield
\begin{eqnarray}
\fl \frac{\partial \chi}{\partial t}(\boldsymbol{X},t;\hbar)=\frac{\partial R}{\partial t}(\boldsymbol{X}',t;\hbar)+(\boldsymbol{X}-\boldsymbol{X}')\cdot \frac{\partial^2 \chi}{\partial t\partial \boldsymbol{X}}(\boldsymbol{X}',t;\hbar)+O(|\boldsymbol{X}-\boldsymbol{X}'|^2)\\ \nonumber 
\fl \frac{\partial \Phi}{\partial t}(\boldsymbol{X},t;\hbar)=\frac{\partial S}{\partial t}(\boldsymbol{X}',t)+(\boldsymbol{X}-\boldsymbol{X}')\cdot \frac{\partial^2 \Phi}{\partial t\partial \boldsymbol{X}}(\boldsymbol{X}',t)+O(|\boldsymbol{X}-\boldsymbol{X}'|^2)
\end{eqnarray}
and
\begin{eqnarray}
\fl \frac{\partial \chi}{\partial \boldsymbol{X}}(\boldsymbol{X},t;\hbar)=\frac{\partial R}{\partial \boldsymbol{X}'}(\boldsymbol{X}',t;\hbar)\frac{\partial \boldsymbol{X}'}{\partial \boldsymbol{X}}+(\textbf{I}-\frac{\partial \boldsymbol{X}'}{\partial \boldsymbol{X}})\cdot \frac{\partial \chi}{\partial \boldsymbol{X}}(\boldsymbol{X}',t;\hbar)\\ \nonumber 
+(\boldsymbol{X}-\boldsymbol{X}')\cdot \frac{\partial^2 \chi}{\partial \boldsymbol{X}\partial \boldsymbol{X}'}(\boldsymbol{X}',t;\hbar) \frac{\partial \boldsymbol{X}'}{\partial \boldsymbol{X}}+O(|\boldsymbol{X}-\boldsymbol{X}'|) \ .
\end{eqnarray}
So we conclude that 
\begin{equation}
\frac{\partial\Phi}{\partial t}\Big|_{\Lambda_t\times\mathbb{R}}=\frac{\partial S}{\partial t} \ , \ \ \frac{\partial\chi}{\partial t}\Big|_{\Lambda_t\times\mathbb{R}}=\frac{\partial R}{\partial t} \ , \ \ \frac{\partial\chi}{\partial \boldsymbol{X}}\Big|_{\Lambda_t\times\mathbb{R}}=\frac{\partial R}{\partial \boldsymbol{X}}\ .
\end{equation}

Based on the above approximations, we derive form $(\ref{eq:sclexp})$ the canonical system on the Lagrangian manifold by considering the restriction of the semi-classical asymptotic expansion on $\Lambda_t$
\begin{equation}
\fl \Bigg(i\hbar\, R ^{-1}\frac{\partial R}{\partial t}-\frac{\partial S}{\partial t}-H-i\hbar\, R ^{-1}\textbf{J}\frac{\partial H}{\partial \boldsymbol{X}}\cdot \frac{\partial R}{\partial \boldsymbol{X}}-\frac{i\hbar}{2}\,{\rm tr}\,\textbf{J}\tilde\mathbfcal{Q}\textbf{J}\frac{\partial^2 H}{\partial \boldsymbol{X}^2}
\Bigg) R \,e^{\frac{i}{\hbar}S}=O(\hbar^2)
\end{equation}
i.e., the real equations
\begin{equation}
\frac{\partial S}{\partial t}+H\Big(\frac{\boldsymbol{q}}{2}-\frac{\partial S}{\partial \boldsymbol{p}},\frac{\boldsymbol{p}}{2}+\frac{\partial S}{\partial \boldsymbol{q}}\Big)=0
\end{equation}
and
\begin{equation} 
\frac{\partial R}{\partial t}+\textbf{J}\frac{\partial H}{\partial \boldsymbol{X}}(\boldsymbol{X})\cdot \frac{\partial R}{\partial \boldsymbol{X}}-\frac{1}{2}\,{\rm tr}\Big(\textbf{J}\tilde\mathbfcal{Q}\textbf{J}\frac{\partial^2 H}{\partial \boldsymbol{X}^2}(\boldsymbol{X})\Big)\, R =0 
\end{equation}
with initial conditions 
\begin{equation}
S(\boldsymbol{q},\boldsymbol{p},0)=S_0(\boldsymbol{q})-\frac{\boldsymbol{p}\cdot \boldsymbol{q}}{2}
\end{equation}
and
\begin{equation}
R(\boldsymbol{q},\boldsymbol{p},0;\hbar)=(\pi\hbar)^{-d/4}\frac{R_0(\boldsymbol{q})}{\sqrt{{\rm det}\Big(\textbf{I}-i
\,\frac{\partial^2S_0}{\partial \boldsymbol{x}^2}(\boldsymbol{q})\Big)}}
\end{equation}
respectively.

It is straightforward to see that the \textit{Weyl-symmetrized phase space action} $S(\boldsymbol{q},\boldsymbol{p},t)$ is related to the solution of the initial value problem for the Hamilton-Jacobi equation
\begin{equation}
\frac{\partial\tilde S}{\partial t}+H\Big(\boldsymbol{q},\frac{\partial\tilde S}{\partial \boldsymbol{q}}\Big)=0 \ , \ \ \tilde S(\boldsymbol{q},0)=S_0(\boldsymbol{q})
\end{equation}
by means of 
\begin{equation}
S(\boldsymbol{q},\boldsymbol{p},t)=\tilde S(\boldsymbol{q},t)-\frac{\boldsymbol{p}\cdot \boldsymbol{q}}{2}
\end{equation}

For the phase space transport equation
\begin{equation}
\frac{\partial R}{\partial t}+\textbf{J}\frac{\partial H}{\partial \boldsymbol{X}}(\boldsymbol{X})\cdot \frac{\partial R}{\partial \boldsymbol{X}}-\frac{1}{2}\,{\rm tr}\Big(\textbf{J}\tilde \mathbfcal{Q}\textbf{J}\frac{\partial^2 H}{\partial \boldsymbol{X}^2}(\boldsymbol{X})\Big)\, R =0 
\end{equation}
we have, along the characteristics, the orbits of the Hamiltonian flow generated by $\mathcal{H}$,
\begin{equation}
\frac{d R}{d t}-\frac{1}{2}\,{\rm tr}\Big(\textbf{J}\tilde\mathbfcal{Q}\textbf{J}\frac{\partial^2 H}{\partial \boldsymbol{X}^2}(\boldsymbol{X})\Big)\, R =0 
\end{equation}
so that 
\begin{eqnarray}
R(\boldsymbol{X},t;\hbar)=R_0(\boldsymbol{X};\hbar)\,\exp\frac{1}{2}\int\displaylimits_0^t{\rm tr}\Big(\textbf{J}\tilde\mathbfcal{Q}\textbf{J}\frac{\partial^2 H}{\partial \boldsymbol{X}^2}(\boldsymbol{X})\Big)\,d\tau\\ \nonumber 
\fl =R_0(\boldsymbol{X};\hbar)\,\exp\frac{1}{2}\int\displaylimits_0^t{\rm tr}\,\mathcal{H}_{\boldsymbol{P}\!\boldsymbol{P}}\,\tilde \mathbfcal{Q}\,d\tau=R_0(\boldsymbol{X};\hbar)\,\exp\frac{1}{2}\int\displaylimits_0^t{\rm tr}\Big(\mathcal{H}_{\boldsymbol{P}\!\boldsymbol{P}}\,\tilde \mathbfcal{Q}+\mathcal{H}_{\boldsymbol{P}\!\boldsymbol{X}}-\mathcal{H}_{\boldsymbol{P}\!\boldsymbol{X}}\Big)\,d\tau\\ \nonumber 
\fl =\Big\{\frac{R_0(\boldsymbol{X};\hbar)}{\sqrt{{\rm det}\,\textbf{C}(\boldsymbol{X}_t(\boldsymbol{\alpha}),t)}}\,\exp\Bigg(-\frac{1}{2}\int\displaylimits_0^t{\rm tr}\,\mathcal{H}_{\boldsymbol{P}\!\boldsymbol{X}}\,d\tau\Bigg)\Big\}_{\boldsymbol{\alpha}=\boldsymbol{\alpha}(\boldsymbol{X},t)}\\ \nonumber 
=\frac{R_0(\boldsymbol{X};\hbar)}{\sqrt{{\rm det}\,\textbf{C}(\boldsymbol{X}_t(\boldsymbol{\alpha}),t)}}\Big|_{\boldsymbol{\alpha}=\boldsymbol{\alpha}(\boldsymbol{X},t)}
\end{eqnarray}
by the variational system and that ${\rm tr}\,\mathcal{H}_{\boldsymbol{P}\!\boldsymbol{X}}=0$.

\section{Illustrations}

In this section we provide illustrations of the two methods for the asymptotic solution of the Cauchy problem for the phase space Schr\"{o}dinger equation, i.e., the Anisotropic Gaussian Approximation, yielding the asymptotic solution $\Psi^{\scriptsize{\mathbfcal{Z}}}$ (eq. $(\ref{eq:sclsol})$), and the phase space narrow beam approximation, yielding the asymptotic solution $\Psi_{\scriptstyle{B}}^{\hbar}$ (eq. $(\ref{psansatz})$). We construct these asymptotic solutions for three different cases, namely $V(q)=0, q , q^2$, which simultaneously serve as a partial test on the accuracy of the two methods against the corresponding exact solutions.

Physically, these examples concern quantum processes relevant in Atomic Physics and Quantum Optics. In particular, we consider motion in one-dimensional physical space for the case of zero potential, corresponding to free motion, the case of linear potential, corresponding to scattering off a constant electrostatic field and the case of parabolic potential, corresponding to bounded motion in a parabolic optical trap. Although these can be considered as special cases of an appropriately parametrized quadratic potential, we consider them separately, as each encapsulates different dynamical properties of a particular significance.

Quadratic and sub-quadratic Hamiltonians share the common characteristic of linearity of the Hamiltonian flow. For the cases of the sub-quadratic potentials, we have identical matrix Riccati dynamics, yielding the --uniform in phase space-- anisotropy co-efficient
\begin{equation}
\mathcal{Z}(t)=\frac{i}{1+2it}
\end{equation}
and phase space anisotropy matrix
\begin{equation}
\mathbfcal{Q}(t) = \frac{i}{2(1+it)}\left(
\begin{array}{rl}
1 & -t \\
-t & 1+2it 
\end{array} \right) \ .
\end{equation}
For the marginal case of the quadratic potential, the Riccati dynamics are trivial; the anisotropy coefficient and phase space anisotropy matrix are constants
\begin{equation}
\mathcal{Z}(t)=i
\end{equation}
and
\begin{equation}
\mathbfcal{Q}(t) = \frac{i}{2}\left(
\begin{array}{rl}
1 & 0 \\
0 & 1 
\end{array} \right) \ .
\end{equation}

As initial data, for all examples, we consider semi-classical WKB initial data for the Schr\"odinger equation, which allows for explicit calculations, 
\begin{equation}
\psi_0^{\hbar}(x)=\pi^{-1/4}\,e^{-\frac{1}{2}x^2}e^{\frac{i}{\hbar}\frac{x^2}{2}} \ ,
\end{equation}
so that the initial phase and amplitude are the real analytic functions $S_0(x)=x^2/2$ and $R_0(x)=\pi^{-1/4}\,e^{-\frac{1}{2}x^2}$. The corresponding stationary condition $(\ref{eq:stat0})$ reads 
\begin{equation}
z-p+i(z-q)=0 \ ,
\end{equation}
which admits the unique solution $z(q,p)=\frac{q-ip}{1-i}$, while the initial data $(\ref{eq:theta0})$ for the Cauchy problem for the phase space Hamilton-Jacobi equation $(\ref{eq:phshj1})$ is 
\begin{equation}
\theta_0(q,p)=\frac{1+i}{4}(q-p)(q-ip) \ .
\end{equation}
The Hessian matrix 
\begin{equation}
\frac{\partial^2\theta_0}{\partial \boldsymbol{X}^2}(\boldsymbol{X})=\frac{1}{2}\left(\begin{array}{ccc} 1+i & -i \\ -i & -1+i \end{array} \right)=:\textbf{N}
\end{equation}
is constant.

The initial Lagrangian plane generated by the above initial data is 
\begin{equation}
\Lambda_0=\{(q,p)\in\mathbb{R}^{2}\,|\, p=q\}
\end{equation}
admitting the global parametric representation 
\begin{equation}
(q_0(\alpha),p_0(\alpha))=(\alpha,\alpha) \ , \ \ \alpha\in\mathbb{R} \ .
\end{equation}

For each case, first we construct the exact phase space wave function $\Psi(q,p,t;\hbar)$, by explicit calculation of the wave packet transform of the exact configuration space wave function $\psi(x,t;\hbar)$. Then, by performing the Gaussian integration in $(\ref{eq:scprop})$ we calculate the propagator $ \mathcal{K}^{\scriptsize{\mathbfcal{Z}}}$ and the semi-classical asymptotic solution $\Psi^{\scriptsize{\mathbfcal{Z}}}(q,p,t;\hbar)$ by means of $(\ref{eq:sclsol})$. The action of the semi-classical phase space propagator on $\Psi_0$ is unsurprisingly identified with the exact solution in all cases. Finally we calculate the phase and the amplitude of the beam solution $\Psi_{\scriptscriptstyle B}^{\hbar}(q,p,t)$, and we check that in all cases they match to leading order the exact phase and amplitude, semi-classically, the discrepancy depending on
the distance between a given point $(q,p)$, external to $\Lambda_t$, and its unique nearest point thereon.

The exact wave functions satisfy of course the phase space Schr\"odinger equation
\begin{equation}
i\hbar\,\frac{\partial\Psi}{\partial t}=\Big(\frac{p}{2}-i\hbar\,\frac{\partial}{\partial q}\Big)^2\Psi+V\Big(\frac{q}{2}+i\hbar\,\frac{\partial}{\partial p}\Big)\Psi \ , \ \ t\in[0,T]
\end{equation}
and the initial condition $\Psi(q,p,0)=\Psi_0(q,p;\hbar):=(\mathcal{W}\psi_0^{\hbar})(q,p;\hbar) $, where 
\begin{equation}
\fl \Psi_0(q,p;\hbar)=\frac{\pi^{-1/4}(\pi\hbar)^{-1/4}}{\sqrt{1-i+\hbar}}\,e^{\frac{i}{\hbar}\Big(\frac{pq}{2}+\frac{i}{2}q^2\Big)} \exp\frac{i}{\hbar}\Big(-\frac{i}{2}\frac{(q-ip)^2}{1-i+\hbar}\Big) \ .
\end{equation}

\subsection{Free Motion}

The Hamiltonian is
\begin{equation}
H(q,p)=p^2 
\end{equation}
which generates the Hamiltonian flow $g_t(q,p)=(q+2tp,p)$, $t\in[0,T]$, while the phase space action along the Hamiltonian flow is $ A (q,p,t)=p^2t$.

The solution of the configuration space problem is obtained by means of the free propagator \cite{FeHi},
\begin{eqnarray}
\psi(x,t;\hbar)=\frac{1}{(4\pi i \hbar t)^{1/2}}\int e^{\frac{i}{\hbar}\frac{(x-y)^2}{4t}}\psi_0^{\hbar}(y)\,dy 
\end{eqnarray}
which, for the particular initial data, is 
\begin{equation}
\psi(x,t;\hbar)=\frac{\pi^{-1/4}}{\sqrt{1+2(1+i\hbar)t}}\exp\frac{i}{\hbar}\frac{1+i\hbar}{1+2(1+i\hbar)t}\frac{x^2}{2} 
\end{equation}
whose wave packet transform is
\begin{eqnarray}\label{eq:sol1}
\Psi(q,p,t;\hbar)=\frac{\pi^{-1/4}(\pi\hbar)^{-1/4}}{\sqrt{1-i+\hbar+2(1+i\hbar)t}}\,e^{\frac{i}{\hbar}\Big(\frac{pq}{2}+\frac{i}{2}q^2\Big)}\\\nonumber
\times \exp\frac{i}{\hbar}\Big(-\frac{i}{2}\frac{1+2(1+i\hbar)t}{1-i+\hbar+2(1+i\hbar)t}(q-ip)^2\Big) \ .
\end{eqnarray}

The semi-classical phase space propagator $(\ref{eq:scprop})$ is 
\begin{eqnarray}
\fl \mathcal{K}^{\scriptsize{\mathbfcal{Z}}}(q,p,\eta,\xi,t;\hbar)=
\frac{1}{2\pi\hbar}\sqrt{\frac{1}{1+it}}\exp\frac{i}{\hbar}\Bigg(\frac{q\eta-p\xi}{2}-tq\xi\\ \nonumber
+\frac{i}{4(1+it)}\left(\begin{array}{ccc}
q-\eta-2t\xi\\
p-\xi 
\end{array}\right)^{{\rm T}}
\left(
\begin{array}{rl}
1 & -t \\
-t & 1+2it 
\end{array} \right)
 \left(
\begin{array}{ccc}
q-\eta-2t\xi\\
p-\xi 
\end{array}\right)\Bigg) \ .
\end{eqnarray}

The semi-classical asymptotic solution --and the solution itself-- is semi-classically concentrated on the transported Lagrangian plane
\begin{equation} 
\Lambda_t=\{(q,p)\in\mathbb{R}^{2}\,|\,p=\frac{q}{1+2t}\} 
\end{equation}
i.e., initially the diagonal straight line $p=q$ asymptotically rotating clockwise to the horizontal straight line $p=0$.

A convenient global parametric representation of the propagated Lagrangian plane is
\begin{equation}
(q_t(\alpha),p_t(\alpha))=((1+2t)\alpha,\alpha) \ , \ \ \alpha\in\mathbb{R} \ .
\end{equation}

As the above is a flat curve, the curvature boundedness condition is met, so that, uniformly in time, there exists a narrow neighborhood of $\Lambda_t$, each point of which possesses a unique nearest point on $\Lambda_t$; for given $(q,p)$ appropriately close to $\Lambda_t$, the unique nearest point on it is given by $(q_t(\alpha),p_t(\alpha))$ for $\alpha\in\mathbb{R}$ such that the condition below holds 
\begin{equation}
(q_t(\alpha)-q)\frac{\partial q_t}{\partial\alpha}+(p_t(\alpha)-p)\frac{\partial p_t}{\partial\alpha}=0
\end{equation}
which yields the unique global solution $\alpha(q,p,t)=\frac{(1+2t)q+p}{1+(1+2t)^2}$.

In order to construct the phase space narrow beam asymptotic solution, it is required to construct the solution of the Cauchy problem for the phase space variational system
\begin{equation}
\frac{d}{dt}\left(
\begin{array}{ccc}
\textbf{C} \\
\textbf{D}
\end{array}\right)=\left(\begin{array}{ccc}
\frac{1}{2}\textbf{J}_2H_{\boldsymbol{X}\!\boldsymbol{X}} & -\textbf{J}_2H_{\boldsymbol{X}\!\boldsymbol{X}}\textbf{J}_2 \\
-\frac{1}{4}H_{\boldsymbol{X}\!\boldsymbol{X}} & \frac{1}{2}H_{\boldsymbol{X}\!\boldsymbol{X}}\textbf{J}_2
\end{array}\right)\left(
\begin{array}{ccc}
\textbf{C} \\
\textbf{D} 
\end{array}\right)
\end{equation}
with initial conditions
\begin{equation} 
\textbf{C}(\boldsymbol{X}_0(\alpha),0)=\textbf{I}_2 \ , \ \ \ \textbf{D}(\boldsymbol{X}_0(\alpha),0)=\textbf{N} \ .
\end{equation}

For the given motion, the matrix of the variational system is 
\begin{equation}
\textbf{M}=\left(\begin{array}{cccc}
0 & 1 & 2 & 0\\
0 & 0 & 0 & 0\\
0 & 0 & 0 & 0\\
0 & -\frac{1}{2} & -1 & 0
\end{array}\right) \ .
\end{equation}
As $\textbf{M}$ is nilpotent, the evolution matrix of the system becomes $\exp\,t\textbf{M}=\textbf{I}_4+t\,\textbf{M}$, yielding the solution of the problem 
\begin{equation}
\fl \left(
\begin{array}{ccc}
\textbf{C}(t) \\
\textbf{D}(t)
\end{array}\right)=\exp\,t\textbf{M}\,\left(
\begin{array}{ccc}
\textbf{C}(0) \\
\textbf{D}(0)
\end{array}\right)=\left(\begin{array}{ccc}
1+(1+i)t & (1-i)t\\
0 & 1\\
\frac{1+i}{2} & -\frac{i}{2}\\
\frac{-i-(1-i)t}{2} & \frac{-1+i-(1-i)t}{2}
\end{array}\right)
\end{equation}
so that the narrow beam quadratic form is given by 
\begin{equation}\label{eq:Q1}
\fl \tilde \mathbfcal{Q}(t)=\textbf{D}(t)\textbf{C}(t)^{-1}=\frac{1}{2}\frac{1}{1+(1+i)t}\left(\begin{array}{ccc}
1+i & -i-(1+i)t\\
-i-(1+i)t & -(1-i)(1+2t)
\end{array}\right) \ .
\end{equation}
It depends only on time $t$ since $\textbf{M}, \textbf{N}$ are constant matrices.

A straightforward calculation yields the narrow beam approximation phase 
\begin{equation}
\Phi(q,p,t)=-\frac{1+i}{4}\frac{p-q+2tp}{1+(1+i)t}\,(q-ip) \ ,
\end{equation}
and the corresponding amplitude
\begin{equation}
\chi(q,p,t;\hbar)=\frac{\pi^{-1/4} (\pi\hbar)^{-1/4}}{\sqrt{1-i+2t}}\,\exp\Bigg(-\frac{1}{2}\Big(\frac{(1+2t)q+p}{1+(1+2t)^2}\Big)^2\Bigg) \ .
\end{equation}
The narrow beam phase matches the phase of the solution \ref{eq:sol1} to leading order, semi-classically, in the sense that 
\begin{equation}
\fl \frac{pq}{2}+\frac{i}{2}q^2-\frac{i}{2}\frac{1+2(1+i\hbar)t}{1-i+\hbar+2(1+i\hbar)t}(q-ip)^2-\Phi(q,p,t)=O(\hbar)
\end{equation}
while the corresponding narrow beam amplitude matches the amplitude of the solution \ref{eq:sol1} to leading order, semi-classically, modulo an error pertaining to the distance from $\Lambda_t$, in the sense that 
\begin{equation}
\fl \frac{\pi^{-1/4}(\pi\hbar)^{-1/4}}{\sqrt{1-i+\hbar+2(1+i\hbar)t}}\,e^{-\frac{i}{4}\frac{(q-ip)^2}{(1+(1+i)t)^2}}=\Big(1+O(\varepsilon(q,p,t)^{1/2})\Big)\Big(1+O(\hbar)\Big)\,\chi(q,p,t)
\end{equation}
where, for given $t\geq 0$, 
\begin{equation}
\fl \varepsilon(q,p,t)=|(q-q_t(\alpha),p-p_t(\alpha))|\Big|_{\alpha=\frac{(1+2t)q+p}{1+(1+2t)^2}}=\frac{1}{\sqrt{2}}\frac{|p-q+2tp|}{\sqrt{1+2t(1+t)}}
\end{equation}
i.e., the distance between a given point, external to $\Lambda_t$, and its unique nearest point thereon.

\subsection{Scattering by a Constant Electrostatic Field}

The Hamiltonian is
\begin{equation}
H(q,p)=p^2+q
\end{equation}
which generates the Hamiltonian flow $g_t(q,p)=(q+2tp-t^2,p-t)$, $t\in[0,T]$, while the phase space action along the Hamiltonian flow is $ A (q,p,t)=(p^2-q)t-2pt^2+\frac{2}{3}t^3$.

The solution of the configuration space problem is obtained by means of the propagator \cite{Hol,DaZh} 
\begin{eqnarray} 
\fl \psi(x,t;\hbar)=\frac{1}{(4\pi i\hbar t)^{1/2}}\int \exp\frac{i}{\hbar}\Big(\frac{(x-y)^2}{4t}-\frac{1}{2}t (x+y)-\frac{1}{12}t^3\Big)\psi_0^{\hbar}(y)\,dy \ .
\end{eqnarray}

For the particular initial data, we obtain,
\begin{equation}
\fl \psi(x,t;\hbar)=\frac{\pi^{-1/4}}{\sqrt{1+2(1+i\hbar)t}}\,\exp\frac{i}{\hbar}\frac{1}{4t}\Big(-\frac{1}{3}(t^4+6t^2-3x^2)-\frac{(t^2+x)^2}{1+2(1+i\hbar)t}\Big)
\end{equation}
whose wave packet transform is 
\begin{eqnarray}\label{eq:sol2}
\fl \Psi(q,p,t;\hbar)=\frac{\pi^{-1/4}(\pi \hbar)^{-1/4}}{\sqrt{1-i+2t+(1+2it)\hbar}}\\ \nonumber 
\fl\times e^{\frac{i}{\hbar}\Big(\frac{pq}{2}+\frac{i}{2}q^2\Big)}\exp\frac{i}{\hbar}\frac{1}{-1-i-i\hbar+2(\hbar-i)t}\Bigg(t^2\Big(\frac{1}{2}+\frac{1+i+i\hbar}{3}t+\frac{i-\hbar}{6}t^2\Big)\\ \nonumber
+t\Big(i+(i-\hbar)t\Big)(q-ip)-\frac{1+2t+2i\hbar t}{2}(q-ip)^2\Bigg) \ .
\end{eqnarray}

The semi-classical phase space propagator is 
\begin{eqnarray}
\fl \mathcal{K}^{\scriptsize{\mathbfcal{Z}}}(q,p,\eta,\xi,t;\hbar)=\frac{1}{2\pi\hbar}\sqrt{\frac{1}{1+it}}\exp\frac{i}{\hbar}\Bigg(\frac{\xi q-p\eta}{2}-\frac{q+\eta+2p\xi}{2}t+\frac{p-\xi}{2}t^2+\frac{t^3}{6}\\ \nonumber 
\fl +\frac{i}{4(1+it)}\left(\begin{array}{ccc}
q-\eta-2t\xi+t^2\\
p-\xi+t 
\end{array}\right)^{T}
\left(
\begin{array}{rl}
1 & -t \\
-t & 1+2it 
\end{array} \right)
\left(
\begin{array}{ccc}
q-\eta-2t\xi+t^2\\
p-\xi+t \end{array}\right)\Bigg) \ .
\end{eqnarray}

The corresponding transported Lagrangian plane is
\begin{equation} 
\Lambda_t=\{(q,p)\in\mathbb{R}^{2}\,|\,p=\frac{q-t(t+1)}{1+2t}\} 
\end{equation}
i.e., initially the diagonal straight line $p=q$, asymptotically rotating and displacing toward the horizontal straight line $p=-\frac{t}{2}$. 

A convenient global parametric representation of the propagated Lagrangian plane is
\begin{equation}
(q_t(\alpha),p_t(\alpha))=((1+2t)\alpha-t^2,\alpha-t) \ , \ \ \alpha\in\mathbb{R} \ .
\end{equation}

As in the example of free motion, the above is a flat curve, the curvature boundedness condition is met, so that, uniformly in time, there exists a narrow neighborhood of $\Lambda_t$, each point of which possesses a unique nearest point on $\Lambda_t$; for given $(q,p)$ appropriately close to $\Lambda_t$, the unique nearest point on it is given by $(q_t(\alpha),p_t(\alpha))$ for $\alpha\in\mathbb{R}$ such that the condition below holds 
\begin{equation}
(q_t(\alpha)-q)\frac{\partial q_t}{\partial\alpha}+(p_t(\alpha)-p)\frac{\partial p_t}{\partial\alpha}=0
\end{equation}
which yields the unique global solution $\alpha=\frac{t+(1+2t)t^2+(1+2t)q+p}{1+(1+2t)^2}$.
 
In order to construct the phase space narrow beam asymptotic solution, it is required to construct the phase space narrow beam anisotropy matrix, which again is identified with $(\ref{eq:Q1})$, the one in the previous example.

A straightforward calculation yields the narrow beam approximation phase 
\begin{eqnarray}
\fl \Phi(q,p,t)=\frac{1}{12(1+(1+i)t)}\Bigg(\Big(3(-1+i)-4t-(1+i)t^2\Big)t^2\\ \nonumber 
-6(1+i)\Big(\frac{p}{1+i}+(1+p)t+t^2\Big)(q-ip)+3(1+i)(q-ip)^2\Bigg)
\end{eqnarray}
and the corresponding amplitude
\begin{eqnarray}
\fl \chi(q,p,t;\hbar)=\frac{\pi^{-1/4} (\pi\hbar)^{-1/4}}{\sqrt{1-i+2t+(1+2it)\hbar}}\\ \nonumber 
\times\,\exp\Bigg(-\frac{1}{2}\Big(\frac{t+(1+2t)t^2+(1+2t)q+p}{1+(1+2t)^2}\Big)^2\Bigg) \ .
\end{eqnarray}
The narrow beam phase matches the phase of the solution \ref{eq:sol1} to leading order, semi-classically, in the sense that 
\begin{eqnarray}
\fl \frac{1}{1-i+\hbar+2(1+i\hbar )t}\Big(\frac{i}{2}t^2-\frac{1-i+\hbar}{3}t^3-\frac{1+i\hbar}{6}t^4\\ \nonumber 
\fl-(t+(1+i\hbar)t^2+(\frac{1-i+\hbar}{2}+(1+i\hbar)t)p)(q-ip)+\frac{1+i\hbar}{2}(q-ip)^2\Big)-\Phi(q,p,t)\\ \nonumber
=O(\hbar)
\end{eqnarray}
while the corresponding narrow beam amplitude matches the amplitude of the solution \ref{eq:sol1} to leading order, semi-classically, modulo an error pertaining to the distance from $\Lambda_t$, in the sense that 
\begin{eqnarray}
\fl \frac{\pi^{-1/4}(\pi\hbar)^{-1/4}}{\sqrt{1+i+2t+(1+2it)\hbar}}\\ \nonumber 
\times \,e^{-\frac{i}{4}\frac{(q-ip-it+t^2)^2}{(1+(1+i)t)^2}}=\Big(1+O(\varepsilon(q,p,t)^{1/2})\Big)\Big(1+O(\hbar)\Big)\,\chi(q,p,t)
\end{eqnarray}
where, for given $t\geq 0$, 
\begin{equation}
\fl \varepsilon(q,p,t)=|(q-q_t(\alpha),p-p_t(\alpha))|\Big|_{\alpha=\frac{t+(1+2t)t^2+(1+2t)q+p}{1+(1+2t)^2}}=\frac{1}{\sqrt{2}}\frac{|p-q+2tp+t^2|}{\sqrt{1+2t(1+t)}}
\end{equation}
i.e., the distance between a given point, external to $\Lambda_t$, and its unique nearest point thereon.

\subsection{Bound Motion in a Parabolic Optical Trap}

The Hamiltonian is
\begin{equation}
H(q,p)=p^2+q^2
\end{equation}
which generates the Hamiltonian flow $g_t(q,p)=R(t)\left(\begin{array}{c}
q\\
p
\end{array}\right)$, $t\in[0,T]$, where $R(t)=\left(
\begin{array}{rl}
\cos\,2t & \sin\,2t \\
-\sin\,2t & \cos\,2t 
\end{array} \right)$, while the phase space action along the Hamiltonian flow is $ A (q,p,t)=\frac{1}{4}(p^2-q^2)\sin\,4t+\frac{1}{2}pq\Big(\cos\,4t-1\Big)$.

The solution of the configuration space problem is obtained by means of the harmonic oscillator propagator \cite{FeHi}, 
\begin{equation}
\fl \psi(x,t;\hbar)=\Big(\frac{1}{2\pi i \hbar\,\sin\,2t}\Big)^{1/2}\int \exp\frac{i}{\hbar}\Big(\frac{(x^2+y^2)\,\cos\,2t-2xy}{2\,\sin\,2t}\Big)\psi_0^{\hbar}(y)\,dy \ .
\end{equation}

For the particular initial data, we obtain,
\begin{equation}
\fl \psi(x,t;\hbar)=\frac{\pi^{-1/4}}{\sqrt{\cos\,2t+(1+i\hbar)\sin\,2t}}\,\exp\frac{i}{\hbar}\Big(\cot\,2t-\frac{\csc^2\,2t}{1+i\hbar+\cot\,2t}\Big)\frac{x^2}{2}
\end{equation}
whose wave packet transform is 
\begin{eqnarray}\label{eq:sol3}
\fl\Psi(q,p,t;\hbar)=\frac{\pi^{-1/4}(\pi\hbar)^{-1/4}}{\sqrt{1-i+\hbar}}\,e^{-it}e^{\frac{i}{\hbar}\Big(\frac{pq}{2}+\frac{i}{2}q^2\Big)}\\ \nonumber 
\times \,\exp\frac{i}{\hbar}\frac{1+i\hbar+\cot\,2t}{(1+i+i\hbar)(i+\cot\,2t)}\,\frac{(q-ip)^2}{2} \ .
\end{eqnarray}

The semi-classical phase space propagator is 
\begin{eqnarray}
\fl \mathcal{K}^{\scriptsize{\mathbfcal{Z}}}(q,p,\eta,\xi,t;\hbar)=\frac{1}{2\pi\hbar}\,e^{-it}\exp\frac{i}{\hbar}\Bigg(\frac{1}{2}\Big((\xi q-p\eta)\,\cos\,2t-(q\eta+p\xi)\,\sin\,2t\Big)\\ \nonumber
+\frac{i}{4}\left(\begin{array}{ccc}
q-\eta\cos\,2t-\xi\sin\,2t\\
p-\xi\cos\,2t+\eta\sin\,2t
\end{array}\right)^{T}
\left(\begin{array}{ccc}
q-\eta\cos\,2t-\xi\sin\,2t\\
p-\xi\cos\,2t+\eta\sin\,2t \end{array}\right)\Bigg) \ .
\end{eqnarray}

The corresponding transported Lagrangian plane is 
\begin{equation} 
\Lambda_t=\{(q,p)\in\mathbb{R}^{2}\,|\,p=\frac{\cos\,4t}{(\cos\,2t+\sin\,2t)^2}q\} 
\end{equation} 
i.e., initially the diagonal straight line $p=q$, rotating clock-wise, becoming the vertical line at $t=\frac{\pi}{2}(n-\frac{1}{8})$ for $n=1,2,\ldots$

The global parametric representation of the propagated Lagrangian plane is
\begin{equation}
\fl (q_t(\alpha),p_t(\alpha))=((\cos\,2t+\sin\,2t)\alpha,(\cos\,2t-\sin\,2t)\alpha) \ , \ \ \alpha\in\mathbb{R} \ .
\end{equation}

As the above is a flat curve, the curvature boundedness condition is met, so that, uniformly in time, there exists a narrow neighborhood of $\Lambda_t$, each point of which possesses a unique nearest point on $\Lambda_t$; for given $(q,p)$ appropriately close to $\Lambda_t$, the unique nearest point on it is given by $(q_t(\alpha),p_t(\alpha))$ for $\alpha\in\mathbb{R}$ such that the condition below holds 
\begin{equation}
(q_t(\alpha)-q)\frac{\partial q_t}{\partial\alpha}+(p_t(\alpha)-p)\frac{\partial p_t}{\partial\alpha}=0
\end{equation}
which yields the unique global solution $\alpha=\frac{1}{2}\Big((\cos\,2t+\sin\,2t)q+(\cos\,2t-\sin\,2t)p\Big)$.

For the given motion, the matrix of the variational system is 
\begin{equation}
\textbf{M}=\left(\begin{array}{cccc}
0 & 1 & 2 & 0\\
-1 & 0 & 0 & 2\\
-\frac{1}{2} & 0 & 0 & 1\\
0 & -\frac{1}{2} & -1 & 0
\end{array}\right) \ .
\end{equation}
As $\textbf{M}$ has the particular algebraic property that, for $k=1,2,\ldots$
\begin{equation}
\textbf{M}^n=\left\{
\begin{array}{rl}
(-4)^{k-1}\textbf{M} \ , & \ \ n=2k-1 \\
(-4)^{k-1}\textbf{M}^2 \ , & \ \ n=2k
\end{array} \right.
\end{equation}
the evolution matrix of the system becomes 
\begin{equation}
\exp\,t\textbf{M}=\textbf{I}_4+\frac{1}{2}\,(\sin\,2t)\,\textbf{M}+\frac{1}{4}\,(1-\cos\,2t)\,\textbf{M}^2
\end{equation}
yielding the solution of the problem 
\begin{eqnarray}
\fl \left(
\begin{array}{ccc}
\textbf{C}(t) \\
\textbf{D}(t)
\end{array}\right)=\exp\,t\textbf{M}\,\left(
\begin{array}{ccc}
\textbf{C}(0) \\
\textbf{D}(0)
\end{array}\right)\\ \nonumber 
\fl =\frac{1}{2}\left(\begin{array}{ccc}
(1+i)(-i+\cos\,2t+\sin\,2t) & (1-i)(-1+\cos\,2t+\sin\,2t)\\
(1+i)(-1+\cos\,2t-\sin\,2t) & (1-i)(i+\cos\,2t-\sin\,2t)\\
\frac{1}{2}(1+i)(1+\cos\,2t-\sin\,2t) & \frac{1}{2}(1-i)(-i+\cos\,2t-\sin\,2t)\\
\frac{1}{2}(-1-i)(i+\cos\,2t+\sin\,2t) & \frac{1}{2}(-1+i)(1+\cos\,2t+\sin\,2t)
\end{array}\right)
\end{eqnarray}
so that the narrow beam quadratic form is given by 
\begin{equation}
\fl \tilde \mathbfcal{Q}(t)=\textbf{D}(t)\textbf{C}(t)^{-1}=\frac{1}{2}\left(\begin{array}{ccc}
i+e^{-4it} & -i\,e^{-4it}\\
-i\,e^{-4it} & i-e^{-4it}
\end{array}\right) \ .
\end{equation}

A straightforward calculation yields the narrow beam approximation phase 
\begin{equation}
\Phi(q,p,t)=\frac{1}{4}\Big(i(p^2+q^2)+e^{-4it}(q-ip)^2\Big)
\end{equation}
and the corresponding amplitude 
\begin{eqnarray}
\fl\chi(q,p,t;\hbar)=\pi^{-1/4} (\pi\hbar)^{-1/4}\,e^{-it}\\ \nonumber 
\times \exp\Bigg(-\frac{1}{8}\Big((\cos\,2t+\sin\,2t)q+(\cos\,2t-\sin\,2t)p\Big)^2\Bigg) \ .
\end{eqnarray}

The narrow beam phase matches the phase of the solution \ref{eq:sol1} to leading order, semi-classically, in the sense that 
\begin{eqnarray}
\fl -\frac{1}{2}p(q-ip)-\frac{1}{2(1-i+\hbar)}(\sin\,2t-(1+i\hbar)\,\cos\,2t)\,e^{-2it}\,(q-ip)^2-\Phi(q,p,t)\\ \nonumber =O(\hbar)
\end{eqnarray}
while the corresponding narrow beam amplitude matches the amplitude of the solution \ref{eq:sol1} to leading order, semi-classically, modulo an error pertaining to the distance from $\Lambda_t$, in the sense that 
\begin{equation}
\fl \frac{\pi^{-1/4}(\pi\hbar)^{-1/4}}{\sqrt{1-i+\hbar}}\,e^{-i\Big(t+\frac{e^{-4it}}{4}\,(q-ip)^2\Big)}=\Big(1+O(\varepsilon(q,p,t)^{1/2})\Big)\Big(1+O(\hbar)\Big)\,\chi(q,p,t)
\end{equation}
where, for given $t\geq 0$, 
\begin{eqnarray}
\fl \varepsilon(q,p,t)=|(q-q_t(\alpha),p-p_t(\alpha))|\Big|_{\alpha=\frac{1}{2}\Big((\cos\,2t+\sin\,2t)q+(\cos\,2t-\sin\,2t)p\Big)}\\ \nonumber 
=\frac{1}{\sqrt{2}}|(p-q)\,\cos\,2t+(p+q)\,\sin\,2t|
\end{eqnarray}
i.e., the distance between a given point, external to $\Lambda_t$, and its unique nearest point thereon.

\section{Discussion and Conclusions}

We have constructed asymptotic approximations of the solution of the Weyl-symmetrized phase space Schr\"{o}dinger equation by employing the wave packet transform and a particular semi-classical approximation for the time evolution of single isotropic Gaussian wave packets, termed the Anisotropic Gaussian Approximation to construct a semi-classical propagator in phase space.

For initial data defined as the wave packet transform of a configuration space WKB function, the action of this propagator leads to an approximate solution in the form of a semi-classical Fourier integral representation which is further approximated by Complex Stationary Phase formula and it suggests a phase space WKB ansatz with complex phase. This ansatz is an asymptotic solution of the phase space Schr\"{o}dinger equation provided that its complex phase and amplitude solve a canonical system of the Weyl-symmetrized Hamilton-Jacobi and transport equations in double phase space. The solution of this system has been constructed by applying Maslov's complex WKB method and it turns out that it has the form of a narrow beam concentrated along the transported Lagrangian manifold defined by the configuration space WKB initial data.

The calculations of certain examples where the phase space wave function can be analytically constructed, show that while the Fourier integral representation leads to the exact solution, the narrow beam solution is a reasonable asymptotic approximation to the corresponding analytical solutions, the accuracy depending on the distance from the Lagrangian manifold, and it is exponentially small far enough from the manifold, as it is naturally anticipated.

The construction method which is conceptually and computationally concise, could be of interest for applications in problems of Atomic Physics, Quantum Optics and Quantum Chemistry. 

To the authors' knowledge there is no prior work focusing on the construction of semi-classical asymptotic solutions of the phase space Schr\"{o}dinger equation for the semi-classical time evolution problem. We consider this work to be a contribution in the understanding of the phase space Schr\"{o}dinger equation and linear representations of Quantum Mechanics in phase space, and hopefully of linear phase space representations of other linear differential equations, such as the wave equation. 

Possible directions for a deeper understanding of the subject could be the explanation of the special geometric structure induced by the Weyl quantization in double phase space, and the efficient transformation of the Fourier integral representation along and near the propagated Lagrangian manifold in order to employ itself as an ansatz of the solution of phase space Schr\"{o}dinger equation, in the spirit of Fourier Integral Operators. Finally, a rigorous treatment on the validity of the semi-classical approximation with respect to the time interval for the Cauchy problem is indeed open and necessary.

An interesting direction of further investigation would be the extension of the theory for the case of compact configuration space, which could provide a more convenient setting for numerical investigations in the special case of compact domains in $\mathbb{R}^3$, with potential applications for billiard or cavity problems in the fields of Quantum Chaos and Optics.

\section{Acknowledgments}

The authors thank Sergey Dobrokhotov, Frederic Faure, Maurice de Gosson, Robert Littlejohn, Vladimir Nazaikinskii, Vesselin Petkov and Roman Schubert for fruitful conversations and enlightening comments. The authors acknowledge financial support of this work in its beginning by the Archimedes Center for Modeling, Analysis and Computation.

\appendix

\section{A Commutation Formula for Weyl Operators in Phase Space}
\label{QuRep}

We derive a commutation formula for Weyl operators in phase space involving exponential functions with complex phase. The standard commutation formula for Weyl operators and real phase exponential functions is given by Karavev et al. \cite{KaMa}, while a generalization for non-analytic complex phases is proven by Kucherenko \cite{Kuc}, which we provide below.
\\

\noindent \textbf{Theorem}(Kucherenko \cite{Kuc}) \textit{Let $f\in C^\infty(\mathbb{R}^{2d})$ be a real-valued symbol, $R\in C^\infty_0 (\mathbb{R}^d)$ be a real-valued amplitude and $S\in C^\infty(\mathbb{R}^{d})$ a complex-valued phase, with $S(\boldsymbol{x})=S_1(\boldsymbol{x})+i\,S_2(\boldsymbol{x})$, where $S_{1,2}$ are real valued. Then, for every $s\in \mathbb{N}_0$ and $\boldsymbol{x}\in{\rm supp}\,R$, as $\hbar\rightarrow 0^+$, the following holds} 
\begin{eqnarray}
\fl f\Big(\stackrel{2}{\boldsymbol{x}},-i\hbar \stackrel{1}{\frac{\partial}{\partial \boldsymbol{x}}}\Big)\Big(R(\boldsymbol{x})\,e^{\frac{i}{\hbar}S(\boldsymbol{x})}\Big)\\ \nonumber 
=R(\boldsymbol{x})\,e^{\frac{i}{\hbar}S(\boldsymbol{x})}\Bigg(f\Big(\boldsymbol{x},\frac{\partial S}{\partial \boldsymbol{x}}(\boldsymbol{x})\Big)+\hbar\,R(\boldsymbol{x})^{-1}V_1 R(\boldsymbol{x})\Bigg)+r_1(\boldsymbol{x};\hbar)
\end{eqnarray}
\textit{where functions with complex arguments imply almost analytic continuation to that argument, and the action of the first order differential operator $V_1$ is given by}
\begin{eqnarray}
\fl V_1 R(\boldsymbol{x})=-i\sum_{\boldsymbol{\mu}\in\mathbb{N}_0^{d}\,:\,|\boldsymbol{\mu}|=1}\partial_{\boldsymbol{p}}^{\boldsymbol{\mu}} f\Big(\boldsymbol{x},\frac{\partial S}{\partial \boldsymbol{x}}(\boldsymbol{x})\Big) \partial_{\boldsymbol{x}}^{\boldsymbol{\mu}} R(\boldsymbol{x})\\ \nonumber 
-\frac{i}{2}R(\boldsymbol{x})\sum_{\boldsymbol{\mu}\in\mathbb{N}_0^d\,:\,|\boldsymbol{\mu}|=2}\partial_{\boldsymbol{x}}^{\boldsymbol{\mu}} S(\boldsymbol{x})\partial_{\boldsymbol{p}}^{\boldsymbol{\mu}} f\Big(\boldsymbol{x},\frac{\partial S}{\partial \boldsymbol{x}}(\boldsymbol{x})\Big) -i\,\frac{\partial f}{\partial \boldsymbol{p}}\Big(\boldsymbol{x},\frac{\partial S}{\partial \boldsymbol{x}}(\boldsymbol{x})\Big)\cdot \frac{\partial R}{\partial \boldsymbol{x}}(\boldsymbol{x})\\ \nonumber 
\fl -\frac{i}{2}R(\boldsymbol{x})\sum_{\boldsymbol{\mu},\boldsymbol{\nu}\in\mathbb{N}_0^{d}\,:\,|\boldsymbol{\mu}|=2,\,|\boldsymbol{\nu}|\leq s}\frac{1}{\boldsymbol{\nu}!}\partial_{\boldsymbol{x}}^{\boldsymbol{\mu}} S(\boldsymbol{x})\Big(i\frac{\partial S_2}{\partial \boldsymbol{x}}(\boldsymbol{x})\Big)^{\boldsymbol{\nu}}\partial_{\boldsymbol{p}}^{\boldsymbol{\mu}+\boldsymbol{\nu}} f\Big(\boldsymbol{x},\frac{\partial S_1}{\partial \boldsymbol{x}}(\boldsymbol{x})\Big)
\end{eqnarray}
\textit{and the remainder $r_1$ as given by the Theorem of Kucherenko \cite{Kuc}. In other words, the following holds}
\begin{eqnarray}
\fl f\Big(\stackrel{2}{\boldsymbol{x}},-i\hbar \stackrel{1}{\frac{\partial}{\partial \boldsymbol{x}}}\Big)\Big(R(\boldsymbol{x})\,e^{\frac{i}{\hbar}S(\boldsymbol{x})}\Big)=R(\boldsymbol{x})\,e^{\frac{i}{\hbar}S(\boldsymbol{x})}\Bigg(f\Big(\boldsymbol{x},\frac{\partial S}{\partial \boldsymbol{x}}(\boldsymbol{x})\Big) \\ \nonumber 
\fl -i\hbar\,R(\boldsymbol{x})^{-1}\,\frac{\partial f}{\partial \boldsymbol{p}}\Big(\boldsymbol{x},\frac{\partial S}{\partial \boldsymbol{x}}(\boldsymbol{x})\Big)\cdot \frac{\partial R}{\partial \boldsymbol{x}}(\boldsymbol{x})-\frac{i\hbar}{2}\,{\rm tr}\,\frac{\partial^2 S}{\partial \boldsymbol{x}^2}(\boldsymbol{x})\,\Big(\frac{\partial^2 f}{\partial \boldsymbol{p}^2}\Big)\Big(\boldsymbol{x},\frac{\partial S}{\partial \boldsymbol{x}}(\boldsymbol{x})\Big)\Bigg)+r_1(\boldsymbol{x};\hbar) \ .
\end{eqnarray}
$${}$$

Based on the above result, we derive a commutation formula for semi-classical Weyl operators in phase space involving complex phase functions.
\\

\noindent \textbf{Theorem} (Complex Phase Commutation Formula for Semi-Classical Phase Space Weyl Operators)\textbf{.}$\,$-- \textit{Let $H\in C^\infty(\mathbb{R}^{2d})$ be a real-valued symbol on classical phase space and $\mathcal{H}(\boldsymbol{X},\boldsymbol{P})$ its Weyl symbol in double phase space $\mathbb{R}^{2d}\oplus \mathbb{R}^{2d}$, meaning, the Weyl symbol of} 
\begin{equation}
\mathcal{W}H\Big(\stackrel{\boldsymbol{\omega}}{\boldsymbol{q}},-i\hbar \stackrel{\boldsymbol{\omega}}{\frac{\partial}{\partial \boldsymbol{q}}}\Big)\mathcal{W}^{-1}=\mathcal{H}\Big(\stackrel{\boldsymbol{\omega}}{\boldsymbol{X}},-i\hbar \stackrel{\boldsymbol{\omega}}{\frac{\partial}{\partial \boldsymbol{X}}}\Big)
\end{equation}
\textit{so that $\mathcal{H}(\boldsymbol{X},\boldsymbol{P})=H\Big(\frac{\boldsymbol{X}}{2}-\textbf{J}\boldsymbol{P}\Big)$. Also, let $\varphi\in C^\infty_0(\mathbb{R}^{2d})$ be a complex-valued amplitude and $F\in C^\infty(\mathbb{R}^{2d})$ a complex-valued phase, with $F(\boldsymbol{X})=F_1(\boldsymbol{X})+i\,F_2(\boldsymbol{X})$, where $F_{1,2}$ are real valued. Then, for every $s\in \mathbb{N}_0$ and $\boldsymbol{X}\in{\rm supp}\,\varphi$, as $\hbar\rightarrow 0^+$, the following holds} 
\begin{eqnarray}
\fl \mathcal{H}\Big(\stackrel{\boldsymbol{\omega}}{\boldsymbol{X}},-i\hbar \stackrel{\boldsymbol{\omega}}{\frac{\partial}{\partial \boldsymbol{X}}}\Big)\Big(\varphi(\boldsymbol{X})\,e^{\frac{i}{\hbar}F(\boldsymbol{X})}\Big)=\varphi(\boldsymbol{X})\,e^{\frac{i}{\hbar}F(\boldsymbol{X})}\Bigg(\mathcal{H}\Big(\boldsymbol{X},\frac{\partial F}{\partial \boldsymbol{X}}(\boldsymbol{X})\Big)\\ \nonumber 
\fl -i\hbar\,\varphi(\boldsymbol{X})^{-1}\frac{\partial\mathcal{H}}{\partial \boldsymbol{P}}\Big(\boldsymbol{X},\frac{\partial F}{\partial \boldsymbol{X}}(\boldsymbol{X})\Big)\cdot \frac{\partial\varphi}{\partial \boldsymbol{X}}(\boldsymbol{X})\\ \nonumber 
-\frac{i\hbar}{2}\sum_{\boldsymbol{\mu}\in\mathbb{N}_0^{2d}\,:\,|\boldsymbol{\mu}|=2}\partial_{\boldsymbol{X}}^{\boldsymbol{\mu}} F(\boldsymbol{X})\partial_{\boldsymbol{P}}^{\boldsymbol{\mu}} \mathcal{H}\Big(\boldsymbol{X},\frac{\partial F}{\partial \boldsymbol{X}}(\boldsymbol{X})\Big)\Bigg)+\tilde r_1(\boldsymbol{X};\hbar)\\ \nonumber
\fl =\varphi(\boldsymbol{X})\,e^{\frac{i}{\hbar}F(\boldsymbol{X})}\Bigg(H\Big(\frac{\boldsymbol{X}}{2}-\textbf{J}\frac{\partial F}{\partial \boldsymbol{X}}(\boldsymbol{X})\Big)-i\hbar\,\varphi(\boldsymbol{X})^{-1}\textbf{J}\frac{\partial H}{\partial \boldsymbol{X}}\Big(\frac{\boldsymbol{X}}{2}-\textbf{J}\frac{\partial F}{\partial \boldsymbol{X}}(\boldsymbol{X})\Big)\cdot \frac{\partial\varphi}{\partial \boldsymbol{X}}(\boldsymbol{X})\\ \nonumber 
\fl -\frac{i\hbar}{2}\sum_{\boldsymbol{\mu},\boldsymbol{\nu}\in\mathbb{N}_0^{2d}\,:\,|\boldsymbol{\mu}|=2,\,|\boldsymbol{\nu}|\leq s}\frac{1}{\boldsymbol{\nu}!}\partial_{\boldsymbol{X}}^{\boldsymbol{\mu}} F(\boldsymbol{X})\Big(i\frac{\partial F_2}{\partial \boldsymbol{X}}(\boldsymbol{X})\Big)^{\boldsymbol{\nu}}(\textbf{J}\partial_{\boldsymbol{X}})^{\boldsymbol{\mu}+\boldsymbol{\nu}} H\Big(\frac{\boldsymbol{X}}{2}-\textbf{J}\frac{\partial F_1}{\partial \boldsymbol{X}}(\boldsymbol{X})\Big)\Bigg)\\ \nonumber 
+\tilde r_1(\boldsymbol{X};\hbar)\\ \nonumber
\fl =\varphi(\boldsymbol{X})\,e^{\frac{i}{\hbar}F(\boldsymbol{X})}\Bigg(H\Big(\frac{\boldsymbol{X}}{2}-\textbf{J}\frac{\partial F}{\partial \boldsymbol{X}}(\boldsymbol{X})\Big)-i\hbar\,\varphi(\boldsymbol{X})^{-1}\textbf{J}\frac{\partial H}{\partial \boldsymbol{X}}\Big(\frac{\boldsymbol{X}}{2}-\textbf{J}\frac{\partial F}{\partial \boldsymbol{X}}(\boldsymbol{X})\Big)\cdot \frac{\partial\varphi}{\partial \boldsymbol{X}}(\boldsymbol{X}) \\ \nonumber
\fl -\frac{i\hbar}{2}\sum_{\boldsymbol{\mu}\in\mathbb{N}_0^{2d}\,:\,|\boldsymbol{\mu}|=2}(\textbf{J}\partial_{\boldsymbol{X}})^{\boldsymbol{\mu}} F(\boldsymbol{X})\partial_{\boldsymbol{X}}^{\boldsymbol{\mu}} H\Big(\frac{\boldsymbol{X}}{2}-\textbf{J}\frac{\partial F}{\partial \boldsymbol{X}}(\boldsymbol{X})\Big)\Bigg)+\tilde r_1(\boldsymbol{X};\hbar) 
\end{eqnarray}
\textit{and the remainder $\tilde r_1$ as given by the Theorem of Kucherenko in \cite{Kuc}. In other words, the following holds}
\begin{eqnarray}
\fl H\Big(\stackrel{\boldsymbol{\omega}}{\frac{\boldsymbol{X}}{2}}+i\hbar\,\textbf{J}\,\stackrel{\boldsymbol{\omega}}{\frac{\partial}{\partial \boldsymbol{X}}}\Big)\Big(\varphi(\boldsymbol{X})\,e^{\frac{i}{\hbar}F(\boldsymbol{X})}\Big)=\varphi(\boldsymbol{X})\,e^{\frac{i}{\hbar}F(\boldsymbol{X})}\Bigg(H\Big(\frac{\boldsymbol{X}}{2}-\textbf{J}\frac{\partial F}{\partial \boldsymbol{X}}(\boldsymbol{X})\Big)\\ \nonumber
\fl -i\hbar\,\varphi(\boldsymbol{X})^{-1}\textbf{J}\frac{\partial H}{\partial \boldsymbol{X}}\Big(\frac{\boldsymbol{X}}{2}-\textbf{J}\frac{\partial F}{\partial \boldsymbol{X}}(\boldsymbol{X})\Big)\cdot \frac{\partial\varphi}{\partial \boldsymbol{X}}(\boldsymbol{X})\\ \nonumber 
\fl +\frac{i\hbar}{2}\,{\rm tr}\,\textbf{J}\frac{\partial^2 F}{\partial \boldsymbol{X}^2}(\boldsymbol{X})\textbf{J}\,\!\Big(\frac{\partial^2 H}{\partial \boldsymbol{X}^2}\Big)\Big(\frac{\boldsymbol{X}}{2}-\textbf{J}\frac{\partial F}{\partial \boldsymbol{X}}(\boldsymbol{X})\Big)\Bigg)+\tilde r_1(\boldsymbol{X};\hbar) \ .
\end{eqnarray}

\section{Dynamics of Semi-Classical Wave Packets}
\label{WPTr}

In the framework of the \textit{Anisotropic Gaussian Approximation}, we consider linear wave packet superpositions, using as a `building block' the semi-classical propagation of a single isotropic Gaussian wave packet, $U_tG_{(\boldsymbol{q},\boldsymbol{p})}$, which retains its Gaussian wave packet form, yet acquires an anisotropy in its phase. In particular, the initial state
\begin{equation}
\fl G_{(\boldsymbol{q},\boldsymbol{p})}(\boldsymbol{x};\hbar) =(\pi\hbar)^{-d/4}\exp\frac{i}{\hbar}\Big(\phi (\boldsymbol{q},\boldsymbol{p})+\boldsymbol{p}\cdot (\boldsymbol{x}-\boldsymbol{q})+\frac{i}{2}|\boldsymbol{x}-\boldsymbol{q}|^2\Big) 
\end{equation}
is evolved semi-classically to the anisotropic Gaussian wave packet moving along the orbit $\boldsymbol{X}=\boldsymbol{X}_t$, on which it micro-localized, on the Heisenberg scale \cite{BaDa,NSS}. 

There exists a multitude of different definitions of semi-classical wave packets, equivalent modulo phase $\phi (\boldsymbol{q},\boldsymbol{p})$, centered \cite{Fol,Lit2} at the origin $\phi (\boldsymbol{0})=0$. Here we have chosen $\phi (\boldsymbol{q},\boldsymbol{p})=\frac{\boldsymbol{p}\cdot\boldsymbol{q}}{2}$ as dictated by the use of Weyl quantization \cite{PaUr, Rob1}.

Semi-classical wave packets are asymptotic solution of the Cauchy problem 
\begin{equation}
\Big\|\Big(i\hbar\,\frac{\partial}{\partial t}-\widehat H \Big)G^{\scriptsize{\mathbfcal{Z}}}_{(\boldsymbol{q},\boldsymbol{p})}(\bullet,t;\hbar)\Big\|=O(\hbar^{3/2}) \ , \ \ t \in [0,T] \ , \ \ \hbar\rightarrow 0^+ \ 
\label{eq:schrapprox}
\end{equation}
with initial condition $G^{\scriptsize{\mathbfcal{Z}}} _{(\boldsymbol{q},\boldsymbol{p})}(\boldsymbol{x},0;\hbar)=G_{(\boldsymbol{q},\boldsymbol{p})}(\boldsymbol{x};\hbar)$

In order to construct such a solution we adopt the ansatz (see, e.g., \cite{NSS,Rob1})
\begin{eqnarray}
\fl G^{\scriptsize{\mathbfcal{Z}}} _{(\boldsymbol{q},\boldsymbol{p})}(\boldsymbol{x},t;\hbar)
=(\pi\hbar)^{-d/4}a(\boldsymbol{q},\boldsymbol{p},t)\\ \nonumber 
\fl \times \,\exp\frac{i}{\hbar}\Big(\frac{\boldsymbol{p}\cdot\boldsymbol{q}}{2}+A (\boldsymbol{q},\boldsymbol{p},t)+\boldsymbol{p}_t\cdot (\boldsymbol{x}-\boldsymbol{q}_t)+\frac{1}{2}(\boldsymbol{x}-\boldsymbol{q}_t)\cdot {\scriptsize \mathbfcal{Z}}(\boldsymbol{q},\boldsymbol{p},t)(\boldsymbol{x}-\boldsymbol{q}_t)\Big)
\end{eqnarray}
and we follow the derivation in Nazaikinksii et al. \cite{NSS}, Section 2.1. For this we need to express the Weyl operator $\widehat H$ in the normal representation which results in a straightforward asymptotic calculus.

In the normal representation we have 
\begin{equation}
{\rm Op}_{\mathbf{n}}(H)= H_{\scriptsize{\textbf{n}}} \Big(\stackrel{2}{\boldsymbol{x}}, -i\hbar \,\stackrel{1}{\frac{\partial}{\partial \boldsymbol{x}}}\Big) 
\end{equation}
where
\begin{equation}
H_{\scriptsize{\textbf{n}}} (\boldsymbol{q},\boldsymbol{p}):=\mathbf{\sigma_n} (\widehat H)(\boldsymbol{q},\boldsymbol{p})= \exp\Big(-\frac{i\hbar}{2}\frac{\partial^2}{\partial \boldsymbol{q}\partial\boldsymbol{p}}\Big) H(\boldsymbol{q},\boldsymbol{p})
\end{equation}
so that $\mathbf{\sigma_w} (\widehat{H}):=H(\boldsymbol{q},\boldsymbol{p})$ and $\widehat H = \mathbf{{\rm Op}_w}(H) = \mathbf{{\rm Op}_n}(H_{\scriptsize{\textbf{n}}})$. The relation between the symbols of operators in different representations is systematically derived in \cite{BeSh}, \cite{NSS} (see also Section $\ref{weylops}$).

We note that in order to satisfy the Schr\"{o}dinger equation $(\ref{eq:schrapprox})$ to the order 
$O(\hbar^{3/2})$ it is sufficient to include only the first two terms in the expansion of the exponential operator, that is
\begin{equation}
H_{\scriptsize{\textbf{n}}}(\boldsymbol{q},\boldsymbol{p})\sim H(\boldsymbol{q},\boldsymbol{p})-\frac{i\hbar}{2}\,{\rm tr}\,\frac{\partial^2 H}{\partial \boldsymbol{q}\partial\boldsymbol{p}}(\boldsymbol{q},\boldsymbol{p}) \ .
\end{equation}

Then, the separation of orders leads to a system of differential equations along the Hamiltonian orbit emanating from $(\boldsymbol{q},\boldsymbol{p})$ \cite{Lit1,NSS,Rob1}, the \textit{characteristic system} 
\begin{eqnarray} 
\fl O(1): & \ \ \ \ \ \ \ \ \frac{d A}{dt}=\boldsymbol{p}_t\cdot \frac{d\boldsymbol{q}_t}{dt}-H \ , & \ \ \ \ \ \ \ \ A (0)=0 \\ \nonumber
\fl O(\hbar^{1/2}): & \ \ \ \ \ \ \ \ \frac{d\boldsymbol{q}_t}{dt}=\frac{\partial H}{\partial \boldsymbol{p}} \ , \ \ \frac{d\boldsymbol{p}_t}{dt}=-\frac{\partial H}{\partial \boldsymbol{q}} \ , & \ \ \ \ (\boldsymbol{q}_0,\boldsymbol{p}_0)=(\boldsymbol{q},\boldsymbol{p})\\ \nonumber
\fl O(\hbar):& \ \ \ \ \ \ \ \ \frac{da}{dt}+\frac{1}{2}\,{\rm tr}\Big(H_{\boldsymbol{pp}}\,\mathbfcal{Z}+H_{\boldsymbol{pq}}\Big)a=0 \ , &\ \ \ \ \ \ \ \ a(0)=1\\ \nonumber
\fl & \ \ \ \ \ \ \ \ \frac{d\mathbfcal{Z}}{dt}+\mathbfcal{Z}\,H_{\boldsymbol{pp}}\,\mathbfcal{Z}+H_{\boldsymbol{qp}}\,\mathbfcal{Z}+\mathbfcal{Z}\,H_{\boldsymbol{pq}}+H_{\boldsymbol{qq}}=\textbf{0} \ , & \ \ \ \ \ \ \ \ \mathbfcal{Z}(0)=i\textbf{I} \ . 
\end{eqnarray}

This estimate induces a particular dynamics for the `parametrizing' functions; $(\boldsymbol{q}_t,\boldsymbol{p}_t)$ are considered to define a simple smooth phase space curve, $A$ is a real valued function, while the matrix $\mathbfcal{Z}={\rm Re}\,\mathbfcal{Z}+i\,{\rm Im}\,\mathbfcal{Z}$ is symmetric, $\mathbfcal{Z}^{T}=\mathbfcal{Z}$, and has positive definite imaginary part, ${\rm Im}\,\mathbfcal{Z}\succ 0$.

The real-valued phase $A$ is the \textit{phase space action}, related to Hamilton's principal function
\begin{equation}
 A (\boldsymbol{q},\boldsymbol{p}, t)=\int\displaylimits_{0}^{t} \boldsymbol{p}_\tau \cdot \frac{d \boldsymbol{q}_\tau}{d\tau} \,d\tau-H(\boldsymbol{q},\boldsymbol{p})\,t
\end{equation}
satisfying
\begin{equation}
\frac{\partial A}{\partial \boldsymbol{q}}=-\boldsymbol{p}+\boldsymbol{p}_t^{T}\,\frac{\partial \boldsymbol{q}_t}{\partial \boldsymbol{q}} \ , \ \ \frac{\partial A}{\partial \boldsymbol{p}}=\boldsymbol{p}_t^{T}\,\frac{\partial \boldsymbol{q}_t}{\partial \boldsymbol{p}} \ .
\end{equation} 

The \textit{amplitude} $a$, satisfying the transport equation, is given by
\begin{equation}
a(\boldsymbol{q},\boldsymbol{p},t)=\exp\Bigg(-\frac{1}{2}\int\displaylimits_0^t{\rm tr}\Big(H_{\boldsymbol{pp}}\,\mathbfcal{Z}+H_{\boldsymbol{pq}}\Big)\,d\tau\Bigg)
\end{equation}
where the Hessian elements are taken along the transported point $(\boldsymbol{q}_t,\boldsymbol{p}_t)$. 
Note that the second term $H_{\boldsymbol{pq}}$ in the exponential term for $a$ arises due to the second term in the approximation of 
the symbol $H_{\scriptsize{\textbf{n}}}$ and it appears since $\widehat H$ is a Weyl operator.

The \textit{anisotropy matrix} $\mathbfcal{Z}={\rm Re}\,\mathbfcal{Z}+i\,{\rm Im}\,\mathbfcal{Z}$,
satisfying the matrix Riccati equation,
 is symmetric, $\mathbfcal{Z}^{T}=\mathbfcal{Z}$, it has positive definite imaginary part, ${\rm Im}\,\mathbfcal{Z}\succ 0$, and it essentially controls the direction, the shape and spreading of the propagated state (\cite{HHL,NSS,Rob1}), which in turn are related to the separation dynamics of initially nearby Hamiltonian orbits. These dynamics are essentially common to all Gaussian narrow beam dynamics, i.e., dynamics for asymptotic solutions concentrated with a Gaussian profile about the characteristics \cite{Kat}.

The wave packet form of $G^{\scriptsize{\mathbfcal{Z}}}_{(\boldsymbol{q},\boldsymbol{p})}$ is guaranteed, as by beginning at $\mathbfcal{Z}(0)=i\textbf{I}$, $\mathbfcal{Z}$ remains symmetric with positive definite imaginary part for all times. The Riccati equation guarantees unitarity of the semi-classical time evolution, in the sense that $\|G^{\scriptsize{\mathbfcal{Z}}}_{(\boldsymbol{q},\boldsymbol{p})}\|=1$ for all times, $t\geq 0$. 

The Riccati equation for the anisotropy matrix is equivalent to the \textit{variational system} governing the stability of the Hamiltonian flow \cite{HHL,NSS,Rob1}, which can be attained by varying Hamilton's equations with respect to the initial points $(\boldsymbol{q},\boldsymbol{p})$ and taking appropriate complex combinations \cite{NSS}, 
\begin{equation}
\frac{d}{dt}\left(
\begin{array}{ccc}
\textbf{A} \\
\textbf{B} 
\end{array}\right)=\left(\begin{array}{ccc}
H_{\boldsymbol{pq}} & H_{\boldsymbol{pp}} \\
-H_{\boldsymbol{qq}} & -H_{\boldsymbol{qp}} 
\end{array}\right)\left(
\begin{array}{ccc}
\textbf{A} \\
\textbf{B} 
\end{array}\right) 
\label{eq:varsys}
\end{equation}
where the Hessian elements evaluated along the orbit, at $(\boldsymbol{q}_t,\boldsymbol{p}_t)$. 

Following Maslov \cite{Mas2}, for the solutions of the variational system, we coin the terms \textit{position variational matrix} for $\textbf{A}$ and \textit{momentum variational matrix} for $\textbf{B}$. The anisotropy matrix decomposes with respect to the variational matrices as 
\begin{equation}
\mathbfcal{Z}=\textbf{B}\textbf{A}^{-1}
\end{equation}
where $\mathbfcal{Z}=\mathbfcal{Z}(\boldsymbol{q},\boldsymbol{p},t)$ or $\mathbfcal{Z}(t)$, and $\textbf{A}=\textbf{A}(\boldsymbol{q},\boldsymbol{p},t)$ or $\textbf{A}(t)$ and $\textbf{B}=\textbf{B}(\boldsymbol{q},\boldsymbol{p},t)$ or $\textbf{B}(t)$. 
As ${\rm Im}\,\mathbfcal{Z}\succ 0$, ${\rm Im}\,\mathbfcal{Z}$ possesses a unique square root, by which that variational matrices $\textbf{A},\textbf{B}$ can be uniquely expressed 
\begin{equation}
\textbf{A}=({\rm Im}\,\mathbfcal{Z})^{-1/2} \ , \ \ \textbf{B}=({\rm Im}\,\mathbfcal{Z}^{-1})^{-1/2} \ .
\end{equation}

Utilizing the relations between the variational matrices we simplify the expression of the semi-classical anisotropic Gaussian wave packet, $G^{\scriptsize{\mathbfcal{Z}}}_{(\boldsymbol{q},\boldsymbol{p})}$. Considering that 
\begin{eqnarray}
\fl \int\displaylimits_0^t{\rm tr}\Big(H_{\boldsymbol{pp}}\,\mathbfcal{Z}+H_{\boldsymbol{pq}}\Big)\,d\tau=\int\displaylimits_0^t{\rm tr}\Big(H_{\boldsymbol{pp}}\,\textbf{B}\textbf{A}^{-1}+H_{\boldsymbol{pq}}\,\textbf{A}\textbf{A}^{-1}\Big)\,d\tau\\ \nonumber
=\int\displaylimits_0^t{\rm tr}\Big(\frac{d \textbf{A}}{d\tau}\textbf{A}^{-1}\Big)\,d\tau={\rm tr}\,\log\,\textbf{A}
\end{eqnarray}
we obtain the equivalent expression of the amplitude \cite{BaDa,BBT}
\begin{equation}
a(\boldsymbol{q},\boldsymbol{p},t)=\frac{1}{\sqrt{\det\,\textbf{A}(\boldsymbol{q},\boldsymbol{p},t)}} \ .
\end{equation}

\section{The Stationary Complex Phase Formula}
\label{CSPT}

In this appendix, we give the fundamental result on the asymptotic expansion of rapidly oscillating integrals with complex phase function, a version of the method of stationary phase for complex phase functions, following the works of Fedoryuk \cite{Fed1,Fed2}, Nazaikinskii et al. \cite{NOSS}, \cite{NSS}, Treves \cite{Tre} and Mishchenko et al. \cite{MSS}. Sj\"{o}strand and Melin \cite{MeSj} reach the same result along somewhat different lines. 

Let $f\in C^{\infty}(\mathbb{R}^m,\mathbb{C})$, $s\in\mathbb{N}_0$ and $\boldsymbol{x},\boldsymbol{y}\in\mathbb{R}^m$. We define the \textit{$s$-analytic extension of $f$ to $\mathbb{C}^m$,} for $s\geq 1$, as
\begin{equation}
\fl {}^s \! f(\boldsymbol{x}+i\boldsymbol{y}):=\sum_{r=0}^s\frac{1}{r!}\left(i\boldsymbol{y}\cdot \frac{\partial}{\partial \boldsymbol{x}}\right)^rf(\boldsymbol{x}) = \sum_{\boldsymbol{\mu}\in\mathbb{N}_0^d\,:\,|\boldsymbol{\mu}|\leq s}\frac{1}{\boldsymbol{\mu}!}(i\boldsymbol{y})^{\boldsymbol{\mu}}\partial_{\boldsymbol{x}}^{\boldsymbol{\mu}} f(\boldsymbol{x})
\end{equation}
while ${}^0 \! f(\boldsymbol{x}+i\boldsymbol{y}):=f(\boldsymbol{x})$.

Note that, for $f$ real analytic, ${}^s \! f(\boldsymbol{x}+i\boldsymbol{y})$ is a segment of the Taylor series of $f(\boldsymbol{z})$ near the set ${\rm Im}\,\boldsymbol{z}=\boldsymbol{0}$.

In the interest on notational simplification, we follow the convention not to denote the order of the analytic extension, leaving it implicit. A function defined over a real domain given with a complex argument is understood as an analytic extension of appropriate order, unless necessary to prevent inconsistencies when considering differentation. This, for complex $\boldsymbol{x}+i\boldsymbol{y}$, $f(\boldsymbol{x}+i\boldsymbol{y})$ is to be understood as $\,{}^s\! f(\boldsymbol{x}+i\boldsymbol{y})$, and will be called simply the \textit{almost analytic extension of $f$}.
\\

\noindent \textbf{Theorem.}$\,$-- \textit{For $s\geq 1$ and $f$ as above, the following hold} 
\begin{eqnarray}
\fl \frac{\partial \ {}^s\! f}{\partial \boldsymbol{z}}(\boldsymbol{z})=\,{}^s\!\Big(\frac{\partial f}{\partial \boldsymbol{x}}\Big)(\boldsymbol{z})-\frac{1}{2}\frac{1}{s!}\Big(i\boldsymbol{y}\cdot \frac{\partial}{\partial \boldsymbol{x}}\Big)^s\frac{\partial f}{\partial \boldsymbol{x}}(\boldsymbol{x})\\ \nonumber 
=\,{}^s\!\Big(\frac{\partial f}{\partial \boldsymbol{x}}\Big)(\boldsymbol{z})-\frac{1}{2}\sum_{|\boldsymbol{\mu}|=s}\frac{1}{\boldsymbol{\mu}!}(i\boldsymbol{y})^{\boldsymbol{\mu}}\partial_{\boldsymbol{x}}^{\boldsymbol{\mu}}\frac{\partial f}{\partial \boldsymbol{x}}(\boldsymbol{x})
\end{eqnarray}
\textit{and} 
\begin{eqnarray}
\fl \frac{\partial^2 \ {}^s\! f}{\partial \boldsymbol{z}^2}(\boldsymbol{z})=\,{}^s\!\Big(\frac{\partial^2 f}{\partial \boldsymbol{x}^2}\Big)(\boldsymbol{z})-\frac{3}{4}\frac{1}{s!}\Big(i\boldsymbol{y}\cdot \frac{\partial}{\partial \boldsymbol{x}}\Big)^s\frac{\partial^2 f}{\partial \boldsymbol{x}^2}(\boldsymbol{x})\\ \nonumber
-\frac{1}{4}\frac{1}{(s-1)!}\Big(i\boldsymbol{y}\cdot \frac{\partial}{\partial \boldsymbol{x}}\Big)^{s-1}\frac{\partial^2 f}{\partial \boldsymbol{x}^2}(\boldsymbol{x})\\ \nonumber 
=\,{}^s\!\Big(\frac{\partial^2 f}{\partial \boldsymbol{x}^2}\Big)(\boldsymbol{z})-\frac{3}{4}\sum_{|\boldsymbol{\mu}|=s}\frac{1}{\boldsymbol{\mu}!}(i\boldsymbol{y})^{\boldsymbol{\mu}}\partial_{\boldsymbol{x}}^{\boldsymbol{\mu}} \frac{\partial^2 f}{\partial \boldsymbol{x}^2}(\boldsymbol{x})\\ \nonumber 
-\frac{1}{4}\sum_{|\boldsymbol{\mu}|=s-1}\frac{1}{\boldsymbol{\mu}!}(i\boldsymbol{y})^{\boldsymbol{\mu}}\partial_{\boldsymbol{x}}^{\boldsymbol{\mu}} \frac{\partial^2 f}{\partial \boldsymbol{x}^2}(\boldsymbol{x})
\end{eqnarray}
\textit{while, as $|\boldsymbol{y}|\rightarrow 0^+$,} 
\begin{equation}
\frac{\partial \ {}^s\! f}{\partial \boldsymbol{z}}(\boldsymbol{z})=\,{}^s\!\Big(\frac{\partial f}{\partial \boldsymbol{x}}\Big)(\boldsymbol{z})+O(|\boldsymbol{y}|^s)=\frac{\partial f}{\partial \boldsymbol{x}}(\boldsymbol{x})+O(|\boldsymbol{y}|)
\end{equation}
\textit{and} 
\begin{equation}
\frac{\partial^2 \ {}^s\! f}{\partial \boldsymbol{z}^2}(\boldsymbol{z})=\,{}^s\!\Big(\frac{\partial^2 f}{\partial \boldsymbol{x}^2}\Big)(\boldsymbol{z})+O(|\boldsymbol{y}|^{s-1})=\frac{\partial^2 f}{\partial \boldsymbol{x}^2}(\boldsymbol{x})+O(|\boldsymbol{y}|) \ .
\end{equation}
$${}$$

It is straightforward to incur that $\frac{\partial \ {}^s \! f}{\partial \bar {\boldsymbol{z}}}(\boldsymbol{z})=O(|\boldsymbol{y}|^s)$ as $|\boldsymbol{y}|\rightarrow 0^+$, point-wise in $\mathbb{C}^m$, and that this equation has a unique solution $\boldsymbol{z}$, modulo $O(|\boldsymbol{y}|^{s-1})$. 

Before we proceed to the main result on the asymptotic behavior of oscillatory integrals, we note the case of Gaussian integrals. For non-singular $\textbf{M}\in\mathbb{C}^{m\times m}$ with $\textbf{M}^{T}=\textbf{M}$, ${\rm Re}\,\textbf{M}\succ 0$ and $\boldsymbol{p}\in\mathbb{C}^m$, we can give a closed form expression for the Gaussian integral, where $\hbar>0$,
\begin{equation}
\fl \bigg(\frac{1}{2\pi\hbar}\bigg)^{m/2}\int_{\mathbb{R}^m} e^{-\frac{1}{2\hbar}\boldsymbol{x}\cdot \textbf{\scriptsize{M}}\boldsymbol{x}+\frac{i}{\hbar}\boldsymbol{p}\cdot \boldsymbol{x}}\,d\boldsymbol{x}=\frac{1}{\sqrt{\det\, \textbf{M}}}\,\exp\Big(-\frac{1}{2\hbar}\,\boldsymbol{p}\cdot \textbf{M}^{-1}\boldsymbol{p}\Big) \ . 
\end{equation}
In the above we take $\sqrt{\bullet}$ to be the principal branch of the square root function.

We now consider the asymptotic expansion of oscillating integrals of the form 
\begin{equation}
I(\boldsymbol{p};\hbar):= \Big(\frac{i}{2\pi\hbar}\Big)^{m/2}\int_{\mathbb{R}^m} a(\boldsymbol{x})\,e^{\frac{i}{\hbar}\Phi(\boldsymbol{p},\boldsymbol{x})}\,d\boldsymbol{x}
\end{equation}
for small $\hbar>0$ and $\boldsymbol{p}\in \mathbb{R}^n$.
\\

\noindent \textbf{Theorem} (Nazaikinksii et al. \cite{NOSS}, \cite{NSS})\textbf{.}$\,$-- \textit{Consider the oscillatory integral $I(\boldsymbol{p};\hbar)$, with $s\geq 1$, amplitude $a:\mathbb{R}^{m}\rightarrow \mathbb{C}$ and phase $\Phi:\mathbb{R}^n\times \mathbb{R}^m \rightarrow \mathbb{C}$ of class $C^s$ in $\boldsymbol{p}$, where $a$ is compactly supported and $\Phi$ possesses an everywhere non-negative imaginary part on ${\rm supp}\,a$, ${\rm Im}\,\Phi\geq 0$, so that, for given $\boldsymbol{p}\in\mathbb{R}^n$, the equations}
\begin{equation}
{\rm Im}\,\Phi(\boldsymbol{p},\boldsymbol{x})=0 \ , \ \ \frac{\partial \Phi}{\partial \boldsymbol{x}}(\boldsymbol{p},\boldsymbol{x})=\textbf{0}
\end{equation}
\textit{have at most a single solution on ${\rm supp}\,a$, while the Hessian matrix}
\begin{equation}
\frac{\partial^2\Phi}{\partial\boldsymbol{x}^2}(\boldsymbol{p},\boldsymbol{x})
\end{equation}
\textit{is non-singular for all $(\boldsymbol{p},\boldsymbol{x})\in\mathbb{R}^n\times {\rm supp}\,a$. Then, for all $r=0,\ldots, s$, the following estimate holds as $\hbar\rightarrow 0^+$}
\begin{equation}
I(\boldsymbol{p};\hbar)=\frac{{}^r\!a(\boldsymbol{p},\boldsymbol{z}(\boldsymbol{p}))}{\sqrt{{\rm det}\,-\frac{\partial^2 \ {}^r\!\Phi}{\partial \boldsymbol{z}^2}(\boldsymbol{p},\boldsymbol{z}(\boldsymbol{p}))}}\,e^{\frac{i}{\hbar}\,{}^r\!\Phi(\boldsymbol{p},\boldsymbol{z}(\boldsymbol{p}))}\Bigg(1+o(\hbar)\Bigg)
\end{equation}
\textit{where $\sqrt{\bullet}$ is the principal branch of the square root function, $\,{}^r\!a(\boldsymbol{p},\boldsymbol{z})$ and $\,{}^r\!\Phi(\boldsymbol{p},\boldsymbol{z})$ are the $r$-analytic extensions of $a(\boldsymbol{p},\boldsymbol{x})$ and $\Phi(\boldsymbol{p},\boldsymbol{x})$, respectively, to the complex variable $\boldsymbol{z}=\boldsymbol{x}+i\boldsymbol{y}$, and $\boldsymbol{z}=\boldsymbol{z}(\boldsymbol{p})$ is the unique complex solution of the equation} 
\begin{equation}
\frac{\partial \ {}^r\!\Phi}{\partial \boldsymbol{z}}(\boldsymbol{p},\boldsymbol{z})=\textbf{0} \ . 
\end{equation}
\textit{In addition, there exists $c>0$ such that for all $\boldsymbol{p}\in\mathbb{R}^n$} 
\begin{equation}
{\rm Im}\,\Phi(\boldsymbol{p},\boldsymbol{z}(\boldsymbol{p}))\geq c\,|{\rm Im}\,\boldsymbol{z}(\boldsymbol{p})|^2 \ .
\end{equation}

\bigskip

{\bf Bibliography}
\\

\end{document}